\documentclass[%
 reprint,
superscriptaddress,
 amsmath,amssymb,
 aps,
]{revtex4-2}

\usepackage{url}
\usepackage{graphicx}% Include figure files
\usepackage{dcolumn}% Align table columns on decimal point
\usepackage{bm}% bold math
\usepackage{slashbox}
\usepackage[caption=false]{subfig}
\usepackage{float}
\usepackage{xcolor}
\usepackage{amsmath}
\usepackage{amssymb}
\usepackage{array}
\newcolumntype{P}[1]{>{\centering\arraybackslash}p{#1}}
\usepackage{multirow}
\usepackage{hhline}
\begin{document}

\preprint{APS/123-QED}

\title{Space-Time Modulated Metasurfaces with Spatial Discretization: Free-Space N-path Systems}% Force line breaks with \\

\author{Zhanni Wu}
\affiliation{%
Department of Electrical Engineering and Computer Science, University of Michigan, Ann Arbor, Michigan 48109-2122, USA %
}%
%\altaffiliation[Also at ]{Department of Electrical Engineering and Computer Science, University of Michigan}%Lines break automatically or can be forced with \\
%\homepage{agrbic@umich.edu}
% \email{agrbic@umich.edu}

%\author{Younes Ra'di}
 %\homepage{younes.radi@utexas.edu}
\author{Cody Scarborough}%
\affiliation{%
Department of Electrical Engineering and Computer Science, University of Michigan, Ann Arbor, Michigan 48109-2122, USA %
}
\author{Anthony Grbic}
%\homepage{agrbic@umich.edu}
\affiliation{%
Department of Electrical Engineering and Computer Science, University of Michigan, Ann Arbor, Michigan 48109-2122, USA %
}%
%\affiliation{
 %Third institution, the second for Charlie Author
%}%
%\author{Anthony Grbic}

%\affiliation{%
% Department of Electrical Engineering and Computer Science, University of Michigan, Ann Arbor, MI, USA% \textbackslash\textbackslash
%}%


\date{\today}% It is always \today, today,

\begin{abstract}
This work theoretically and experimentally studies metasurfaces with spatially-discrete, traveling-wave modulation (SD-TWM). A representative metasurface is considered consisting of columns of time-modulated subwavelength unit cells, referred to as stixels. SD-TWM is achieved by enforcing a time delay between temporal waveforms applied to adjacent columns. In contrast to the continuous traveling-wave modulation commonly assumed in studies of space-time metasurfaces, here the modulation is spatially discretized. In order to account for the discretized spatial modulation, a modified Floquet analysis is introduced based on a new boundary condition that has been derived for SD-TWM structures. The modified Floquet analysis separates the scattered field into its macroscopic and microscopic variations. The reported theoretical and experimental results reveal that the electromagnetic behavior of an SD-TWM metasurface can be categorized into three regimes. For electrically-large spatial modulation periods, the microscopic field variation across each stixel can be neglected. In this regime, the space-time metasurface allows simultaneous frequency translation and angular deflection. When the spatial modulation period on the metasurface is electrically small, the microscopic variation results in new metasurface capabilities such as subharmonic mixing. When the spatial modulation period of the metasurface is wavelength-scale, the metasurface allows both subharmonic mixing and angular deflection to be achieved simultaneously. To verify our analysis, a dual-polarized, spatio-temporally modulated metasurface, is developed and measured at X-band frequencies. 
%\begin{description}
%You may use the \texttt{description} %environment to structure your abstract;
%use the optional argument of the \verb+\item+ %command to give the category of each item. 
%\end{description}
\end{abstract}

\pacs{Valid PACS appear here}% PACS, the Physics and Astronomy
\maketitle

%\tableofcontents

%%%%%%%%%%%%%%%%%%%%%%%%%%%%%%%%%%%%%%%%%%%%%%%%%%
%%%%%%%%%%%%%%%%%%%%%%%%%%%%%%%%%%%%%%%%%%%%%%%%%%
%%%%%%%%%% Section 1
%%%%%%%%%%%%%%%%%%%%%%%%%%%%%%%%%%%%%%%%%%%%%%%%%%
%%%%%%%%%%%%%%%%%%%%%%%%%%%%%%%%%%%%%%%%%%%%%%%%%%

\section{\label{sec:Intro}INTRODUCTION%\protect\\ The line
%break was forced \lowercase{via} \textbackslash\textbackslash
}

Metasurfaces are two dimensional structures textured at a subwavelength scale to achieve tailored control of electromagnetic waves. Developments in tunable electronic components have allowed dynamic control over the electromagnetic properties of metasurfaces. Electronic devices such as varactors, transistors and MEMS \cite{manjappa2018reconfigurable, roy2018dynamic, cong2017active, wu2019tunable, lonvcar2018reflective}, as well as 2D and phase change materials \cite{wang2016optically, folland2018reconfigurable, fan2015tunable,yao2014electrically} can be integrated into metasurfaces to tune their electric, magnetic and magneto-electric responses. Often, the properties of a metasurface are spatially modulated to shape electromagnetic wavefronts and achieve focusing, beam-steering, and polarization control \cite{estakhri2016recent, kildishev2013planar, pfeiffer2014bianisotropic, yu2014flat}. By incorporating tunable elements into their design, the properties of metasurfaces can also be modulated in time \cite{liu2018huygens, 8854331, 8888738}. While spatial modulation redistributes the plane-wave spectrum of the scattered field, temporal modulation provides control over the frequency spectrum. Simultaneous spatial and temporal variation is known as spatio-temporal modulation, and has recently been applied to metasurfaces \cite{shaltout2015time, shaltout2019spatiotemporal, hadad2015space, caloz2019spacetime, salary2018electrically, zang2019nonreciprocal, taravati2019generalized, shi2016dynamic, wang2020theory,taravati2018advanced}. Space-time modulation can simultaneously allow frequency conversion and beam steering and shaping. It can also be used to break Lorentz reciprocity and enable magnetless nonreciprocal devices including gyrators, circulators and isolators \cite{taravati2017nonreciprocal, estep2014magnetic, sounas2013giant, doerr2014silicon,taravati2020full}.

Most papers to date have examined space-time metasurfaces with a traveling-wave modulation that is a continuous function of space. The unit cell size of metasurfaces is assumed to be deeply subwavelength and the effect of the unit cell size on the space-time modulated system is neglected. Here, metasurfaces with spatially-discrete, space-time modulation are considered. In particular, spatially-discrete traveling-wave modulation (SD-TWM) of metsurfaces is explained. SD-TWM structures have a spatial modulation period made up of a finite number ($N$) of unit cells. The unit cells within each spatial modulation period will be referred to as stixels \cite{stixel}: space-time pixels of the structure. They are the smallest indivisible element of SD-TWM structures. In SD-TWM, the modulation of adjacent stixels is staggered by a time delay of $T_p/N$, where $T_p$ is the temporal modulation period of each stixel.

The SD-TWM scheme is reminiscent of that used in N-path circuit networks \cite{scar2020,richards2012analysis, reiskarimian2016magnetic, ghaffari2011tunable, shrestha2006multipath, franks1960alternative}. These circuit networks have gained strong interest in recent years within the circuits community for their ability to realize high fidelity filters \cite{ghaffari2011tunable,von1983integrated,ghaffari2010differential} and non-reciprocal devices \cite{weiss2017low,zhou2016receiver,doerr2014silicon,reiskarimian2016magnetic}. Leveraging this close relation, and exploiting our earlier work on N-path circuits \cite{scar2020} and the interpath relation \cite{cody_meta}, a modified Floquet boundary condition governing fields on SD-TWM metasurfaces is reported here. The modified Floquet boundary condition accounts for the frequency dependent Bloch wavenumber induced across a stixel. 

The electromagnetic fields on linear electromagnetic structures that vary periodically in space and time can be modeled with a double Floquet expansion in both time and space. This generalized method of analysis can be used to solve the field distribution on any linear, periodic space-time modulated structure. It has been used recently to compute the fields scattered from space-time gratings \cite{taravati2019generalized}. For SD-TWM structures, the interpath relation can be used to compress the double Floquet expansion, leading to a number of simplifications. Applying the interpath relation to the double Floquet expansion of fields reveals that a number of the field amplitudes become zero. Specifically, it is shown that the interpath relation reduces the number of unknown Floquet harmonics by the number of paths ($N$): stixels within a spatial modulation period. Therefore, the interpath relation dramatically reduces the numerical complexity of analyzing these structures. 
Applying the interpath relation to the double Floquet expansion also allows the fields scattered by the SD-TWM structure to be separated into its macroscopic and microscopic spatial variations. The macroscopic variation describes the field variation between respective points in stixels separated by a spatial modulation wavelength. This macroscopic variation is consistent with the field variation of continuously modulated structures and is characterized by the modulation frequency $\omega_p$ and modulation wavenumber $\beta_p=2\pi/\lambda_p$. The microscopic field variation accounts for the field variation across a stixel. The microscopic variation is characterized by spatial harmonics of the modulation wavenumber $\beta_d =2N\pi/\lambda_p=2\pi/d_0$, where $d_0$ is the stixel (unit cell) size.

Separating the scattered field into its macroscopic and microscopic variation, through the interpath relation, also reveals the underlying physics of SD-TWM structures. The field relations reveal three regions of operation for SD-TWM metasurfaces, those with small, large and wavelength-scale  spatial modulation periods. When the spatial modulation period is small with respect to the wavelength of radiation, low-order temporal harmonics (frequencies) are evanescent and adhere to the metasurface, while higher-order temporal harmonics specularly scatter due to aliasing. This can lead to specular subharmonic mixing. Such a phenomenon cannot be predicted by continuous space-time analysis of metasurfaces. SD-TWM metasurfaces with large spatial modulation periods behave much like spatially continuous traveling wave modulated structures. Scattered waves can be deflected to different angles, and non-reciprocal responses observed. Finally, wavelength-scale spatial periods allow both deflection/retroreflection (scattering to non-specular directions) and subharmonic frequency mixing. In other words, large and small spatial period effects can be observed simultaneously. By clearly defining each regime and the effect of unit cell size, this research can provide guidance on practically designing traveling-wave modulated metasurfaces with spatially discrete unit cells.

\begin{figure}[t]
\centering
\includegraphics[width=0.95\columnwidth]{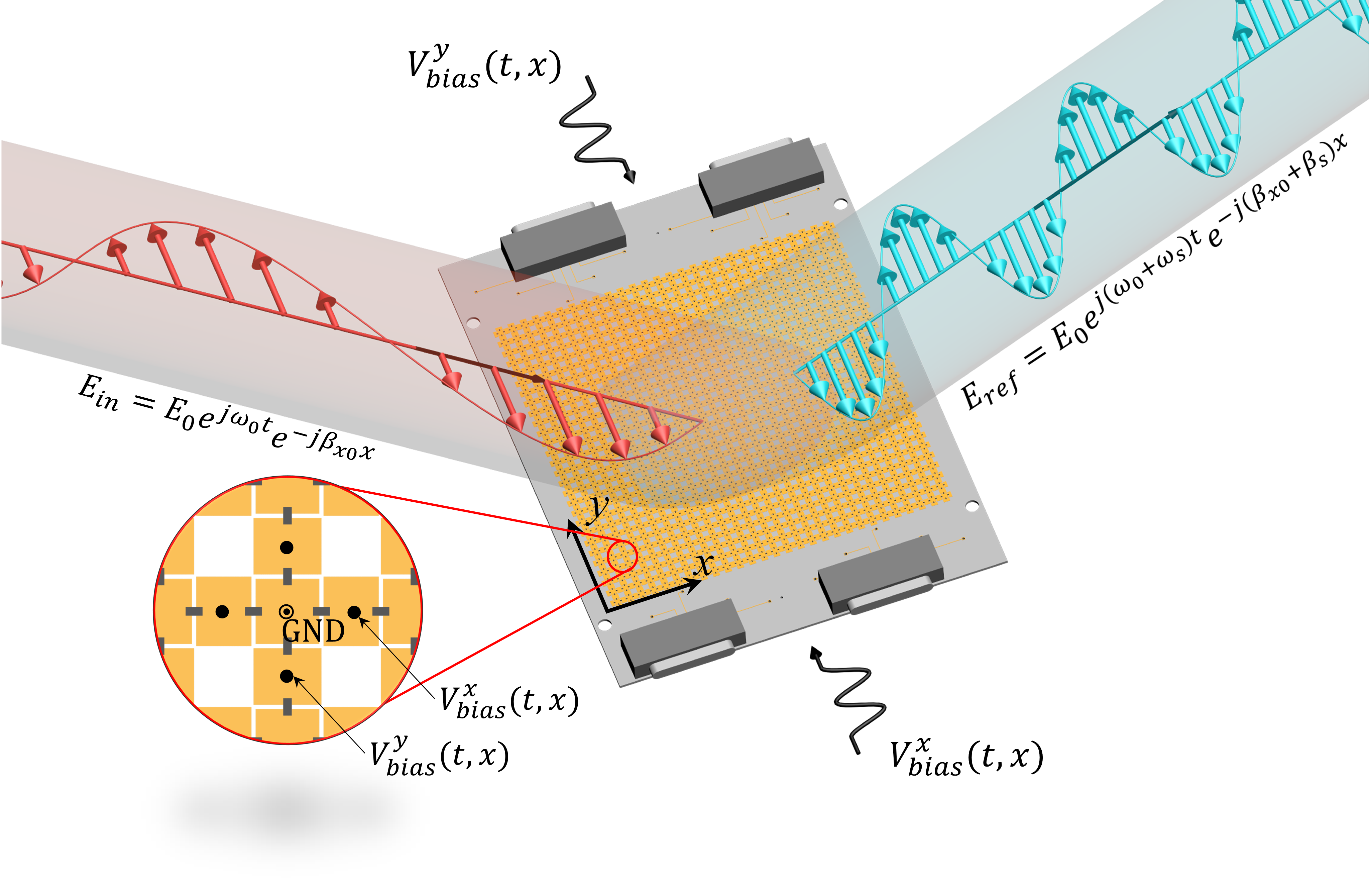}
% where an .eps filename suffix will be assumed under latex, 
% and a .pdf suffix will be assumed for pdflatex; or what has been declared
% via \DeclareGraphicsExtensions.
\caption{A spatially-discrete, traveling-wave modulated (SD-TWM) metasurface. Based on the spatial modulation period, three regimes of operations are supported by the metasurface.}
\label{fig:DPRscheme}
\end{figure}

To aid in the discussion of SD-TWM metasurfaces, a representative example is considered throughout the paper. The reflective metasurface is shown in Fig. \ref{fig:DPRscheme}, and is based on the high impedance surface \cite{sievenpiper2003two}. Varactor diodes are surface-mounted onto the metasurface, acting as tunable capacitances. This metasurface was first presented in \cite{8888738} where the varactor diodes on each column were temporally modulated with the same bias signal. In \cite{8888738}, a sawtooth reflection phase in time was applied to the metasurface, resulting in Doppler-like (serrodyne) frequency translation. Structures of a similar design have been subsequently reported in \cite{8901456}. The metasurface consists of discrete unit cells. Each column of unit cells on the metasurface can be independently temporally modulated, allowing space-time modulation of its reflection phase. Here, only spatially-discrete, traveling-wave modulations of the metasurface are considered. The metasurface is homogenized within each stixel to allow for semi-analytical scattering analysis. Measurements of an experimental metasurface prototype are presented in order to verify the theory (spectral domain analysis) used to predict the electromagnetic response of SD-TWM metasurfaces.

%%%%%%%%%%%%%%%%%%%%%%%%%%%%%%%%%%%%%%%%%%%%%%%%%%
%%%%%%%%%%%%%%%%%%%%%%%%%%%%%%%%%%%%%%%%%%%%%%%%%%
%%%%%%%%%% Section 2
%%%%%%%%%%%%%%%%%%%%%%%%%%%%%%%%%%%%%%%%%%%%%%%%%%
%%%%%%%%%%%%%%%%%%%%%%%%%%%%%%%%%%%%%%%%%%%%%%%%%%

\section{\label{sec:Analysis} Analysis of a free-space N-path modulated metasurface}

% Overview

A representative metasurface is introduced in this section. The SD-TWM metasurface consists of tunable capacitive sheet above a grounded dielectric substrate, as shown in Fig. \ref{fig:DPRscheme}. It is a reflective, electrically-tunable impedance surface \cite{sievenpiper2003two}, which allows independent control of the reflection phase for two orthogonal polarizations. The capacitive sheet is realized as an array of metallic patches interconnected by varactor diodes.  It can be modulated in both space (discretely) and time with a bias signal that is applied through the metallic vias that penetrate the substrate.

\begin{figure}[t]
\subfloat[\label{sfig:UnitCell}]{%
  \includegraphics[clip,width=0.56\columnwidth]{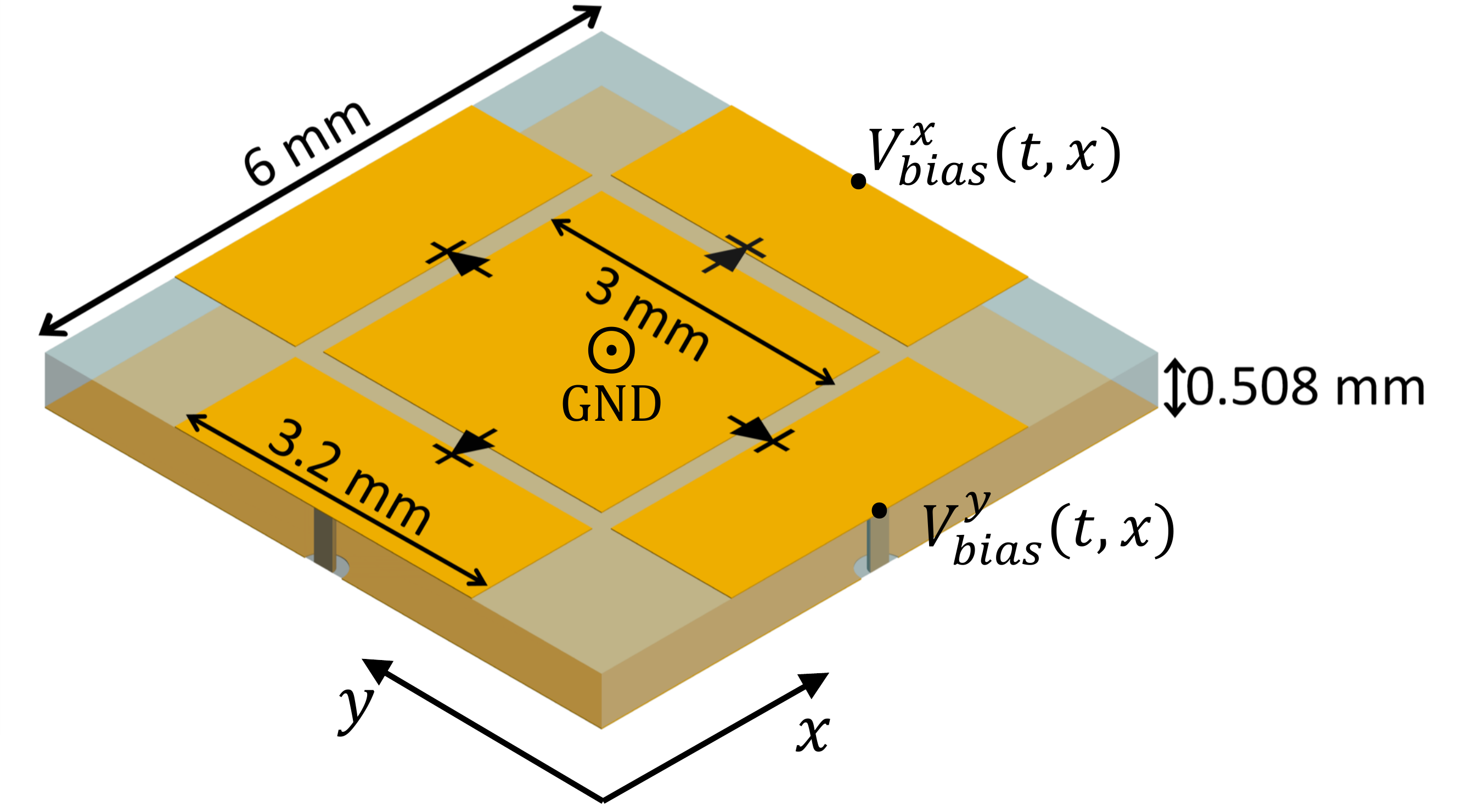}%
}\hfill
\subfloat[\label{sfig:ECircuit}]{%
  \includegraphics[clip,width=0.4\columnwidth]{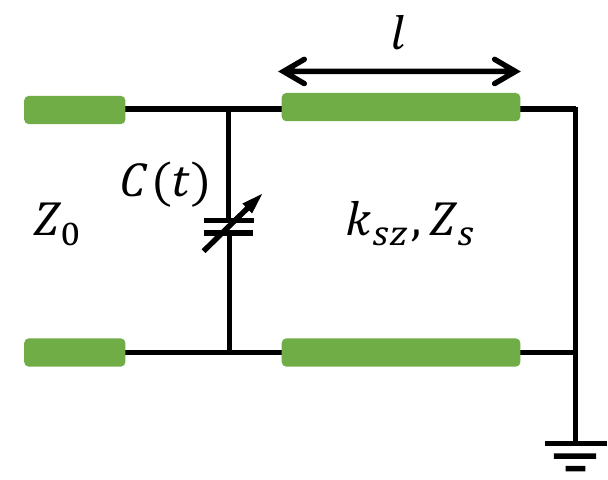}%
}\quad
\caption{(a) A stixel (unit cell) of the dual-polarized, spatio-temporally modulated metasurface. (b) The equivalent circuit model for each polarization.}
\label{fig:BFSCircuit}
\end{figure}

\begin{figure}[t]
\subfloat[\label{sfig:TE}]{%
  \includegraphics[clip,width=0.49\columnwidth]{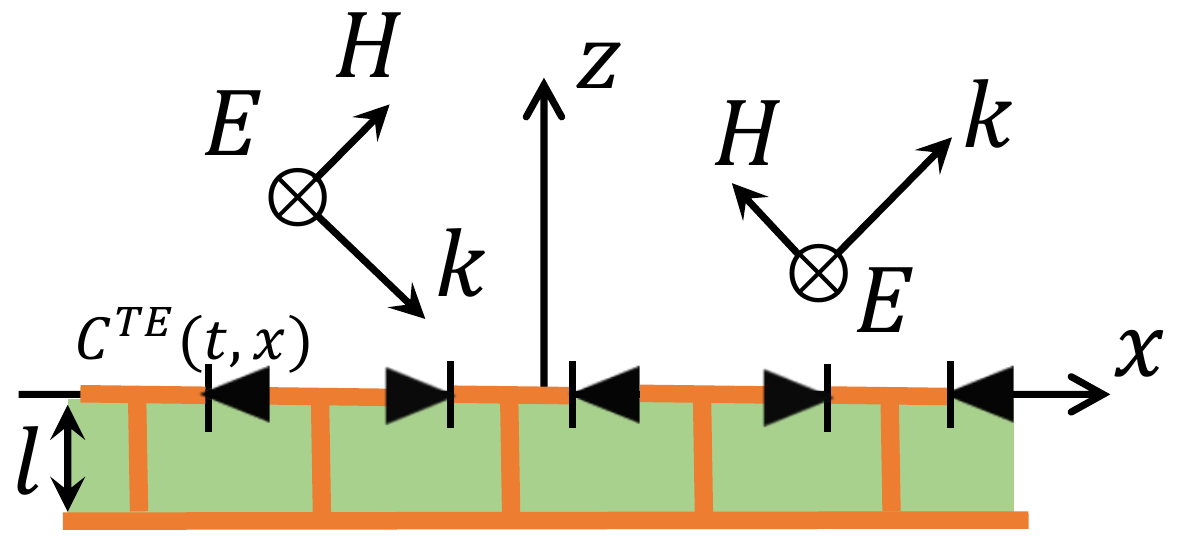}%
}\hfill
\subfloat[\label{sfig:TM}]{%
  \includegraphics[clip,width=0.49\columnwidth]{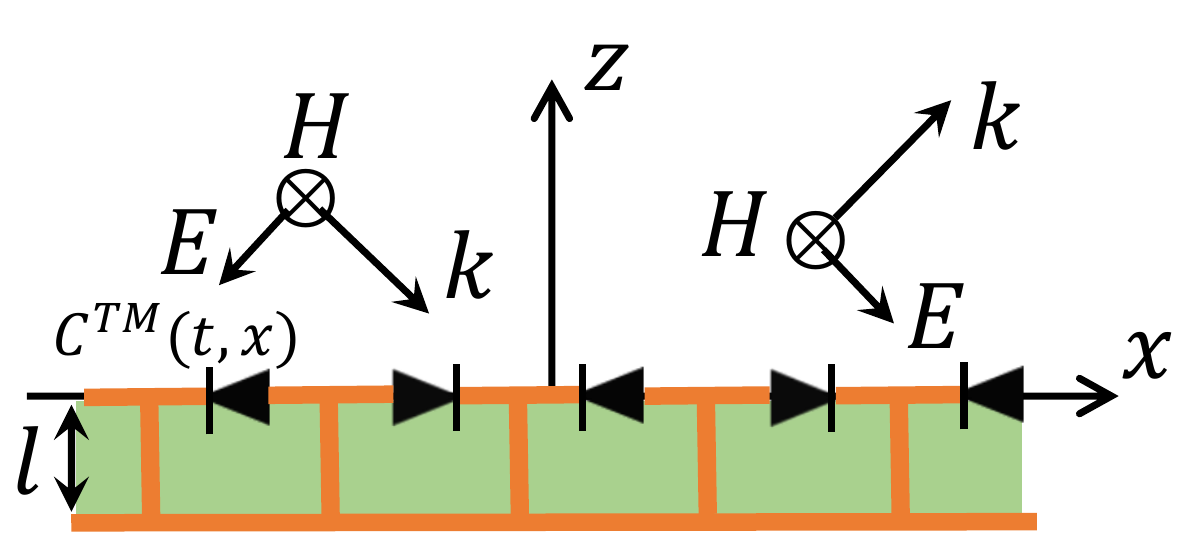}%
}\hfill
\caption{Cross sections of the obliquely illuminated time-modulated metasurface, under (a) TE polarization, and (b) TM polarization.}
\label{fig:CrossSection}
\end{figure}

% Physical unit cell description
The stixel (unit cell) of the designed metasurface is shown in Fig. \ref{sfig:UnitCell}. The varactor diodes connecting the metallic patches are biased through the vias located at the edges of the unit cells, while the the via at the central patch is connected to ground. The remainder of the biasing network is shielded behind the ground plane. The biasing network and diode orientations allow the reflection phase of the metasurface to be independently tuned for two orthogonal (TE and TM) polarizations.  Bias waveforms $V_{bias}^x(t,x)$ and $V_{bias}^y(t,x)$, shown in Fig. \ref{fig:DPRscheme}, control the sheet capacitance for the two orthogonal polarizations. A detailed description of the fabrication and biasing network are provided in Section \ref{sec:Fab}. A cross section of the metasurface is shown in Fig. \ref{fig:CrossSection} under TE and TM excitations. The biasing vias can be seen perforating the dielectric substrate.

In this section, we derive a semi-analytical procedure for computing the response of the SD-TWM metasurface shown in Fig. \ref{fig:DPRscheme}. This includes the homogenization of the representative metasurface; an introduction to the interpath relation that relates the fields of each stixel within a spatial period of SD-TWM metasurface; and the resulting modified Floquet expansion of the field for an SD-TWM structure. 

\subsection{Homogenization of the proposed SD-TWM metasurface} \label{sec:analysis_homo}

In the analysis that follows, the stixels of the metasurface are homogenized to simplify the analysis and allow conclusions to be made about the general behavior of SD-TWM systems. Within each stixel, the metallic patches interconnected by varactor diodes will be treated as a capacitive sheet. This sheet can be modulated in time, independently of adjacent stixels. The dielectric substrate, that is perforated by vias ($-l<z<0$), will be treated as a uniaxial anisotropic material, with a relative permittivity tensor \cite{luukkonen2009effects}:
\begin{equation}\label{Substrate}
% \bf{\epsilon_r}=\begin{pmatrix}
 \overline{\overline{\epsilon_r} }
=\begin{pmatrix}
\epsilon_h & 0 & 0\\
0 & \epsilon_h & 0\\
0 & 0 & \epsilon_{zz}
\end{pmatrix},
\end{equation}
where $\epsilon_h$ is the relative permittivity of the host medium and $\epsilon_{zz}$ is the effective relative permittivity along the vias. Since the metasurface is electrically thin at the operating frequency of 10 GHz ($l=0.016\lambda=0.508$ mm), a local model can be used to describe the perforated substrate \cite{luukkonen2009effects}:
\begin{equation}\label{LocalModel}
\epsilon_{zz}=\epsilon_h(1-\frac{k_p^2}{k_0^2\epsilon_h}),
\end{equation}
where $k_p=541.81$ rad$\cdot$m$^{-1}$ is the plasma wavenumber of the wire medium extracted from a full-wave simulation of the unit cell shown in Fig. \ref{sfig:UnitCell}, and $k_0$ is the free-space wavenumber of the incident wave. The anisotropic substrate supports TE (ordinary mode) and TM (extraordinary mode) polarizations. The normal wavenumber for each polarization in the substrate can be written as,
\begin{align}
k_{sz}^{TE}&=\sqrt{k_0^2\epsilon_h-k_x^2}, \label{KsubTE} \\
k_{sz}^{TM}&=\sqrt{k_0^2\epsilon_h-k_x^2\frac{\epsilon_h}{\epsilon_{zz}}},
\label{TKsubTM} 
\end{align}
where $k_x$ is the tangential wavenumber of the incident wave. 

% Circuit Equivalence
Each stixel can be modeled with the shunt resonator circuit depicted in Fig. \ref{sfig:ECircuit}. The circuit model consists of a tunable capacitance (representing the capacitive sheet) backed by shorted transmission-line section (representing the conductor-backed dielectric substrate) that acts as an inductance. As a result, the bias voltage applied to the varactor diodes can be used to tune the reflection phase. The phase range of this topology is $2\pi - \Delta \phi$, where $\Delta \phi$ is the round trip phase delay through the substrate. Details on the phase range of the realized metasurface are provided in Section \ref{sec:Fab}. In this paper, two different reflection phase waveforms are considered. A sawtooth reflection phase with respect to time is studied, which allows serrodyne frequency translation  \cite{8854331, 8888738}, as well as a sinusoidal reflection phase with respect to time.

As mentioned earlier, each column of unit cells can be biased independently, allowing for space-time modulation along a single ($x$) axis.
As a result, the homogenized model consists of capacitive strips whose widths are given by the stixel size $d_0=\lambda_0/5=6$ mm, where $\lambda_0$ is the wavelength in free space at 10 GHz. The capacitance seen by each polarization can be controlled independently and is uniform over the strip.

\subsection{Interpath relation for SD-TWM metasurfaces} \label{sec:analysis_stm}

\begin{figure}
\subfloat[\label{sfig:TopoDiode}]{%
  \includegraphics[clip,width=0.8\columnwidth]{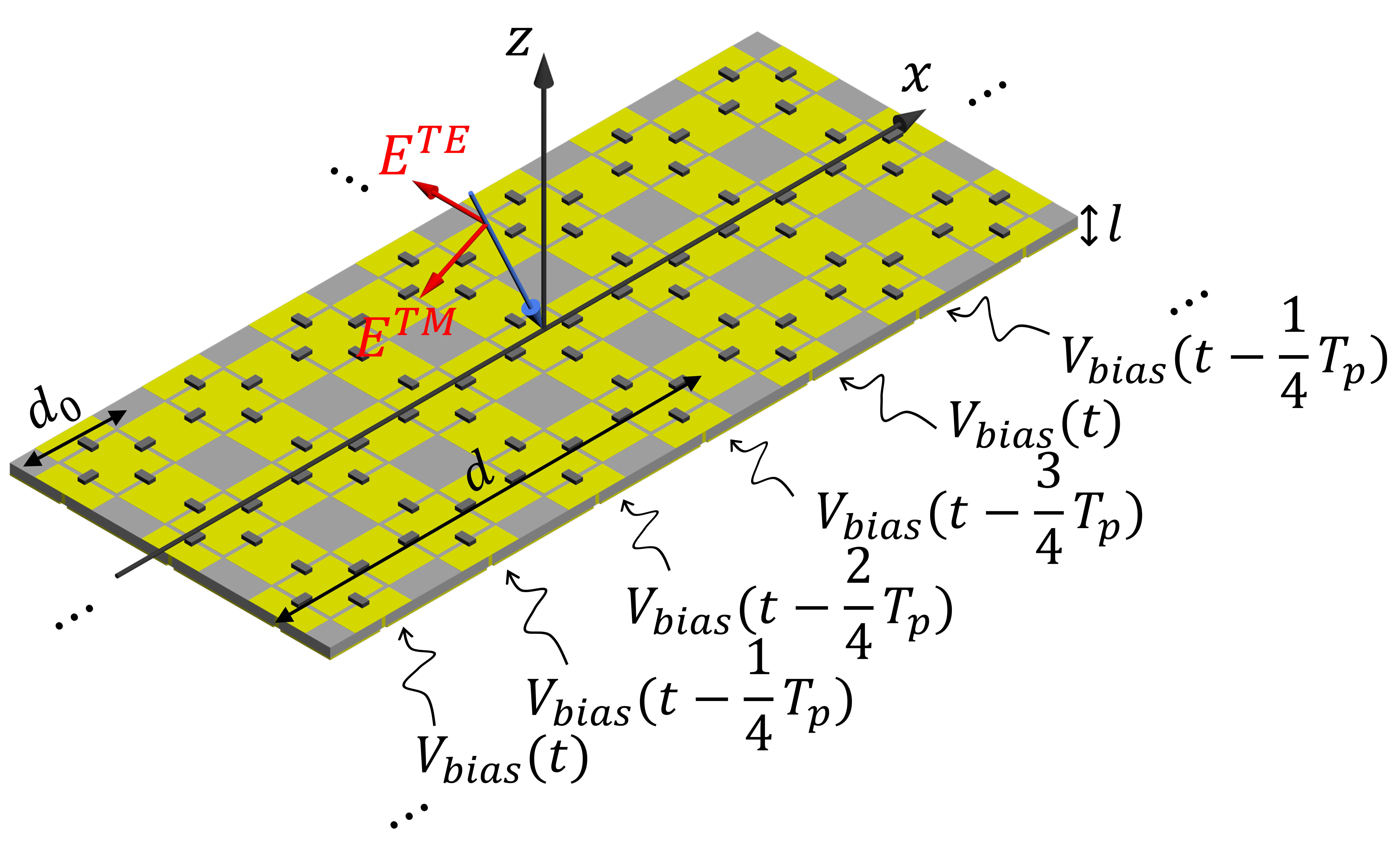}%
}\hfill
\subfloat[\label{sfig:TopoCap}]{%
  \includegraphics[clip,width=0.8\columnwidth]{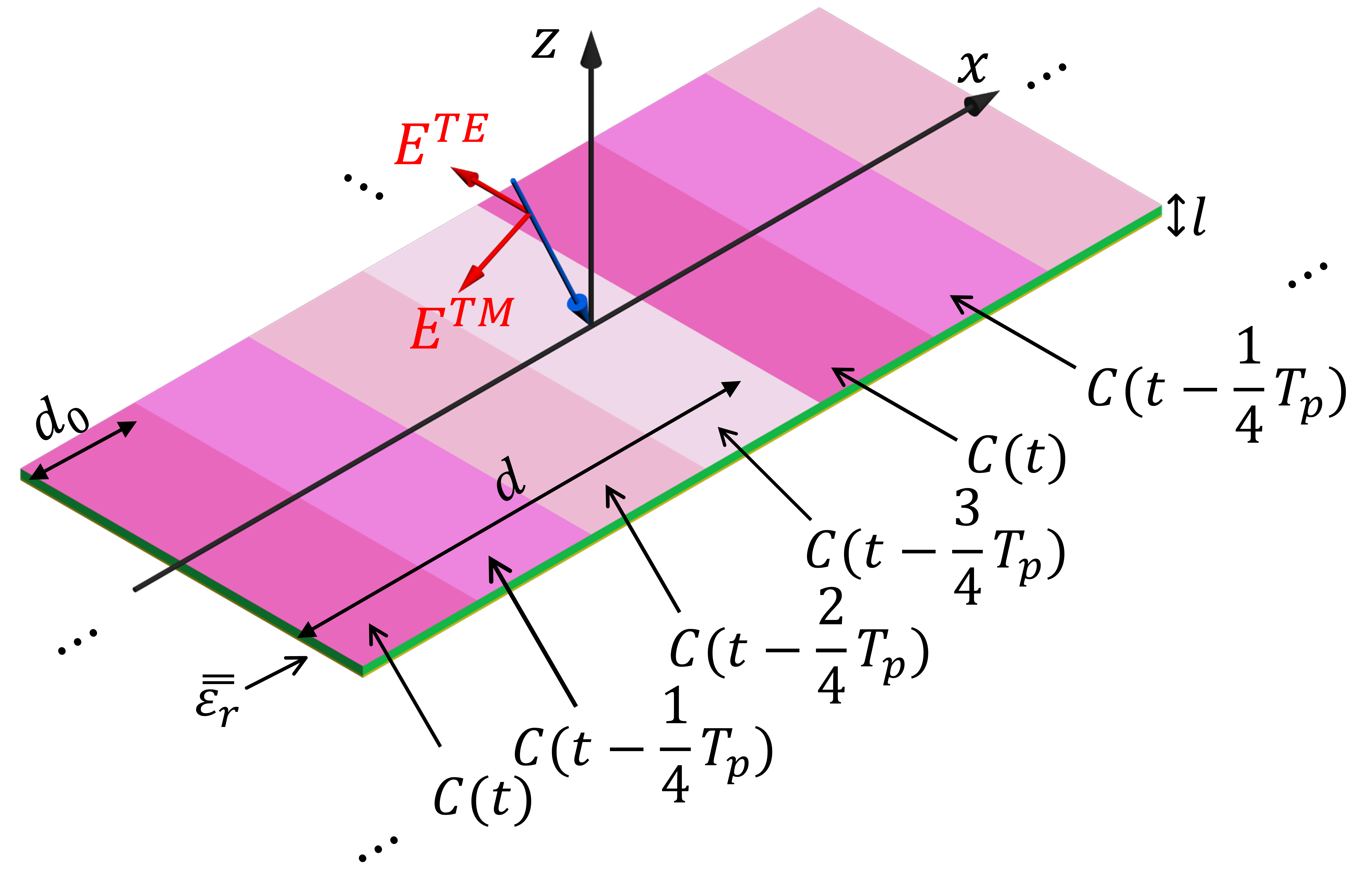}%
}\quad
\caption{The modulation scheme of the SD-TWM metasurface. (a) The designed metasurface with an SD-TWM bias. The stixel size is $d_0=6$ mm and the spatial modulation period is $d$. (b) Homogenized model of the SD-TWM metasurface. The substrate is modeled as a uniaxial, anisotropic material.}
\label{fig:Topoloty}
\end{figure}

\begin{figure}[t]
\subfloat[\label{sfig:SpaceMod2}]{%
  \includegraphics[clip,width=0.49\columnwidth]{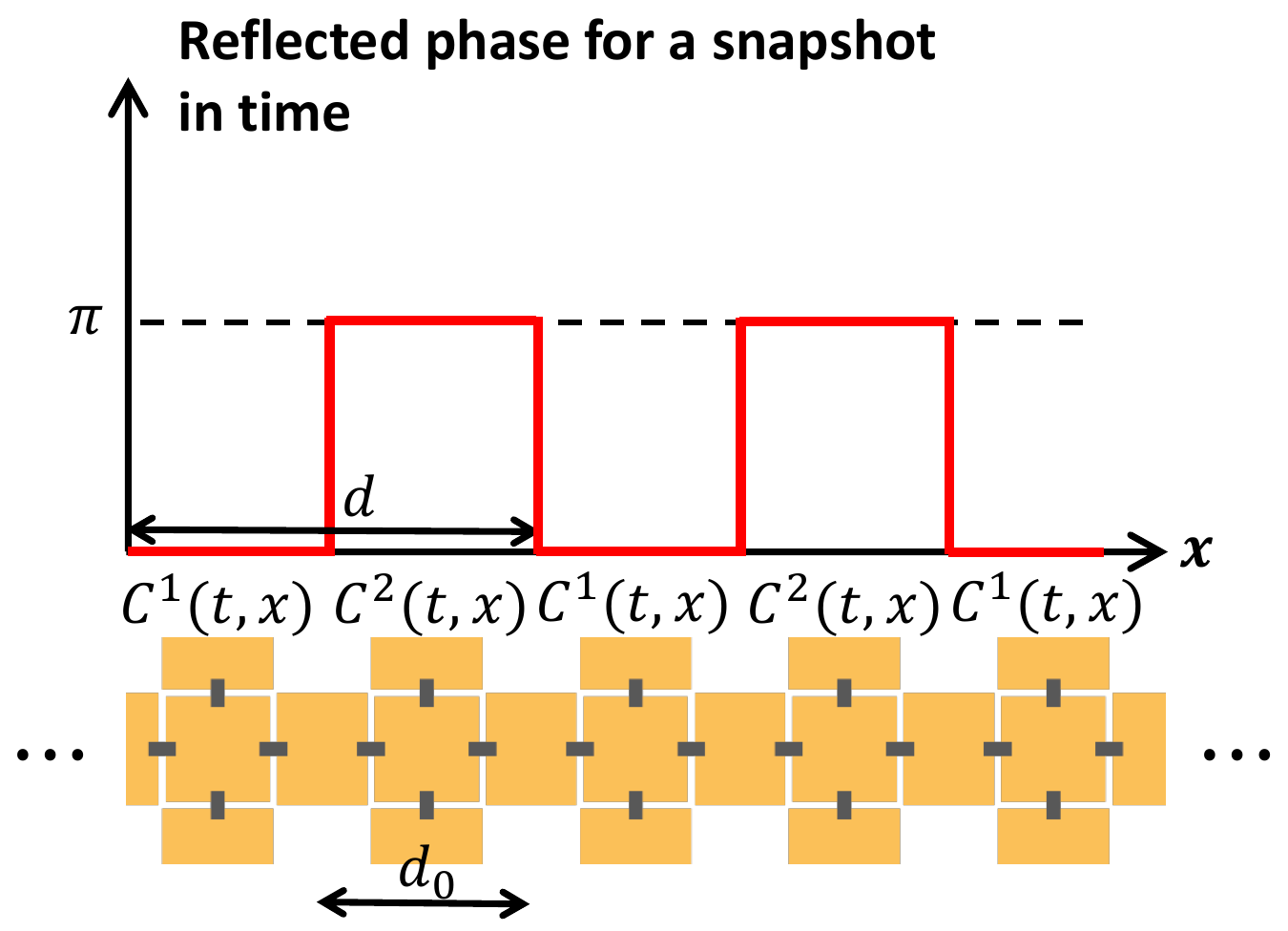}%
}\hfill
\subfloat[\label{sfig:SpaceMod3}]{%
  \includegraphics[clip,width=0.49\columnwidth]{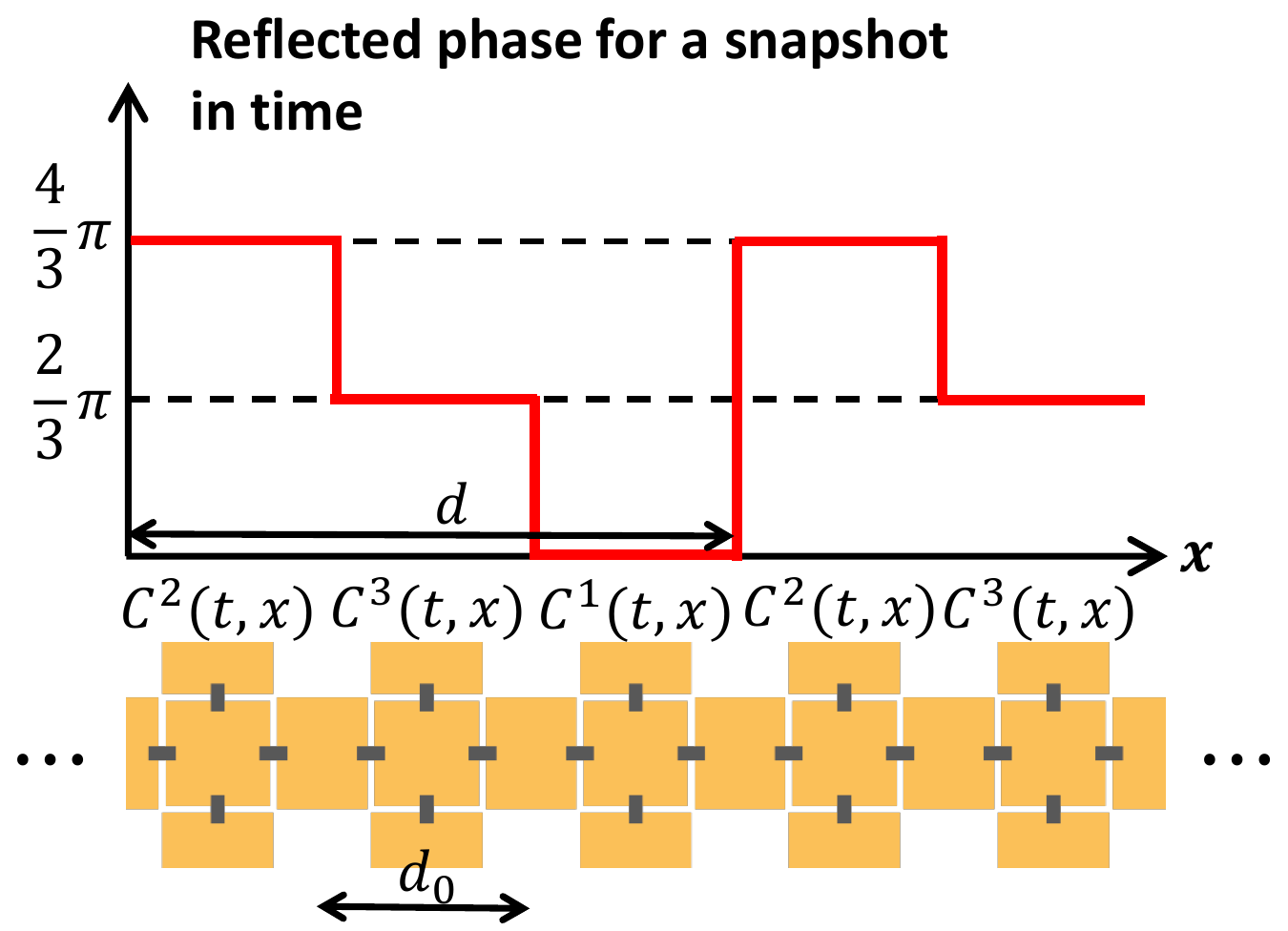}%
}\quad
\caption{Modulation scheme for the SD-TWM metasurface. The stixel (unit cell) dimension is $d_0 = 6$ mm and the spatial modulation period is $d = Nd_0$. (a) A 2-path modulation scheme ($d = 2d_0$). (b) A 3-path modulation scheme ($d = 3d_0$).}
\label{fig:SpaceMode}
\end{figure}

The SD-TWM modulation can be applied to the metasurface by introducing a time delay between the capacitance modulation applied to adjacent columns. This modulation scheme is shown in Fig. \ref{fig:Topoloty}. There are $N$ columns of stixels within one spatial modulation period $d$. Since the stixel size of the metasurface is fixed ($d_0=\lambda_0/5$), the total spatial period $d=Nd_0$ can be controlled by changing the path number $N$. This impresses a modulation wavenumber $\beta_p=2\pi/d=2\pi/(N\lambda_0/5)$ onto the metasurface. $N$ adjacent stixels are modulated with bias signals staggered in time by an interval $T_p/N$, where $T_p=1/f_p$ is the temporal modulation period. In other words, the capacitance modulation of the metasurface satisfies the following relationship:
\begin{equation}\label{CapRelation}
C\left(t,x\right)=C\left(t-\frac{T_p}{N},x-\frac{d}{N}\right).
\end{equation}
The spatial variation across one stixel is approximated to be uniform. Examples of 2- and 3-path spatio-temporal modulation schemes are shown in Fig. \ref{fig:SpaceMode}. At any given time, the spatial variation of the reflection phase is a discretized sawtooth (blazed grating) ranging from $0$ to approximately $2\pi$ over a period $d=Nd_0$. In this paper, the capacitance modulation on each stixel is chosen to either produce a sawtooth or sinusoidal reflection phase with respect to time. The procedure for obtaining the capacitance modulation based on the desired reflection phase is outlined in Supplemental Material I. An example is shown in Fig. \ref{fig:SpaceTimeMode}, where each stixel generates a staggered sawtooth reflection phase in time. 

As mentioned, one can view each column of stixels as a path in a N-path network. In contrast to an N-path circuit, the paths (columns of stixels) are not connected to a common input and output. Instead, each path is displaced by a subwavelength distance $d_0=6$ mm ($d_0=\lambda_0/5$) from its adjacent paths. The N-path symmetry of the SD-TWM metasurface establishes an interpath relation between the fields on adjacent paths \cite{cody_meta}. Accounting for the incident tangential wavenumber, the total electric field on the metasurface must satisfy
\begin{equation}\label{NpathRelationshipE}
E(t,x,y,z)=e^{j(\frac{\omega_0T_p}{N}-\frac{k_xd}{N})}E(t-\frac{T_p}{N},x-\frac{d}{N},y,z),
\end{equation}
where $\omega_0$ is the radial frequency of the incident wave, and $k_x$ is the tangential wavenumber of the incident wave.

\begin{figure}[t]
\subfloat[\label{sfig:SpaceTime2}]{%
  \includegraphics[clip,width=0.49\columnwidth]{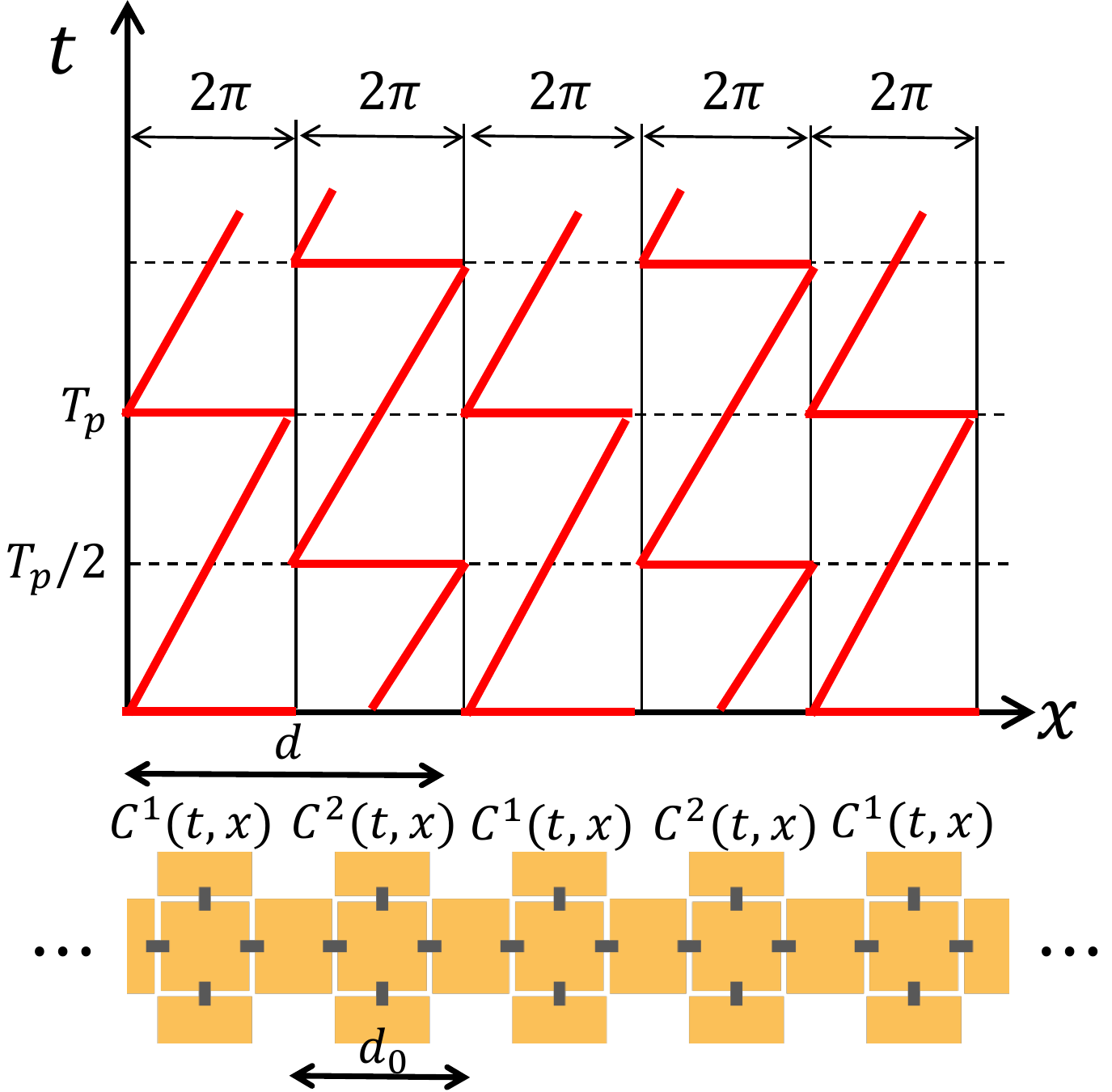}%
}\hfill
\subfloat[\label{sfig:SpaceTime3}]{%
  \includegraphics[clip,width=0.49\columnwidth]{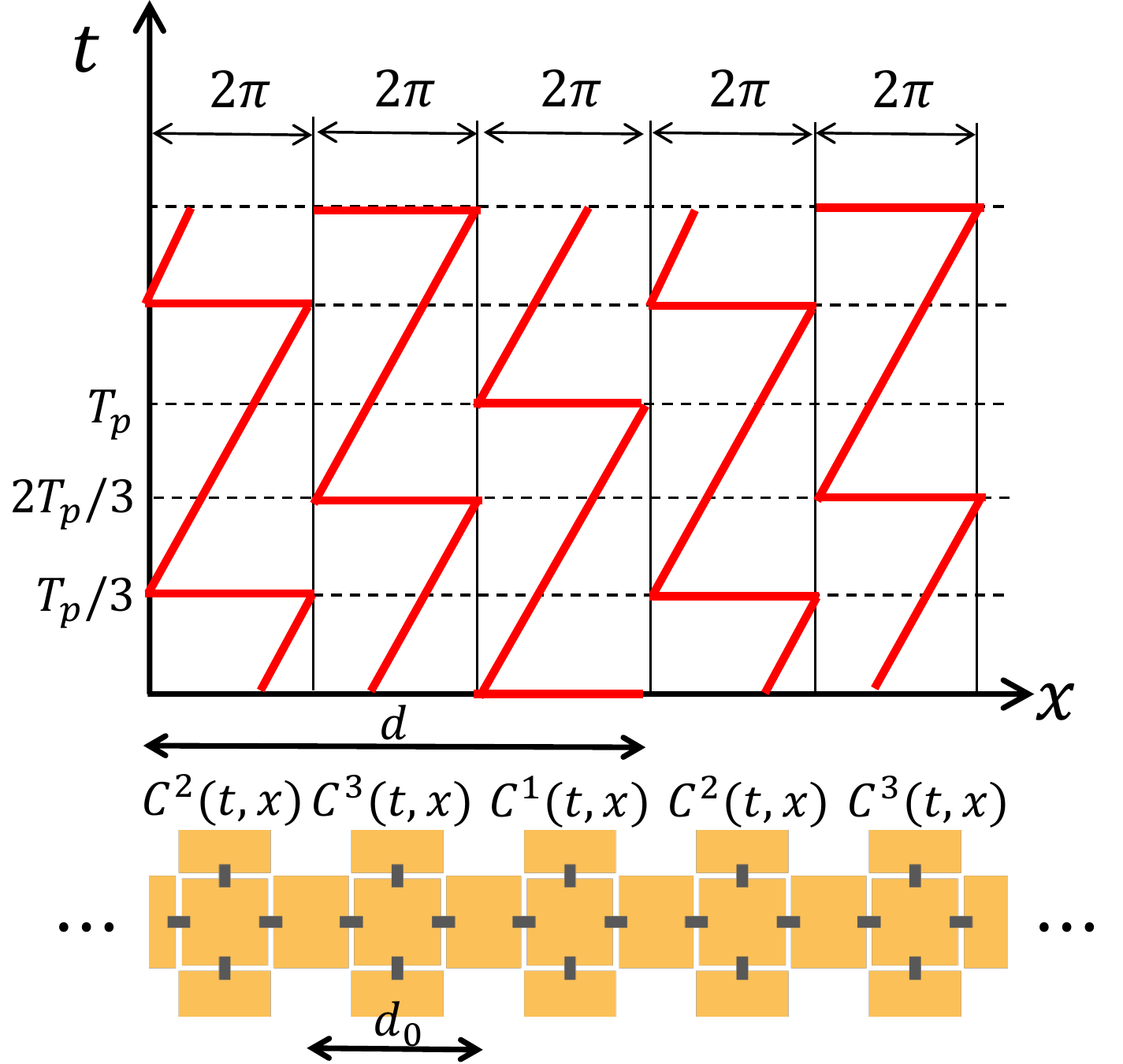}%
}\hfill
\caption{Space-time modulation scheme for the SD-TWM metasurface. The temporal modulation on each path is chosen to generate a sawtooth reflection phase varying from $0$ to $2\pi$ over each period. The $N$ adjacent paths of the metasurface are modulated by bias signals staggered in time by $T_p/N$. (a) A 2-path modulation scheme ($d = 2d_0$). (b)A 3-path modulation scheme ($d = 3d_0$).}
\label{fig:SpaceTimeMode}
\end{figure}

\subsection{Modified Floquet expansion of SD-TWM metasurfaces} 

For any space-time modulated waveform, the capacitance modulation on the metasurface can be expanded as a double Floquet expansion in space and time \cite{taravati2019generalized}, 
\begin{align}\label{Cap2DF}
C(t,x)=\sum_{m=-\infty}^{\infty}\sum_{q=-\infty}^{\infty} C_{mq}e^{-jm\beta_p x}e^{jq\omega_p t},
\end{align}
where $\omega_p=2\pi/T_p$ is the radial frequency (temporal modulation wavenumber) of the modulation. Applying the capacitance relationship given by Eq. (\ref{CapRelation}) to the double Floquet expansion in Eq. (\ref{Cap2DF}) reveals that the only non-zero values for coefficient $C_{mq}$ are with $m=q+rN,r\in \mathbb{Z}$ (see supplemental material II). This allows the capacitance expansion to be recast as the modified Floquet expansion,
\begin{align}\label{CapMod}
C(t,x)=\sum_{r=-\infty}^{\infty}\sum_{q=-\infty}^{\infty} C_{rq}e^{jq(\omega_p t-\beta_px)}e^{-jr\beta_d x},
\end{align}
where $\beta_p=2\pi/d$ is the modulation wavenumber, and  $\beta_d=2\pi/d_0=N\beta_p$ is an additional wavenumber which results from the discretization of the spatial modulation into stixels (paths). The summation over $r$ accounts for the discontinuity in capacitance at the the boundary of each path as well as the microscopic variation of capacitance within the paths (which in this case is uniform). The summation over $q$ accounts for the macroscopic capacitance variation over one spatial modulation period $d$.
%The sheet capacitance of each path, is capable of generating a staggered sawtooth reflection phase in time, as shown in Fig. \ref{fig:SpaceTimeMode}.

Similarly substituting a double Floquet expansion of the field into interpath relation given by Eq. (\ref{NpathRelationshipE}) reveals that the total tangential field distribution above the metasurface can also be expressed in terms of the modified Floquet expansion, 
\begin{align}\label{FieldMod2}
E_t(t,x) &=\sum_{r=-\infty}^{\infty}\sum_{q=-\infty}^{\infty} V_{r q}e^{-jr\beta_d x}e^{jq(\omega_p t-\beta_p x)}e^{j(\omega_0t-k_xx)}.
\end{align}
The spatio-temporal harmonic pair ($r,q$) of the electromagnetic field has a tangential wavenumber 
\begin{align}\label{Kx}
k_{xrq}&=q\beta_p+r\beta_d+k_x\nonumber\\
&=(q+rN)\beta_p+k_x,
\end{align}
and a corresponding radial frequency
\begin{align}\label{W}
\omega_{rq}=\omega_0+q\omega_p,
\end{align}
This implies that the $q^{\rm{th}}$ frequency harmonic, $\omega_0+q\omega_p$, of the field is associated with an infinite number of spatial harmonics which arise from the spatially-discrete stixels.
The reflected angle of each scattered harmonic pair is equal to
\begin{align}\label{Angle}
\theta_{rq}=\arcsin \frac{k_{xrq}}{\omega_{rq} / c}=\arcsin \frac{((q+rN)\beta_p+k_0\sin \theta_i)}{\omega_{rq} / c}
\end{align}

Compared to the standard double Floquet expansion in Eq. (\ref{Cap2DF}), the inclusion of the interpath relation significantly reduces the number of unknowns. It also reveals the underlying physics associated with the SD-TWM system. In fact, Eq. (\ref{FieldMod2}) allows one to separate the field into its macroscopic and microscopic variation. In addition, the field expansion in Eq. (\ref{FieldMod2}) indicates that new physical phenomena, such as subharmonic mixing, can be achieved by an SD-TWM metasurface, much like in an N-path circuit. Subharmonic mixing occurs when the spatial modulation period of the metasurface is electrically small, which we will discuss in detail in Section \ref{sec:SmallPeriod}.

\subsection{Scattered field calculation of SD-TWM metasurfaces} 

The scattered field of the SD-TWM metasurface for an oblique, monochromatic incident wave can be solved using the boundary condition at $z=0$,
\begin{equation}\label{BoundaryT}
H_t|_{z=0^+}-H_t|_{z=0^-}=\frac{d}{dt}(C(t,x)E_t).
\end{equation}
The space-time tangential fields as well as the capacitance modulation can be expressed in the form of the modified Floquet expansion given by Eq. (\ref{CapMod}) and (\ref{FieldMod2}). In addition, we can separate the total tangential field on the metasurface ($z=0^+$) into incident and reflected tangential fields,
\begin{equation} \label{ESPInc}
E_t^{\rm{inc}}=V_{00}^{\rm{inc}}e^{j(\omega_0 t-k_{x} x)},
\end{equation}
\begin{equation} \label{ESPRef}
E_t^{\rm{ref}}=\sum_{r,q=-\infty}^{\infty} V_{rq}^{\rm{ref}}e^{-jr\beta_dx}e^{jq(\omega_p t-\beta_p x)}e^{j(\omega_0 t-k_{x} x)}.
\end{equation}
To solve for the scattered field, the incident and reflected space-time harmonics are organized into vectors as $\bm{V^{\rm{inc}}}$ and $\bm{V^{\rm{ref}}}$ respectively. Each entry corresponds to a unique spatio-temporal harmonic pair $(r,q)$. The vector $\bm{V^{\rm{inc}}}$ contains only a single entry since the incident field is a monochromatic plane wave. Based on the detailed derivation in Supplemental Material III, the reflected electric field can be calculated for each polarization \cite{wang2020theory},
\begin{align}\label{TemporalRefCalTE}
\bm{V^{\rm{ref}}}=&\bm{(Y^{TX}+Y_0^{TX} )^{-1}}\bm{(Y_0^{TX}-Y^{TX} ) V^{\rm{inc}}},
\end{align} 
where the superscript ``X" is ``E" for TE polarized waves and  ``M" for TM polarized waves. $\bm{Y_0^{TX}}$ is the free-space tangential wave admittance matrix. It is a diagonal matrix containing entries of the free-space admittances at the corresponding frequency harmonics. $\bm{Y^{TX}}$ is the input admittance matrix of the time-modulated metasurface. It is not a diagonal matrix since the space-time modulated capacitive sheet introduces coupling between different harmonic pairs. 

\section{\label{sec:ThreeRegime} Three operating regimes of an SD-TWM metasurface}

\begin{table}[t]
\renewcommand{\arraystretch}{1.3}
\caption{Simulated conversion loss and sideband suppression for desired reflected frequency harmonic $f^\prime$ given: $N$ - the path number, $\theta_i$ - the incident angle, $\theta_{\rm{obs}}$ - the observation angle, and the temporal phase modulation waveform (either a sawtooth or a sinusoid). Note that positive values of $\theta_i$ and $\theta_{\rm{obs}}$ correspond to waves traveling along the positive $x$ direction.  }
\label{tab:sim}
\centering
\begin{tabular}{|c|c|c|c|c|c||c|c|c|c|}
\hline
\multirow{2}{*}{Ex.} &
\multirow{2}{*}{N} &
\multirow{2}{*}{$\theta_i$} &
\multirow{2}{*}{$\theta_{\rm{obs}}$} &
\multirow{2}{*}{\begin{tabular}[c]{@{}c@{}}Wave-\\form\end{tabular}} &
\multirow{2}{*}{$f^\prime$} &
\multicolumn{2}{c|}{\begin{tabular}[c]{@{}c@{}}Conversion\\ Loss\\ (dB)\end{tabular}} &
\multicolumn{2}{c|}{\begin{tabular}[c]{@{}c@{}}Sideband\\ Suppression\\ (dB)\end{tabular}} \\
\cline{7-10} 
& & & & & & TE & TM & TE & TM \\
\hhline{|=|=|=|=|=|=||=|=|=|=|}
0 & 1 & 25$^\circ$ & 25$^\circ$ & saw & $f_0+f_p$
& 0.11 & 0.13 & 22.83 & 21.64 \\
\hhline{|-|-|-|-|-|-||-|-|-|-|}
1 & 2 & 25$^\circ$ & 25$^\circ$ & saw & $f_0+2f_p$
& 0.34 & 0.50 & 17.32 & 13.19 \\
\hhline{|-|-|-|-|-|-||-|-|-|-|}
2 & 3 & 25$^\circ$ & 25$^\circ$ & saw & $f_0+3f_p$
& 0.45 & 0.87 & 14.14 & 10.46 \\
\hhline{|-|-|-|-|-|-||-|-|-|-|}
3 & 20 & 25$^\circ$ & 42$^\circ$ & saw & $f_0+f_p$
& 0.13 & 0.62 & 22.83 & 11.75 \\
\hhline{|-|-|-|-|-|-||-|-|-|-|}
4 & 20 & -42$^\circ$ & -25$^\circ$ & saw & $f_0+f_p$
& 0.13 & 0.62 & 24.17 & 9.04 \\
\hhline{|-|-|-|-|-|-||-|-|-|-|}
5 & 20 & -7.2$^\circ$ & 7.2$^\circ$ & saw & $f_0+f_p$
& 0.12 & 0.23 & 23.01 & 19.58 \\
\hhline{|-|-|-|-|-|-||-|-|-|-|}
6 & 4 & 39$^\circ$ & -39$^\circ$ & saw & $f_0+3f_p$
& 2.24 & 1.45 & 9.59 & 11.69 \\
\hhline{|-|-|-|-|-|-||-|-|-|-|}
7 & 4 & -39$^\circ$ & 39$^\circ$ & saw & $f_0+f_p$
& 1.24 & 0.95 & 10.63 & 11.89 \\
\hhline{|-|-|-|-|-|-||-|-|-|-|}
8 & 4 & 39$^\circ$ & -39$^\circ$ & sin & $f_0-f_p$
& 0.50 & 0.31 & 14.94 & 17.71 \\
\hhline{|-|-|-|-|-|-||-|-|-|-|}
9 & 4 & -39$^\circ$ & 39$^\circ$ & sin & $f_0+f_p$
& 0.49 & 0.31 & 14.95 & 17.65 \\
\hline
\end{tabular}
\end{table}

As mentioned earlier, the modified Floquet expansion given by Eq. (\ref{FieldMod2}) separates the field into macroscopic and microscopic spatial variations. It is clear that when the the number of stixels within a spatial period (the path number $N$) increases, the field variation across one stixel decreases. As a result, the microscopic spatial variations on the metasurface will decrease. On the other hand, when the number of stixels within a spatial period is small, the microscopic spatial variation dominates. This enables the same SD-TWM metasurface to achieve different functions depending on the spatial modulation period $d$. In fact, the electromagnetic behavior of the metasurface can be categorized into three regimes.

In this section, we will discuss the scattering performance of the SD-TWM metasurface, for electrically small, large and wavelength-scale spatial modulation periods. In addition, scattering for temporal modulation only (1 path, $N=1$) is given in Section \ref{sec:TModOnly} as reference, since it represents the limiting case where the modulation period approaches infinity. Computed results are given in this section for various space-time modulation examples in the three regimes. For convenience, the conversion loss and sideband suppression at the prescribed observation angle are provided in Table \ref{tab:sim}, for each of the examples that follow. The table will be referred to throughout this section.  In all of the examples studied, the incident signal frequency is $f_0=10$ GHz. The modulation frequency, $f_p=25$ kHz, which is the maximum frequency that could be experimentally validated using the multi-channel digital to analog converters available to the authors (see Section \ref{sec:Fab}). For each angle of incidence, the capacitance modulation is calculated based on Eq. (S.4) to achieve the desired time-varying reflection phase. Unless specifically stated otherwise, the reflection phase of each stixel (path) is a sawtooth function in time. For both polarizations, the field is expanded into $141\times 141$ $(r,q)$ harmonic pairs. The temporal capacitance modulation on each path is truncated to 101 temporal harmonics.

\subsection{Temporal modulation ($\beta_p=0$): serrodyne frequency translation} 
\label{sec:TModOnly}

\begin{figure}[b]
\subfloat[\label{sfig:WaveTE_C}]{%
  \includegraphics[clip,width=0.49\columnwidth]{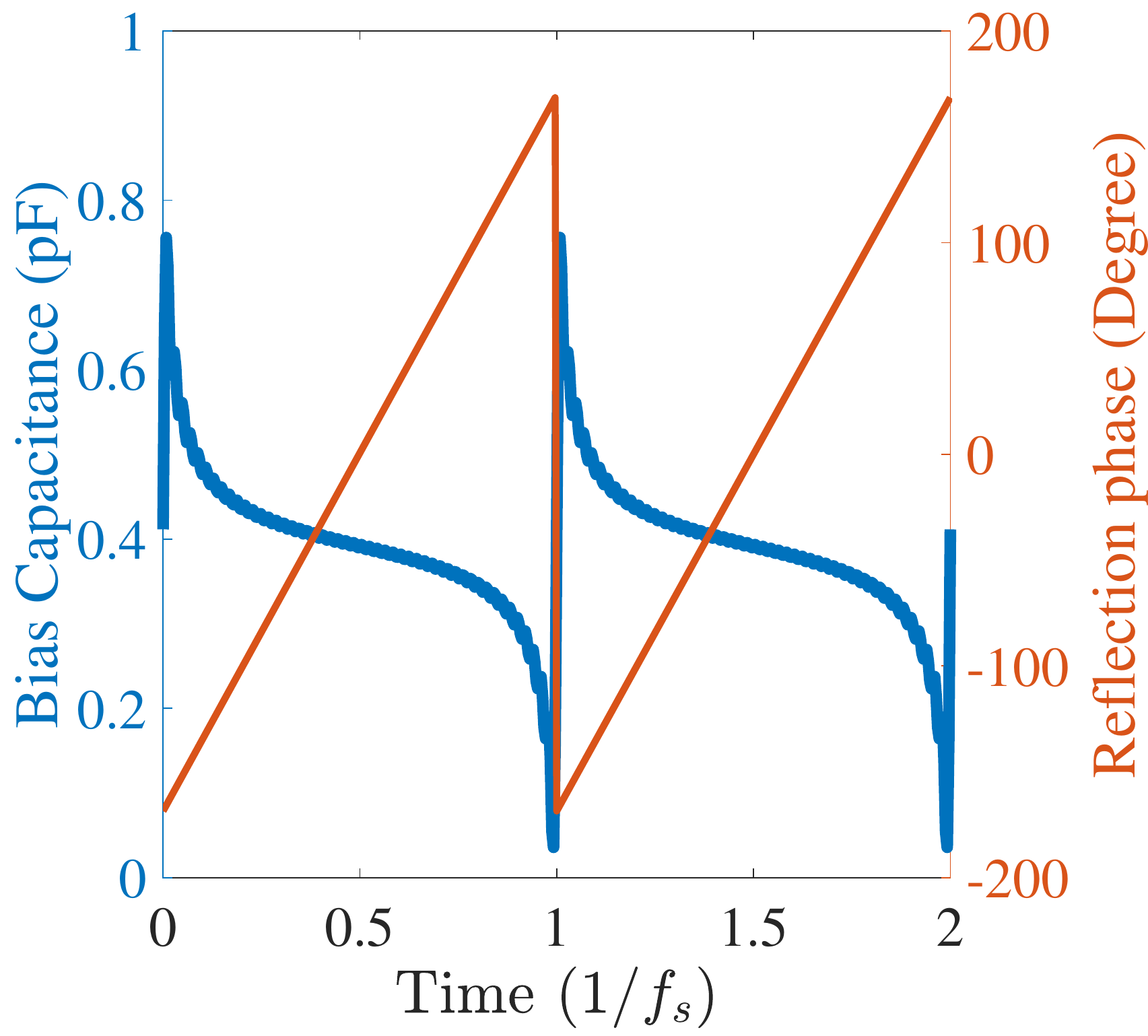}%
}\hfill
\subfloat[\label{sfig:1pathTE_C}]{%
  \includegraphics[clip,width=0.49\columnwidth]{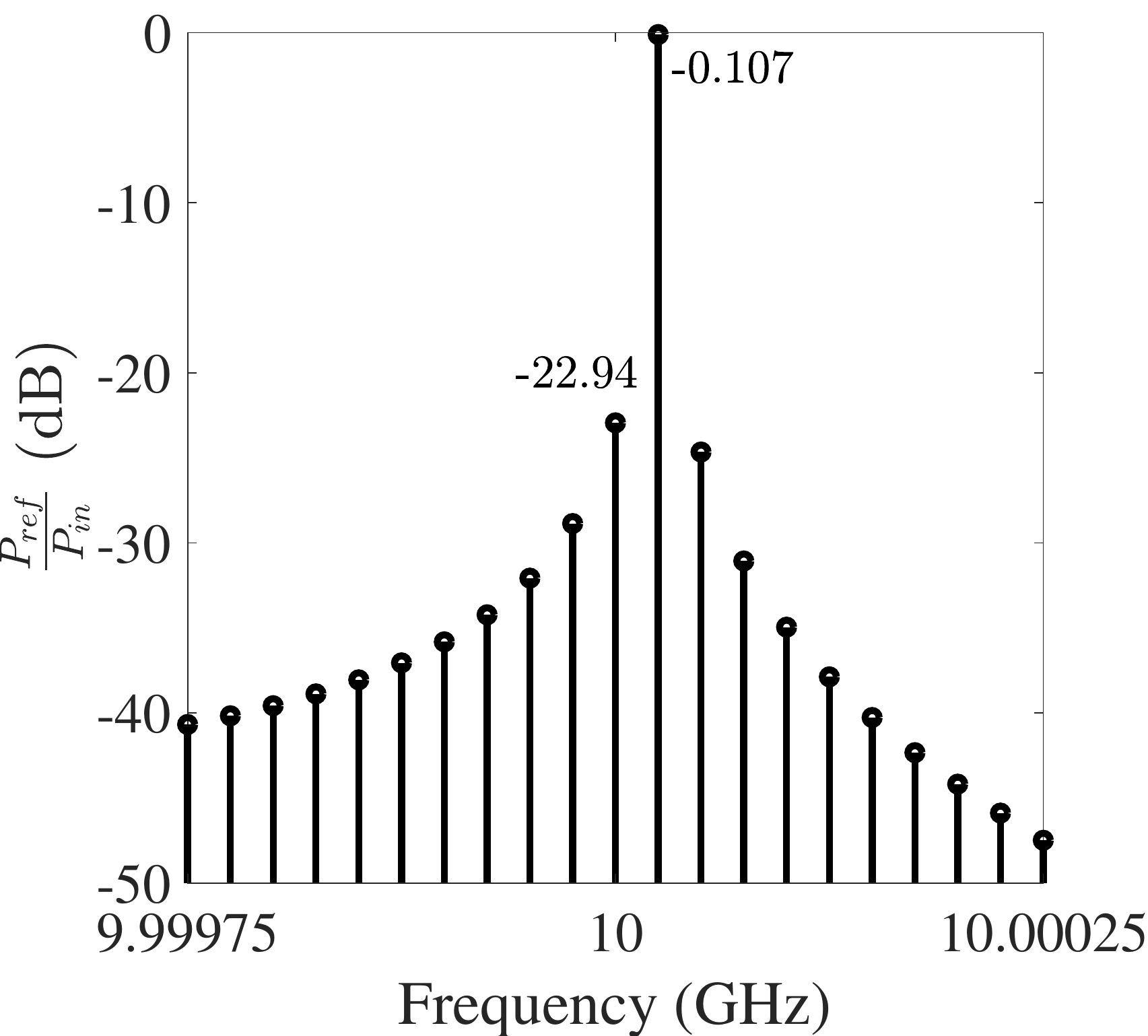}%
}\quad
\subfloat[\label{sfig:WaveTM_C}]{%
  \includegraphics[clip,width=0.5\columnwidth]{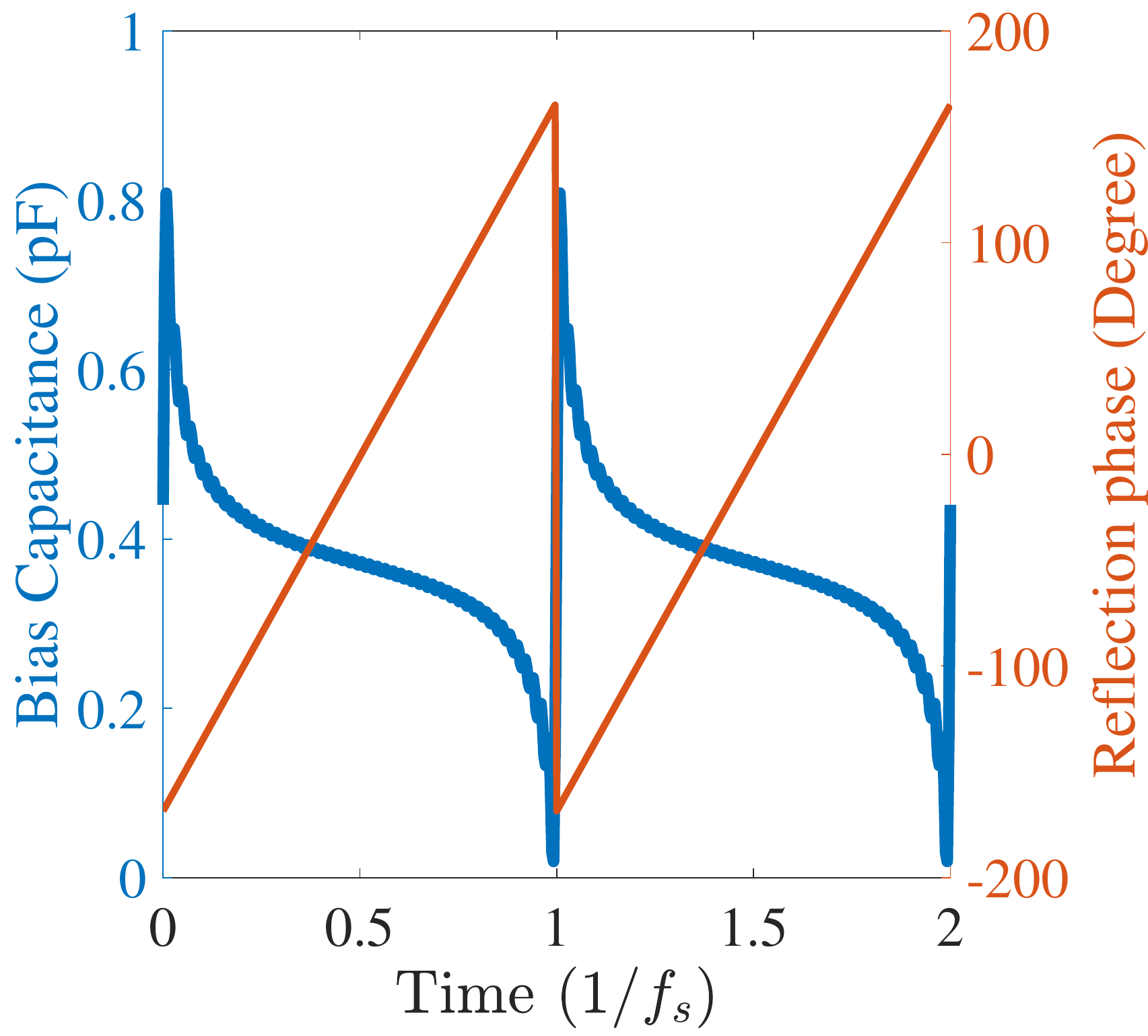}%
}\hfill
\subfloat[\label{sfig:1pathTM_C}]{%
  \includegraphics[clip,width=0.5\columnwidth]{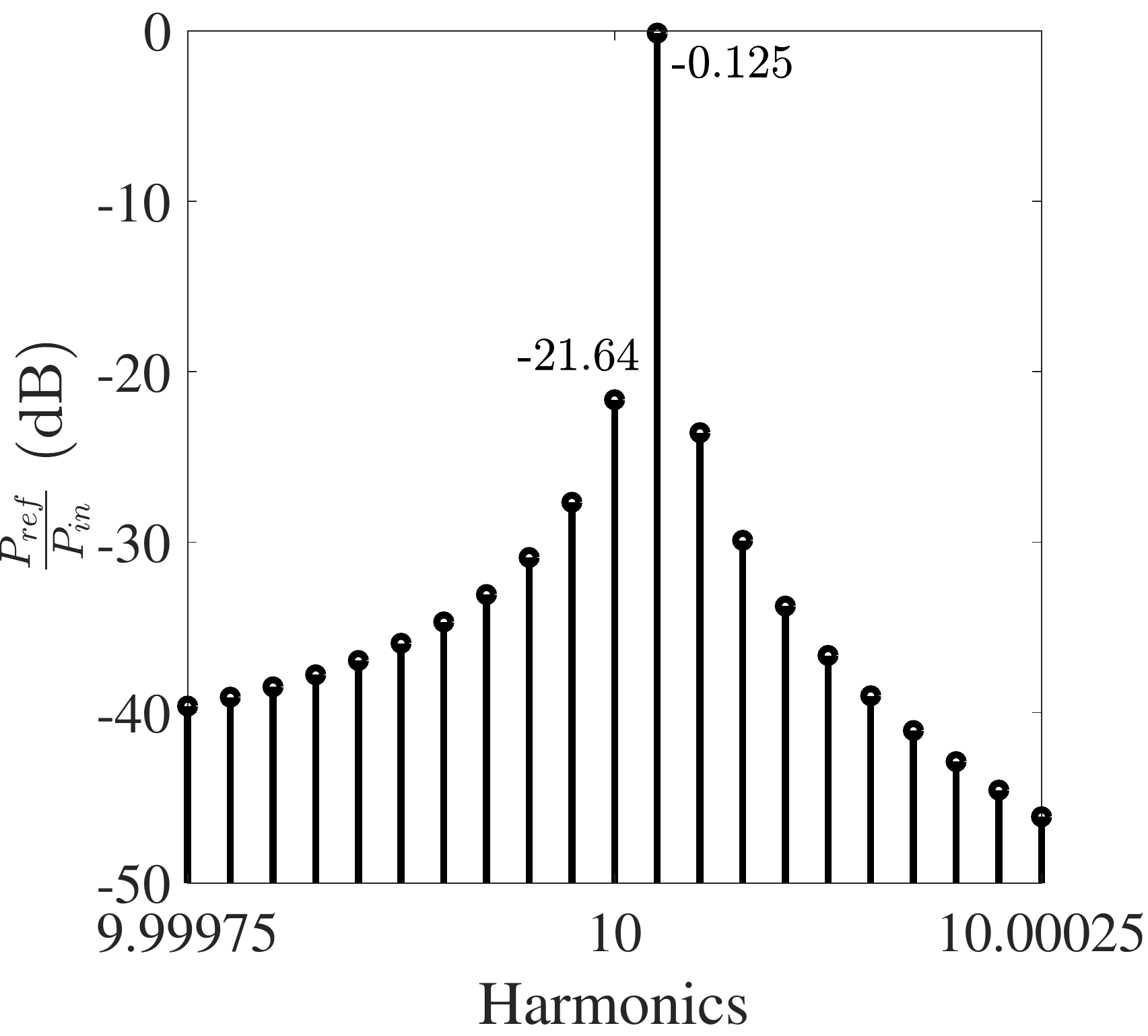}%
}
\caption{(a) Calculated capacitance modulation of the time-modulated metasurface for TE polarization. (b)  Analytical reflection spectrum of the homogenized, lossless, time-modulated metasurface for TE polarization. (c) Calculated capacitance modulation of the time-modulated metasurface for TM polarization. (d)  Analytical reflection spectrum of the homogenized, lossless, time-modulated metasurface for TM polarization.}
\label{fig:1path_C}
\end{figure} 

It is instructive to first consider the behavior of the metasurface when it is uniformly biased across all stixels. In this case, there is no spatial variation in the homogenized model and the reflected power spreads into discrete frequency harmonics due to the periodic time variation of the reflection phase. 

Suppose the reflection phase is modulated by a sawtooth waveform. In this case, serrodyne frequency translation is expected \cite{8854331}. The incident wave is assumed to impinge on the metasurface at an oblique angle of $25^\circ$. The capacitance modulation needed to upconvert the wave to $f_0 + f_p$ is calculated in Supplemental Material I, and shown in Fig. \ref{fig:1path_C} for each polarization. The reflected spectra for the two orthogonal polarizations (shown in Fig. \ref{fig:1path_C}) clearly show a Doppler shift to a frequency of $f_0+f_p$. For each polarization, the conversion loss and sideband suppression are provided in example 0 of Table \ref{tab:sim}. As mentioned earlier, the equivalent circuit model of the unit cell provides a phase range that is slightly less that $2\pi$ ($1.6 \pi$ for TE polarization and $1.54 \pi$ for TM polarization used in the analysis), resulting in undesired sidebands. For TM polarization, the reflection phase range is slightly less than for TE at the oblique angle of $25^\circ$, resulting a slightly higher conversion loss.

\subsection{Small spatial modulation period ($|k_x\pm\beta_p|>k_0$)} 
\label{sec:SmallPeriod}

\begin{figure}[b]
\centering
\includegraphics[width=0.6\columnwidth]{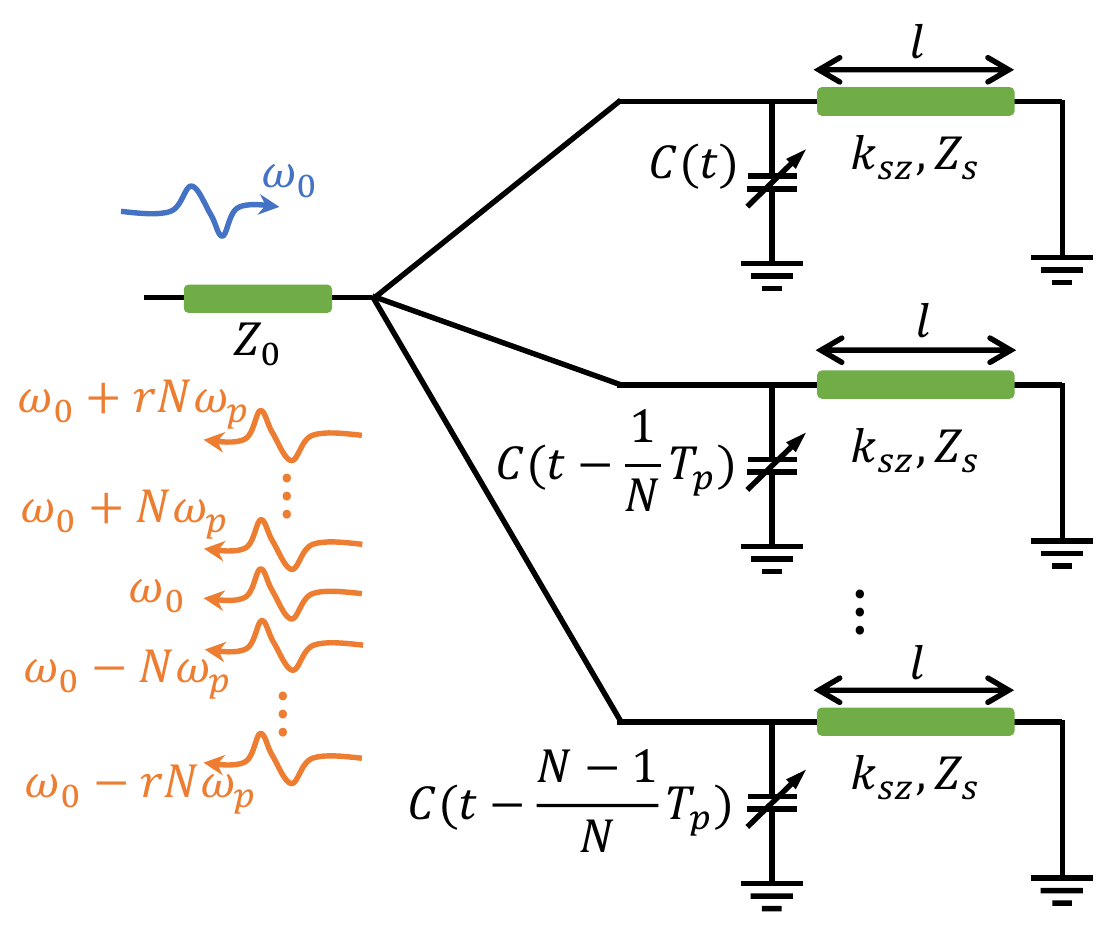}
\caption{Equivalent circuit model of the SD-TWM metasurface when the spatial modulation period is much smaller than the wavelength of radiation.}
\label{fig:NpathCircuit}
\end{figure}

In this section, we consider a spatial modulation period $d$ that is electrically small ($N$ is small). In this case, the paths can be viewed as collocated and a N-path circuit model can be used to approximate the physical structure. The equivalent circuit model of the spatio-temporally modulated metasurface for $d \ll \lambda_0$ is depicted in Fig. \ref{fig:NpathCircuit}. Since the time variation of each path is staggered, certain harmonics mixing products are are suppressed in the reflected signal, allowing subharmonic mixing.

\begin{figure}[t]
\subfloat[\label{sfig:LightConeHigh}]{%
  \includegraphics[clip,width=0.5\columnwidth]{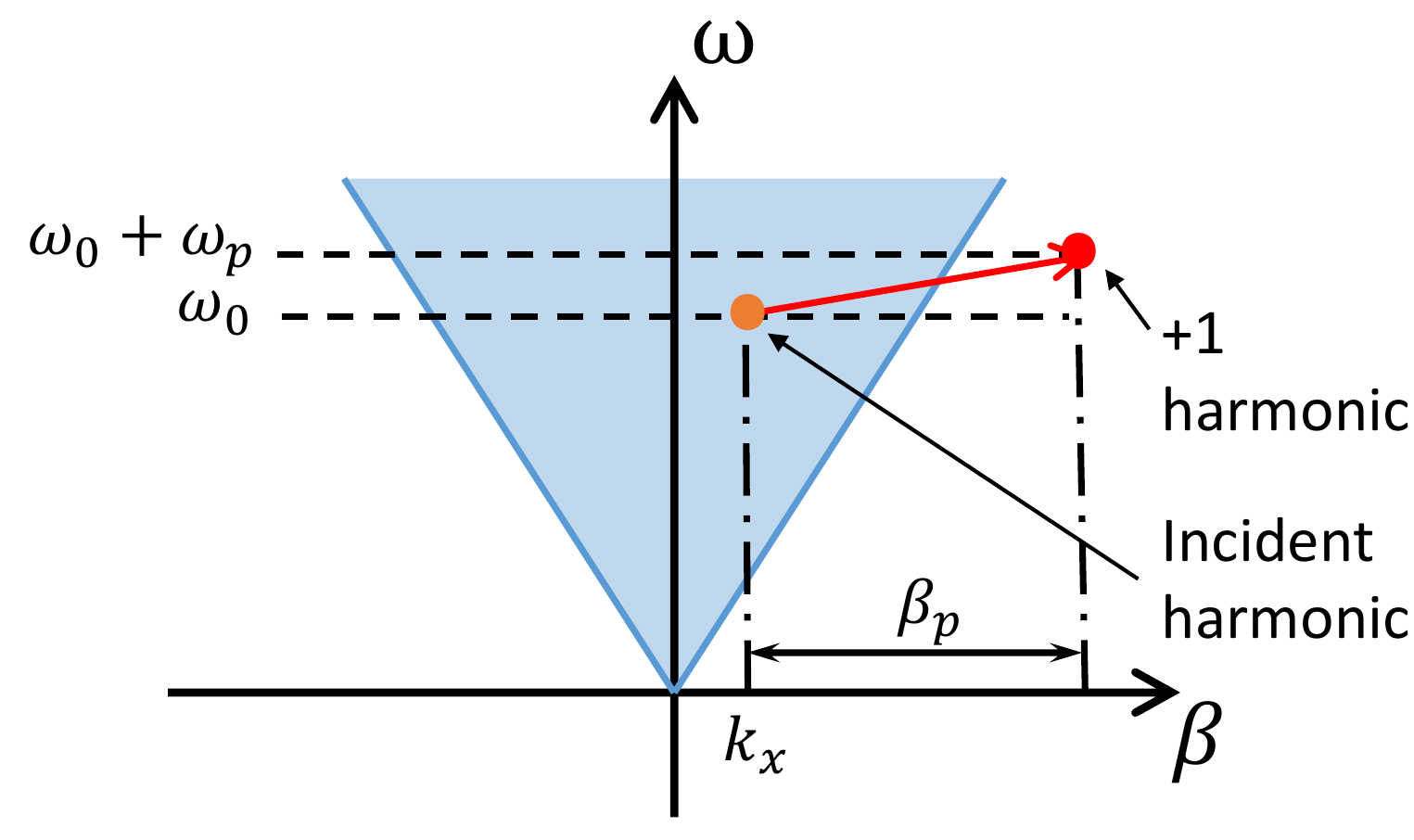}%
}\hfill
\subfloat[\label{sfig:LightConeLow}]{%
  \includegraphics[clip,width=0.5\columnwidth]{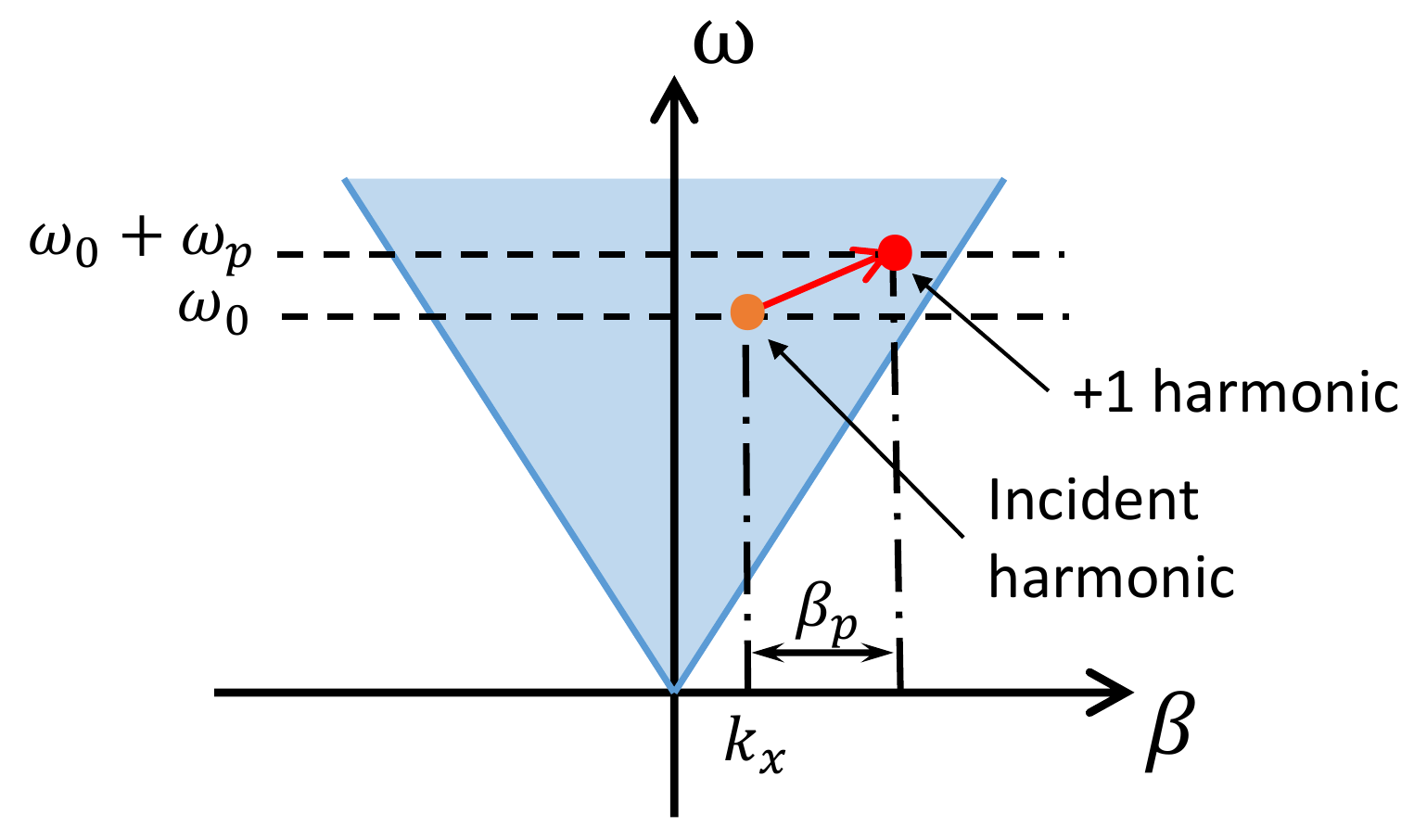}%
}\hfill
\caption{Graphic representation of the spatial and temporal frequency shifts for different modulation wavenumbers. (a) The modulation wavenumber $\beta_p$ is large. (b) The modulation wavenumber $\beta_p$ is small.}
\label{fig:LightCone}
\end{figure}

Another way of understanding the subharmonic mixing behavior of the metasurface is depicted in Fig. \ref{sfig:LightConeHigh}. If the modulation period $d$ of the SD-TWM metasurface is electrically small ($|k_x\pm\beta_p|>k_0$), then both the $+1$ ($k_x+\beta_p$) and $-1$ ($k_x-\beta_p$) spatial harmonics are outside of the light cone. The large modulation wavenumber $\beta_p$ in this regime can lead to a number of higher order harmonics existing outside the light cone. In this case, the metasurface can convert an incident wave to a surface wave, provided that the corresponding surface wave is supported by the metasurface. However, when the corresponding surface wave is not supported, the power can only couple to radiating harmonics: those within the light cone. Based on Eq. (\ref{Kx}), the radiating harmonics are those with:
\begin{align}\label{NpathHarmonics}
q+rN=0.
\end{align} 
Since $r\in \mathbb{Z}$, Eq. (\ref{NpathHarmonics}) implies that propagating harmonics correspond to $q=0,\pm N,\pm 2N\dots$. Therefore, the radiated reflected wave only contains frequency harmonics at $f_0+rNf_p$, where $r\in \mathbb{Z}$. As a result, the metasurface can achieve subharmonic frequency translation.

More generally, for harmonic pairs with tangential wavenumbers larger than free space ($k_{xrq}>\omega_{rq}/c$), the transverse resonance condition can be used to judge if the corresponding surface waves are supported by the SD-TWM metasurface.
\begin{align}\label{TransverseResonance}
\det{(\bm{Y^{TX}+Y_0^{TX}})}=0,
\end{align}
Solving Eq. (\ref{TransverseResonance}) yields the $\omega_0-k_x$ dispersion relationship for the supported surface wave. Note that when a surface wave is supported, the reflection coefficient $V_{ref}/V_{inc}$ in Eq. (\ref{TemporalRefCalTE}) diverges. This is not the case for the incident angle and path number $N$ combinations considered in this paper. Thus, for all the proceeding examples presented in this paper, a surface wave is not supported by the metasurface.

\begin{figure}[t]
\centering
\includegraphics[width=0.5\columnwidth]{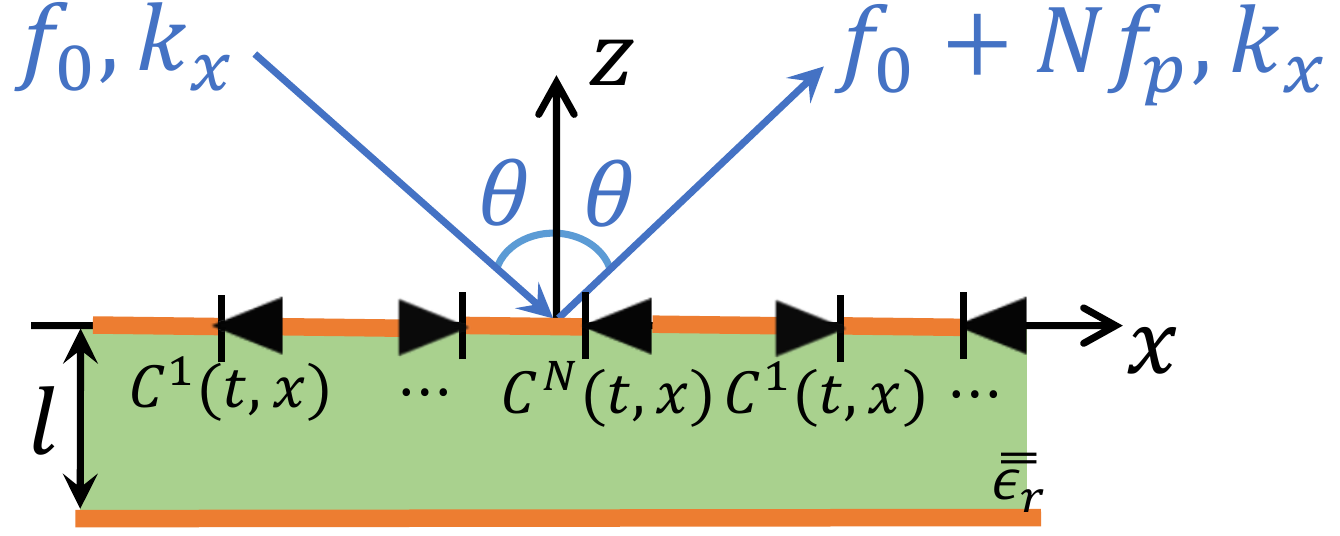}
\caption{Specular subharmonic frequency translation of the SD-TWM metasurface. For the presented metasurface, this can be achieved when $N=2$ or $3$.}
\label{fig:NpathReflectionSub}
\end{figure}

Note that, the subharmonic mixing phenomena can only be observed when the spatial discretization of the metasurface is considered. In the continuum limit, sub-wavelength spatial modulation results in specular reflection at the same frequency as the incident wave. However, the spatial discretization introduces additional spatial harmonics (the summation over $r$ in (\ref{FieldMod2}) that can couple to the incident wave). In addition, all the reflected propagating harmonics share the same tangential wavenumber as the incident wave $($since $k_{xrq}=k_x,$ when $q+Nr=0)$. Since the modulation frequency (25 KHz) is much lower than the incident frequency (10 GHz), $f_p \ll f_0$, all the harmonics are at a reflection angle of $25^\circ$, as depicted in Fig \ref{fig:NpathReflectionSub}. If the modulation frequency $f_p$ is comparable with $f_0$, then each of the reflected, propagating frequency harmonics will have different radiated angles due to their substantially different free space wavenumbers. 

\begin{figure}[b]
\subfloat[\label{sfig:2pathTEA}]{%
  \includegraphics[clip,width=0.5\columnwidth]{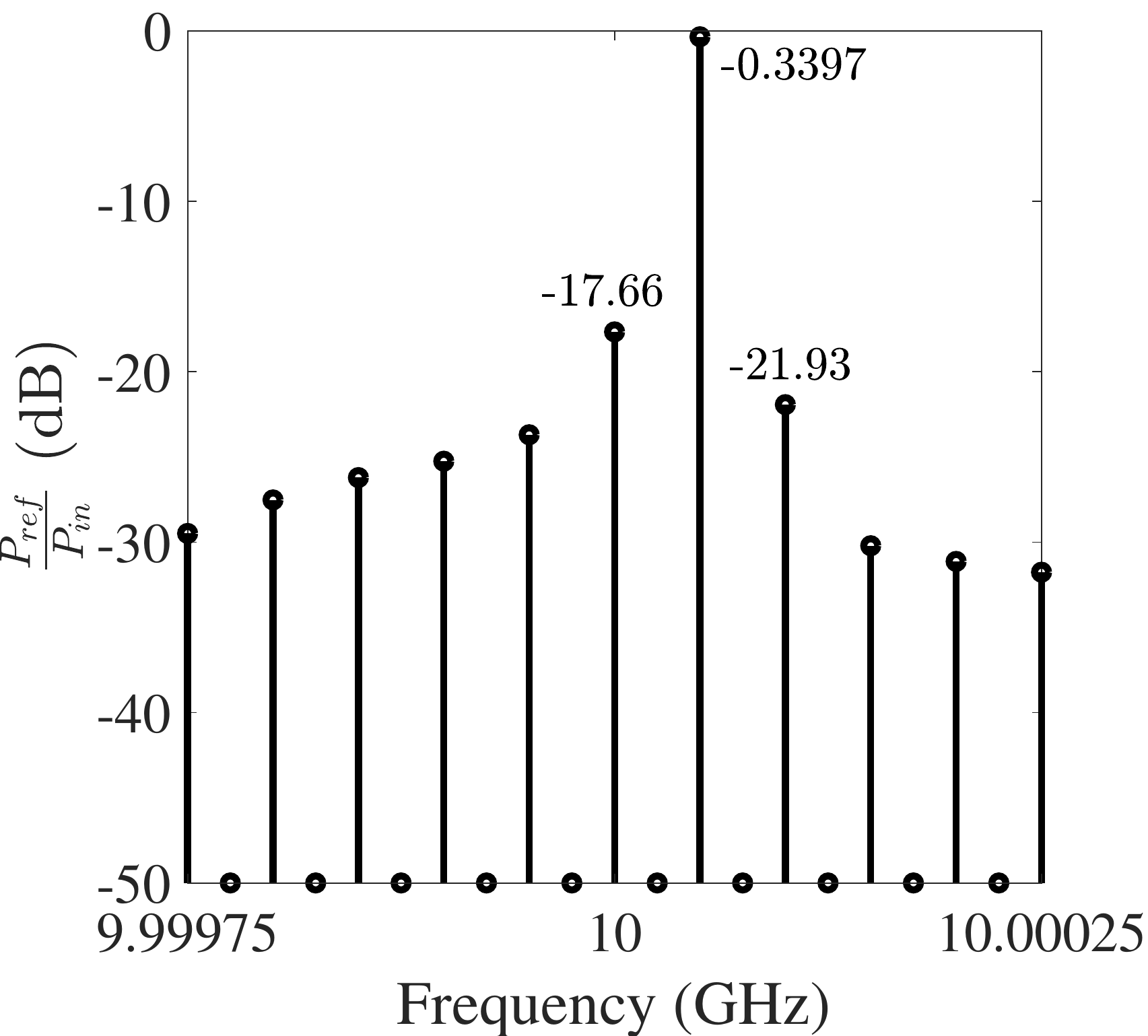}%
}\hfill
\subfloat[\label{sfig:3pathTEA}]{%
  \includegraphics[clip,width=0.5\columnwidth]{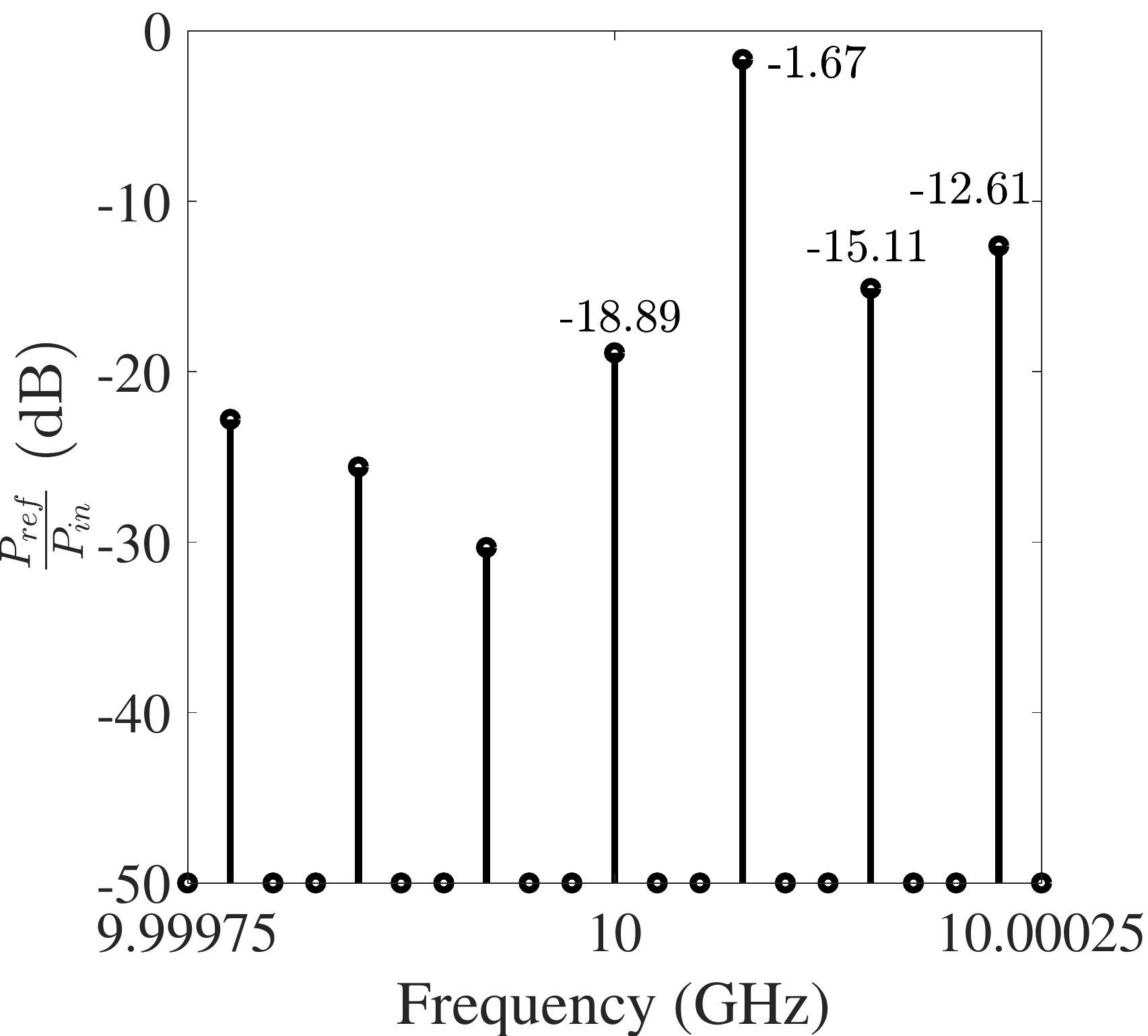}%
}\quad
\subfloat[\label{sfig:2pathTMA}]{%
  \includegraphics[clip,width=0.5\columnwidth]{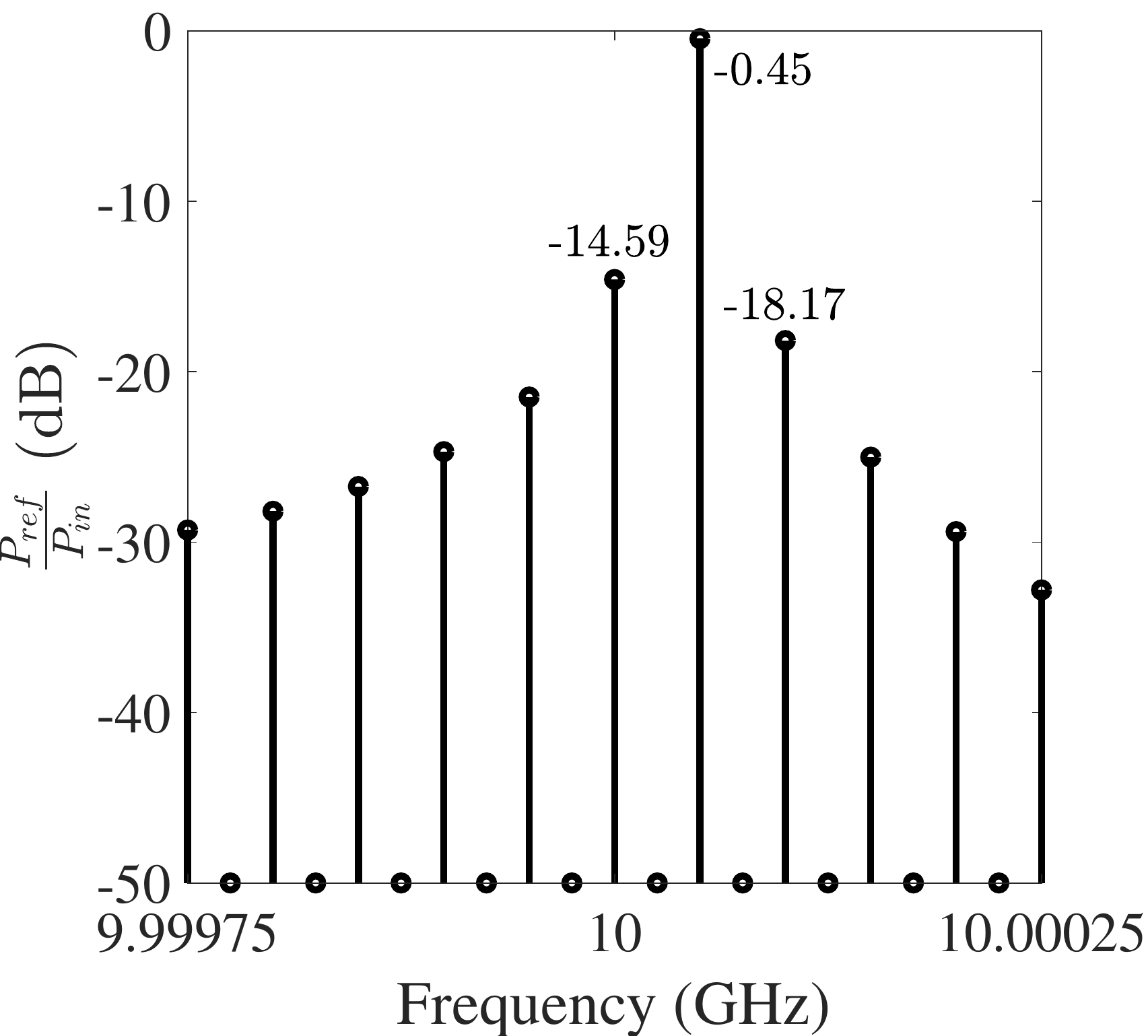}%
}\hfill
\subfloat[\label{sfig:3pathTMA}]{%
  \includegraphics[clip,width=0.5\columnwidth]{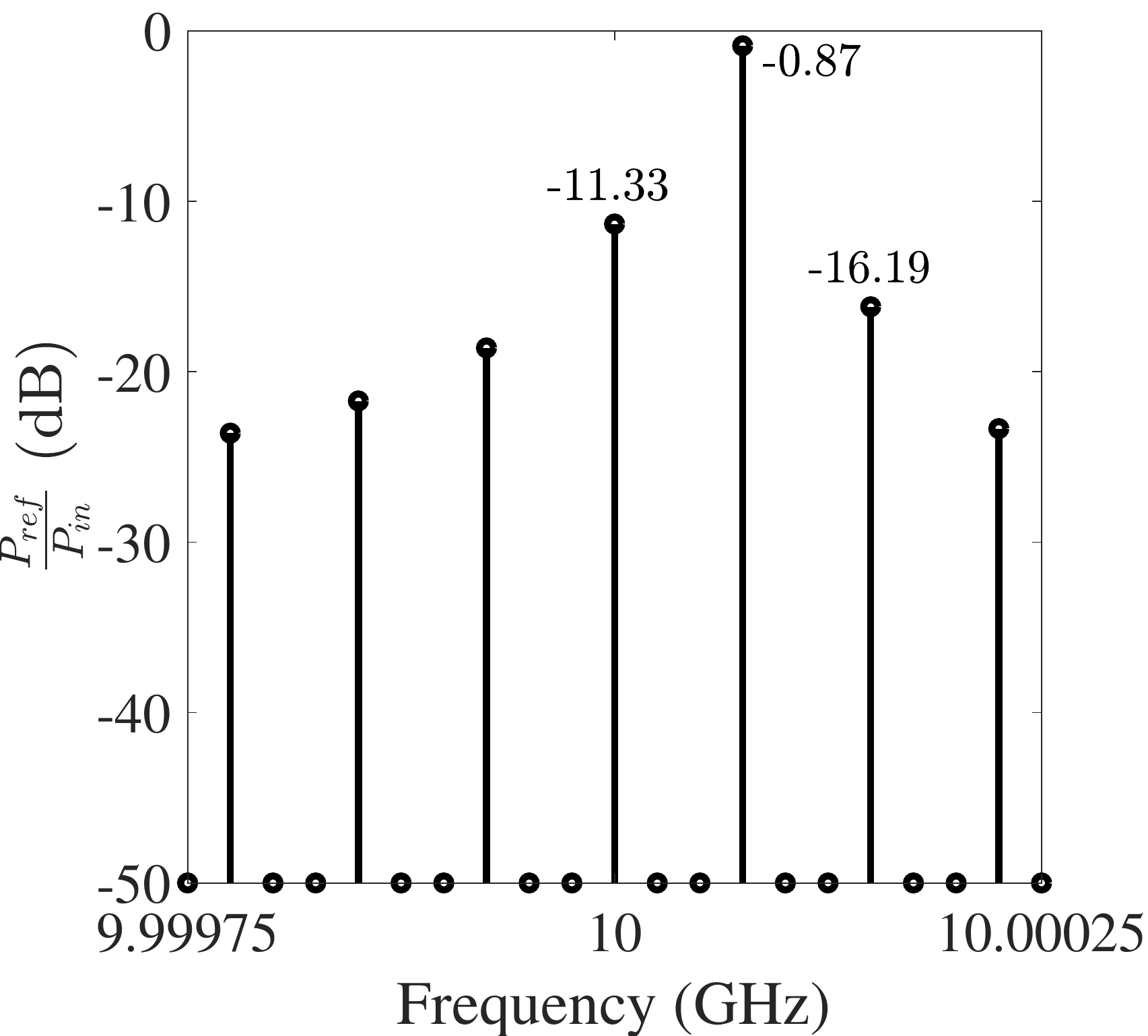}%
}
\caption{Theoretical reflection spectrum of the homogenized, lossless, SD-TWM metasurface. (a) 2-path ($N=2$) modulation for TE polarization. (b) 3-path ($N=3$) modulation for TE polarization. (c)  2-path ($N=2$) modulation for TM polarization. (d) 3-path ($N=3$) modulation for TM polarization.}
\label{fig:23pathCal}
\end{figure}

With the capacitance modulation shown in Fig. \ref{sfig:WaveTE_C} and \ref{sfig:WaveTM_C}, the sawtooth reflection phase on each path enables the metasurface to upconvert the frequency to the first propagating frequency harmonic. In this case, the metasurface performs subharmonic frequency translation from $f_0$ to $f_0+Nf_p$. Since the stixel size of the presented metasurface is fixed to $d_0=\lambda_0/5$, examples of 2- and 3-path modulation ($N=2,3$) are chosen to satisfy the small period condition ($|k_x\pm\beta_p|>k_0$). The incident wave impinges on the metasurface with an oblique angle of $25^\circ$. The computed reflection spectra for both polarizations are shown in Fig. \ref{fig:23pathCal}, which clearly demonstrate subharmonic frequency translation. The reflected harmonics are only radiated at $f=f+rNf_s$ with $r\in \mathbb{Z}$. Doppler-like frequency translations are observed for both polarizations, where the dominant propagating reflected wave is at frequency $f_0+Nf_p$. The conversion loss and sideband suppression for both polarizations using 2-path and 3-path modulation are provided in examples 1 and 2 of Table \ref{tab:sim}.

As mentioned earlier, each stixel provides a phase range that is slightly smaller than $2\pi$, resulting in conversion loss and undesired sidebands. It can be seen that, as the converted frequency harmonic (which is equal to the path number $N$ in this case) is increased, the conversion loss increases and the sideband suppression decreases. This is because the N-path metasurface upconverts the frequency to the first propagating harmonic pair. The higher the upconverted frequency, the longer this process takes and the larger the conversion loss due to the formation of sidebands that results from the imperfect reflection phase range.

\subsection{Large spatial modulation period ($|k_x\pm\beta_p|<k_0$)} 
\label{sec:LargePeriod}

When the modulation period $d$ is electrically large ($N$ is large), the spatial modulation wavenumber $\beta_p$ is small, as depicted in Fig. \ref{sfig:LightConeLow}. In this operating regime, both the $+1$ ($k_x+\beta_p$) and $-1$ ($k_x-\beta_p$) spatial harmonics are inside the light cone. When $N$ is a very large value, the stixel size can be seen as infinitesimally small compared to the spatial modulation period ($d_0 \ll d$). As a result, the discontinuity in the capacitance of neighboring stixels vanishes and the modulation can be approximated as a continuous function of space. According to Eq. (\ref{CapMod}), the capacitance coefficient $C_{rq}$ is zero for $r\neq 0$. For this case, the field variation across each stixel is small, and the capacitance modulation waveform is simplified to the continuum limit:
\begin{equation}\label{CapSmall}
C(t,x)=\sum_{q=-\infty}^{\infty} C_{q}e^{jq(\omega_p t-\beta_px)}.
\end{equation}
Note that Eq. (\ref{CapSmall}) is of the form of a traveling wave, $C(t,x)=C(t-x/v_p)$, where $v_p=\omega_p/\beta_p$. For such a modulation, the metasurface supports harmonics at frequency $f_0+qf_p$, with a corresponding wavenumber $k_x+q\beta_p$. In other words, the SD-TWM metasurface with large spatial modulation period shows a similar performance to a continuous traveling-wave modulated structure. Serrodyne frequency translation to a deflected angle can be achieved using the sawtooth waveform given in Fig. \ref{fig:1path_C}. 

\begin{figure}[b]
\subfloat[\label{sfig:NpathRef25}]{%
  \includegraphics[clip,width=0.5\columnwidth]{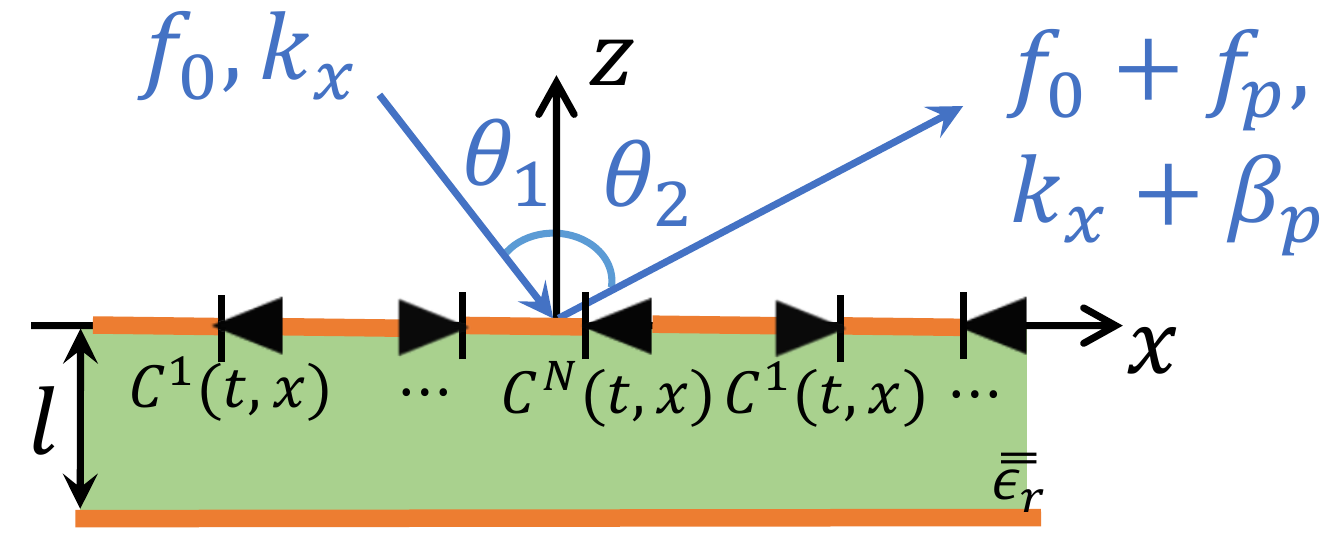}%
}\hfill
\subfloat[\label{sfig:NpathRef42}]{%
  \includegraphics[clip,width=0.5\columnwidth]{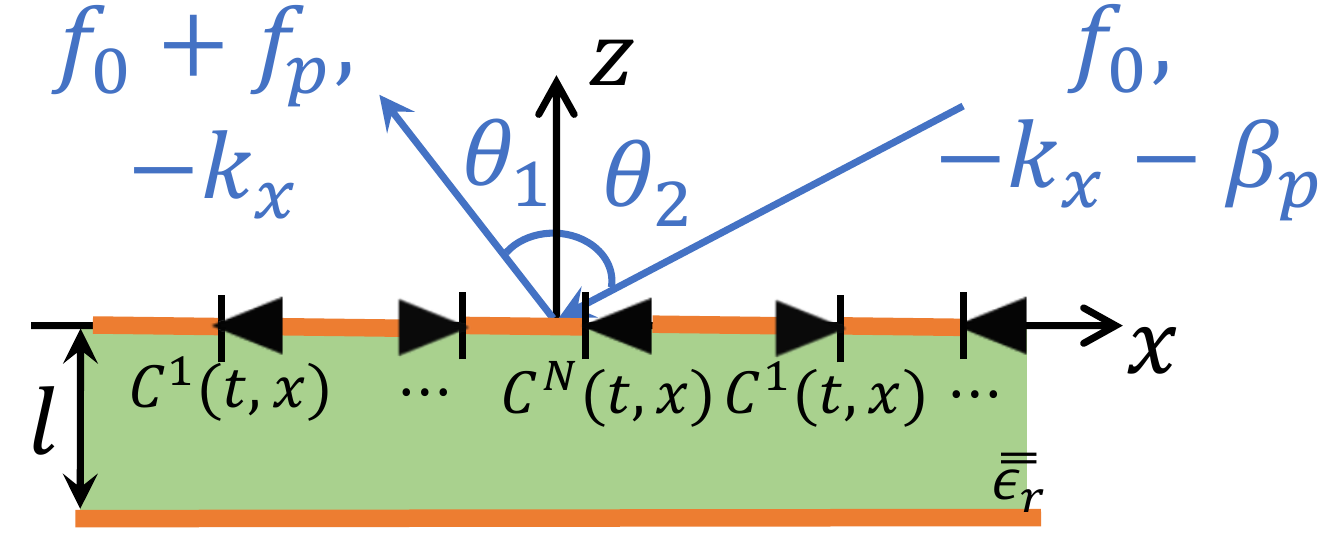}%
}\quad
\caption{The SD-TWM metasurface performing serrodyne frequency translation to a deflected angle. The path number $N>5$. The modulation frequency $f_p$ is much lower than the incident frequency $f_0$. (a) Wave is incident at an oblique angle $\theta_1$. (b) Wave is incident at an oblique angle $-\theta_2$.}
\label{fig:NpathReflectionDeflect}
\end{figure}

\begin{figure}[t]
\subfloat[\label{sfig:25steeringTEA}]{%
  \includegraphics[clip,width=0.5\columnwidth]{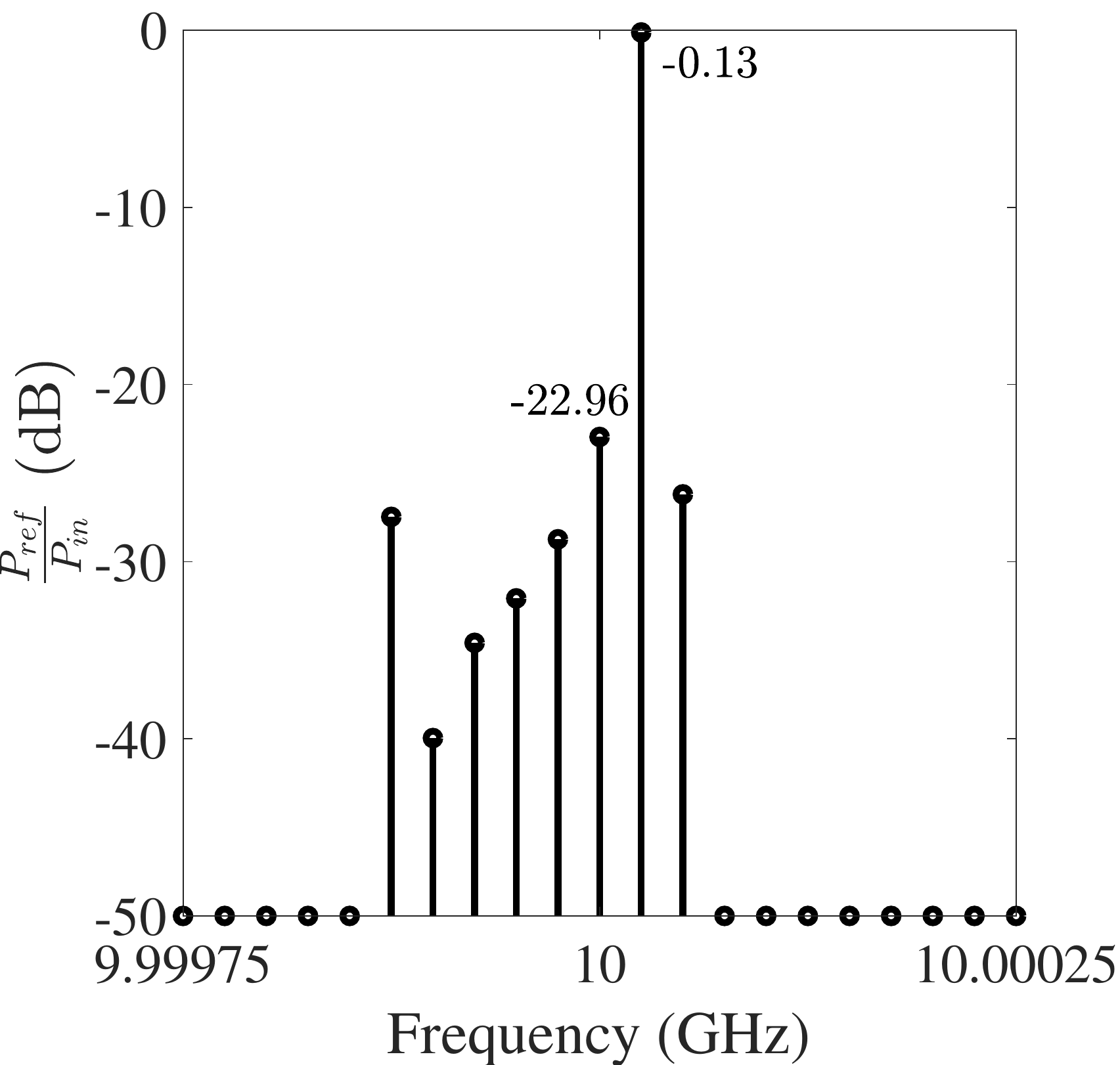}%
}\hfill
\subfloat[\label{sfig:25steeringTMA}]{%
  \includegraphics[clip,width=0.5\columnwidth]{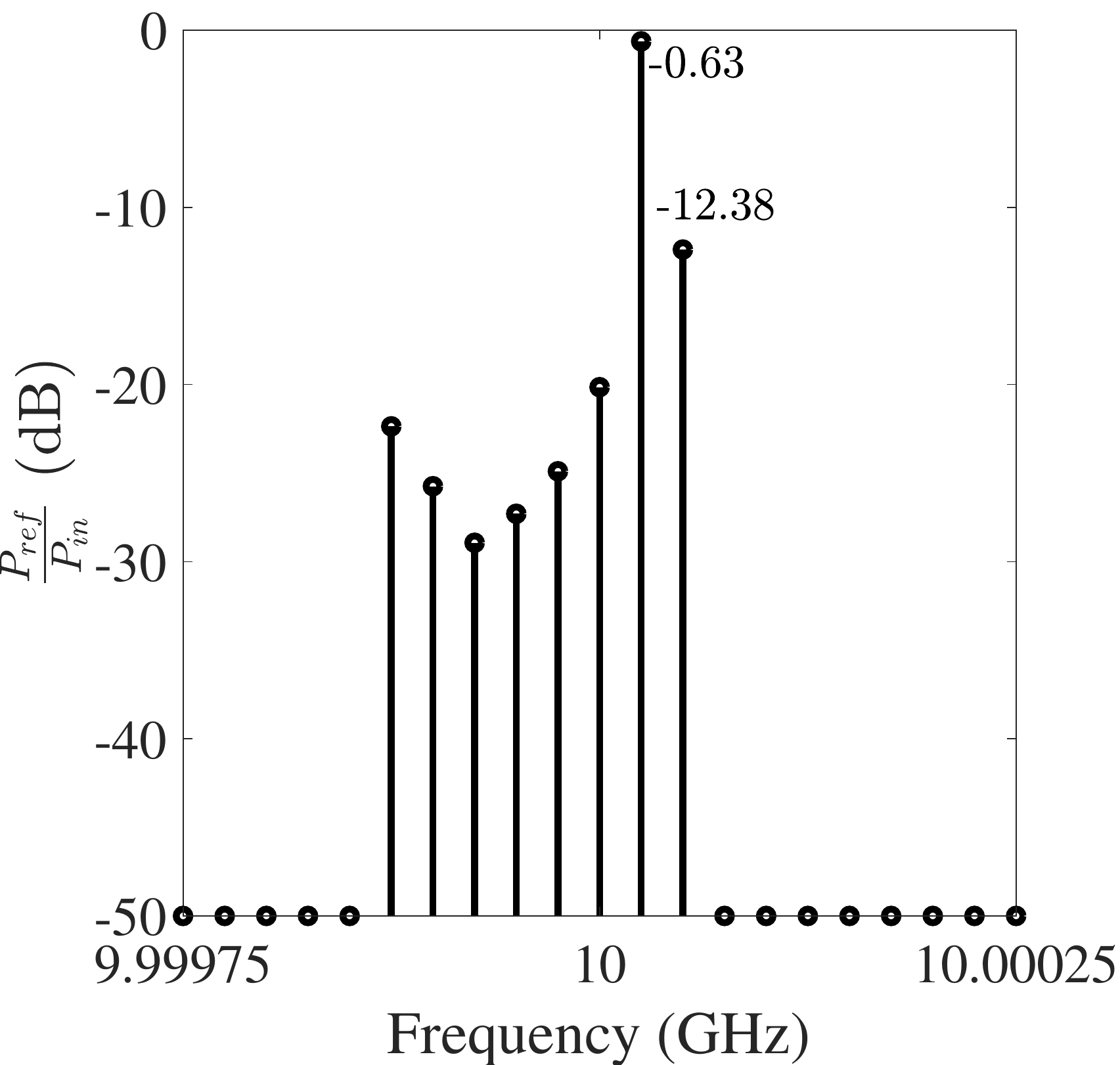}%
}\quad
\caption{Theoretical reflection spectrum of the homogenized, lossless, SD-TWM metasurface with an incident angle of $25^\circ$.(a) 20-path ($N=20$) modulation for TE polarization. (b) 20-path ($N=20$) modulation for TM polarization.}
\label{fig:25steeringA}
\end{figure}

Let us consider the example shown in Fig. \ref{sfig:NpathRef25}, where a wave is incident at an angle $\theta_1=25^\circ$ and the number of paths is large, $N=20$. 
From Eq. (\ref{Kx}), the tangential wavenumbers of the reflected harmonic pairs are given by
% The tangential wavenumbers of the reflected harmonic pairs are given by Eq. (\ref{Kx}), which can be rewrite as
\begin{align}\label{KxSteering25}
{\left. {{k_{xrq}}} \right|_{r = 0}}&=q\beta_p + k_0\sin(25^\circ)\nonumber\\
&=\frac{q}{4}k_0+k_0\sin(25^\circ) ,
\end{align}
given that $d=20 d_0= 4 \lambda_0$. The harmonics located inside the light cone (propagating harmonics) are those with $q=0,\pm 1,\pm 2,-3,-4,-5$. For the capacitance variation shown in Fig. \ref{sfig:WaveTE_C} and Fig. \ref{sfig:WaveTM_C}, the metasurface acts as a  serrodyne frequency translator. It upconverts the incident wave to the harmonic pair $(r=0,q=1)$ with frequency $f=f_0+f_p$. In this case, each scattered harmonic has its own tangential wavenumber, and thus reflects to a different angle given by Eq. (\ref{Angle}). The strongest harmonic ($f_0+f_p$) reflects to $\theta_2=42^\circ$, as shown in Fig. \ref{sfig:NpathRef25}. The reflected spectra for both polarizations are given in Fig. \ref{fig:25steeringA}.
The conversion loss and sideband suppression for both polarizations are provided in example 3 of Table \ref{tab:sim}.

\begin{figure}[t]
\subfloat[\label{sfig:N42steeringTEA}]{%
  \includegraphics[clip,width=0.5\columnwidth]{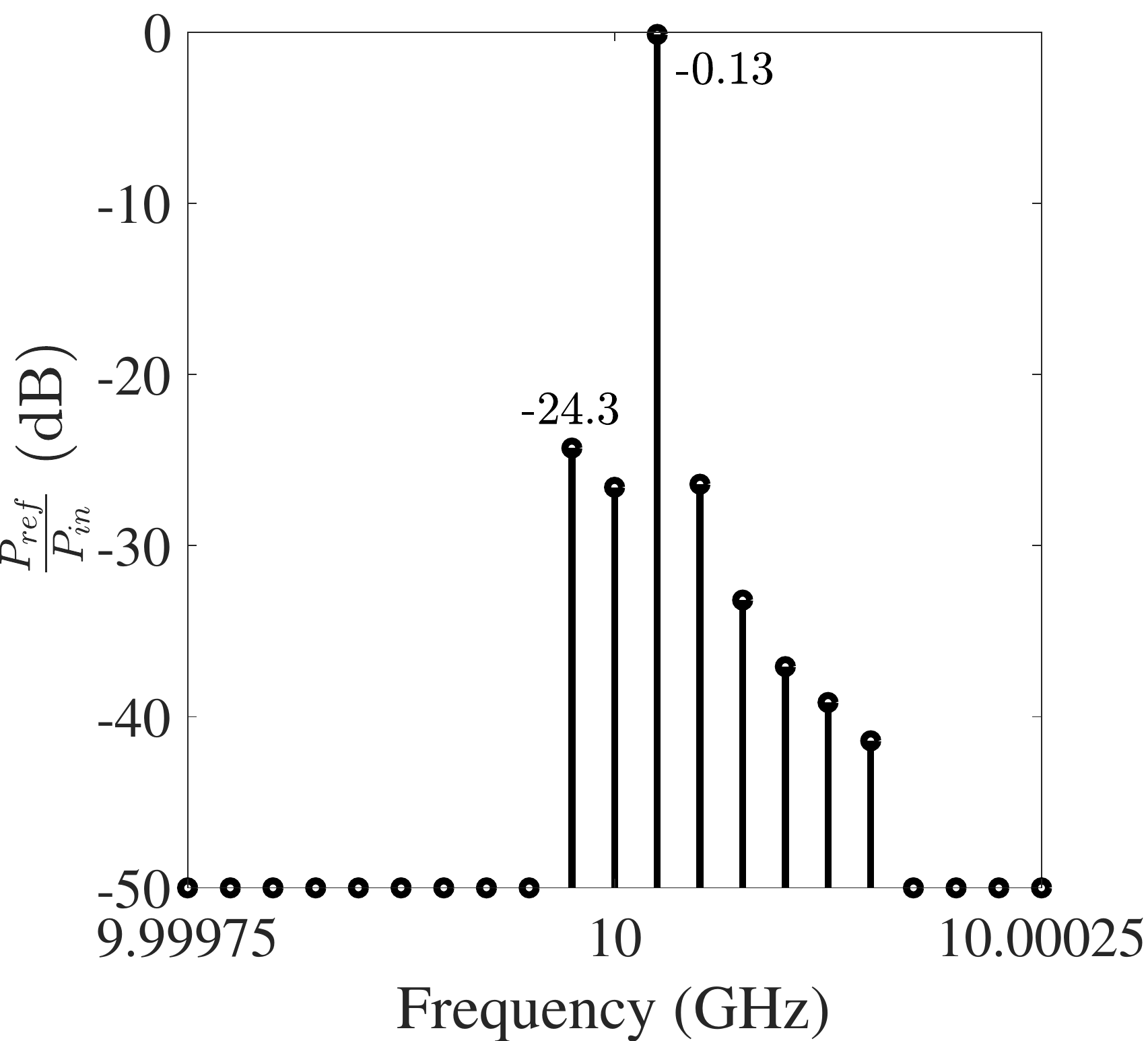}%
}\hfill
\subfloat[\label{sfig:N42steeringTMA}]{%
  \includegraphics[clip,width=0.5\columnwidth]{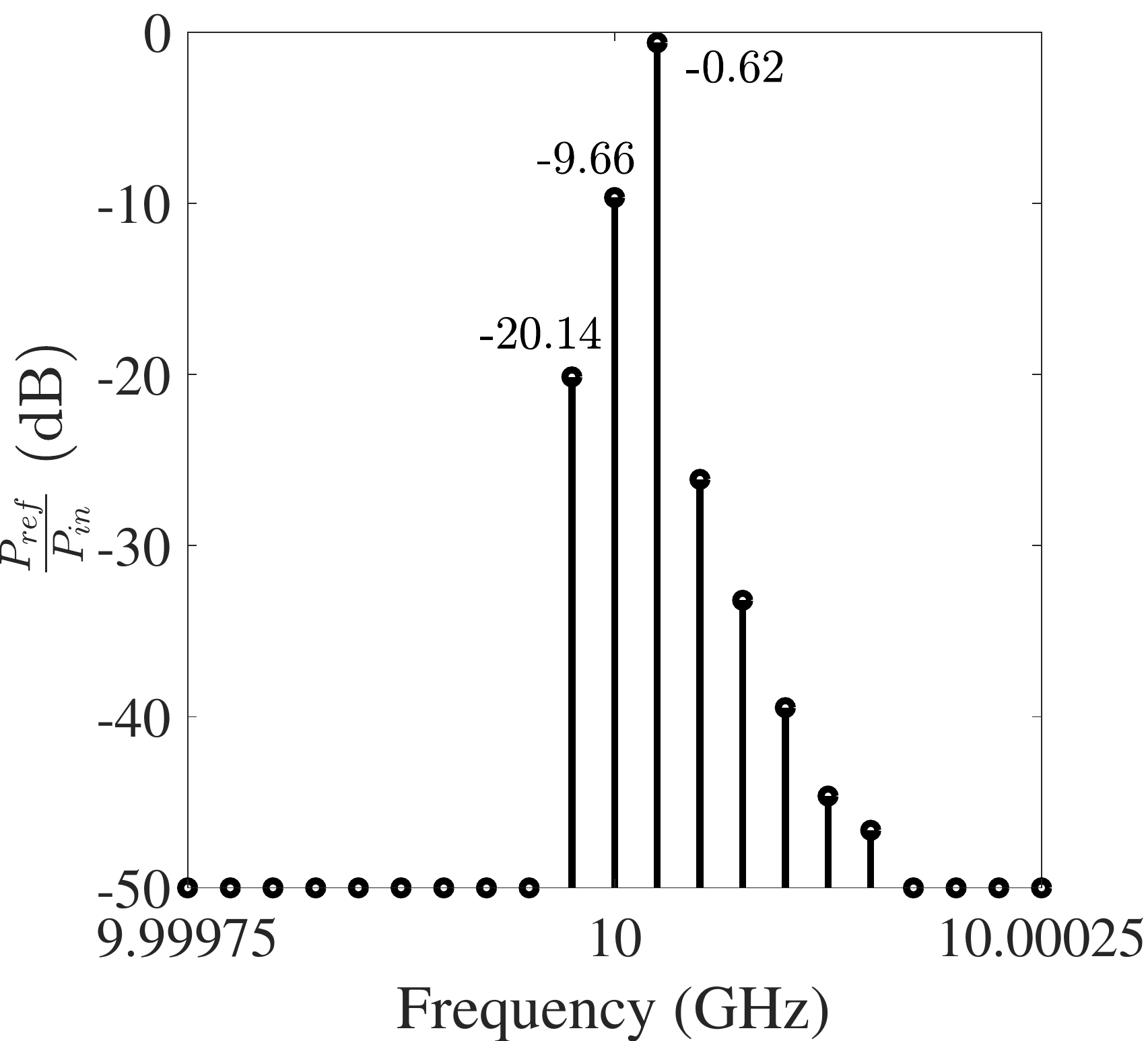}%
}\quad
\caption{Theoretical reflection spectrum of the homogenized, lossless, SD-TWM metasurface with an incident angle of $-42^\circ$. (a) 20-path ($N=20$) modulation for TE polarization. (b) 20-path ($N=20$) modulation for TM polarization.}
\label{fig:N42steeringA}
\end{figure}

Let's consider another example (shown in Fig. \ref{sfig:NpathRef42}) where the spatio-temporal modulation of the metasurface and incident frequency are kept the same, but the incident and reflected angles are swapped. The incident angle is $\theta_2=-42^\circ$. Each scattered harmonic pair of the reflected field has tangential wavenumber 
\begin{align}\label{KxSteering42}
{\left. {{k_{xrq}}} \right|_{r = 0}}&=\frac{q}{4}k_0-k_0\sin(42^\circ).
\end{align}
where $q=0,\pm 1, 2, 3, 4, 5, 6$. The metasurface frequency translates the incident signal at $f_0$ to the harmonic pair $(r=0,q=1)$, which is at frequency $f=f_0+f_p$. %The topology is shown in Fig. \ref{sfig:NpathRef42}.
The tangential wavenumber of the harmonic pair $(r=0,q=1)$ can be easily calculated as $-k_0\sin (25^\circ)$. When modulation frequency $f_p$ is comparable to incident frequency $f_0$, the reflection angle can differ from $\theta_1=-25^\circ$ due to the significant change in the free-space wavenumber at the reflected frequency \cite{shaltout2015time}, as shown in Supplemental Material IV. However, modulation frequency here is $f_p=25$ kHz, which is far smaller than the incident frequency of $f_0=10$ GHz. As a result, the reflection angle is $-25^\circ$. %and the metasurface can be seen as %reciprocal. 
The reflection spectra for both polarizations are given in Fig. \ref{fig:N42steeringA}.
From example 4 of Table \ref{tab:sim}, it can be seen that the conversion loss and sideband suppression are practically identical to those of the previous example (shown in example 3 of Table \ref{tab:sim}).

\begin{figure}[b]
\subfloat[\label{sfig:RetroCone1}]{%
  \includegraphics[clip,width=0.5\columnwidth]{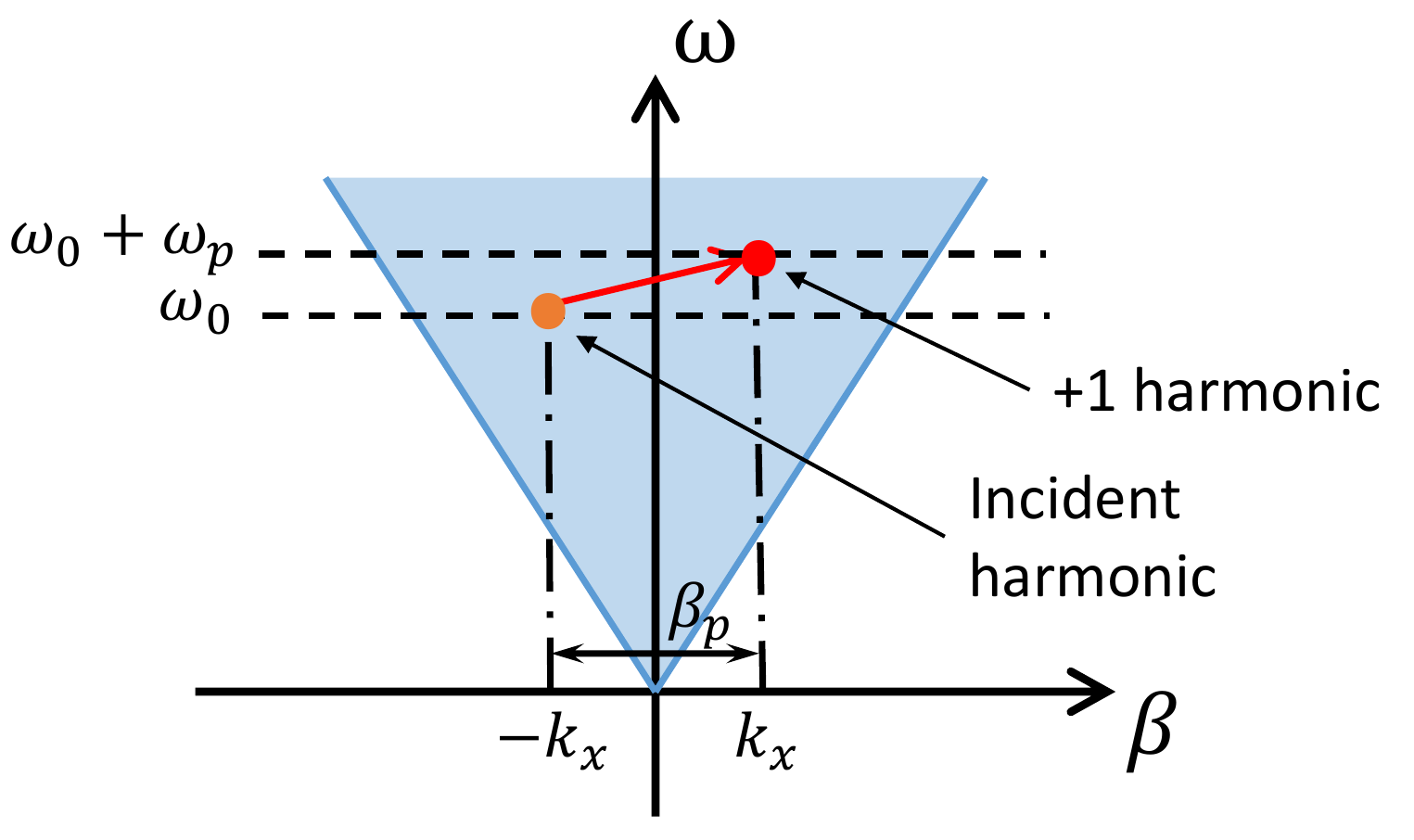}%
}\hfill
\subfloat[\label{sfig:RetroReflect1}]{%
  \includegraphics[clip,width=0.5\columnwidth]{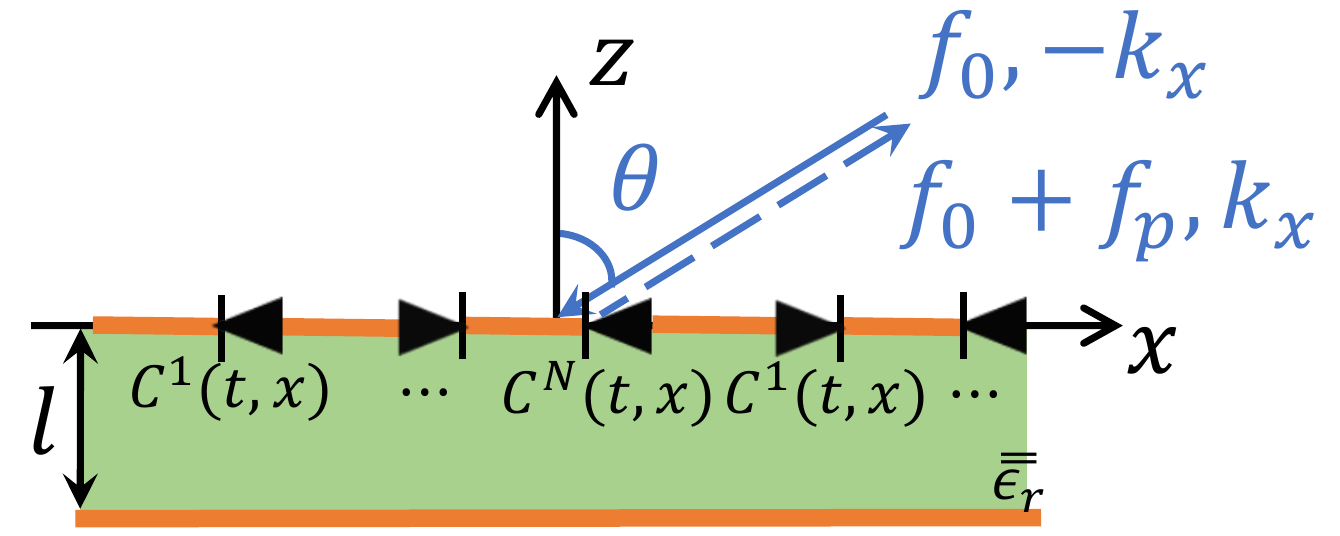}%
}\quad
\caption{(a) Graphic representation of the spatial and temporal frequency  shift for a relatively large path number $N$. (b) Corresponding retroreflection performance for a relatively large path number $N$.}
\label{fig:Retro1}
\end{figure} 

\begin{figure}[t]
\subfloat[\label{sfig:Retro1TEA}]{%
  \includegraphics[clip,width=0.5\columnwidth]{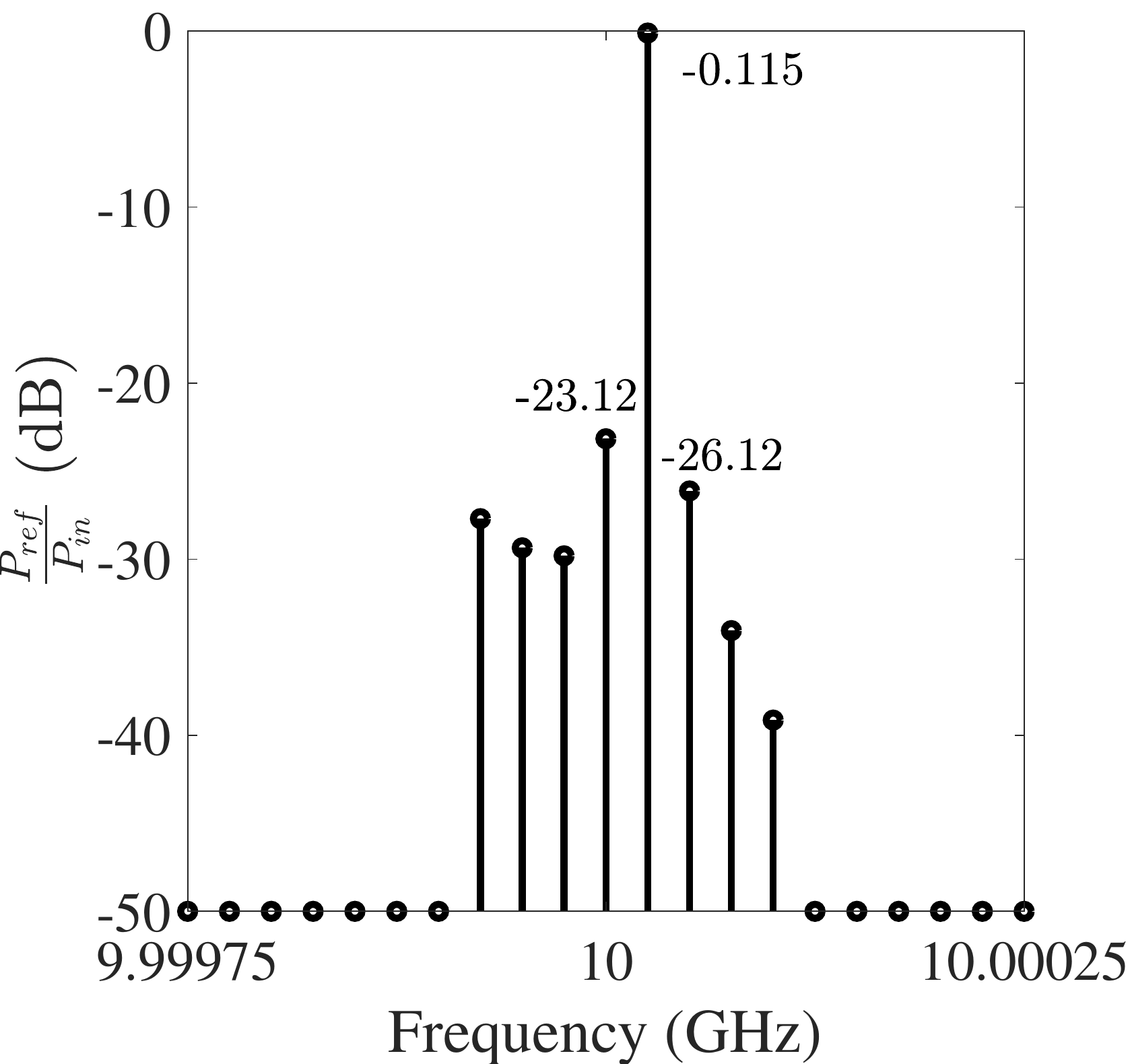}%
}\hfill
\subfloat[\label{sfig:Retro1TMA}]{%
  \includegraphics[clip,width=0.5\columnwidth]{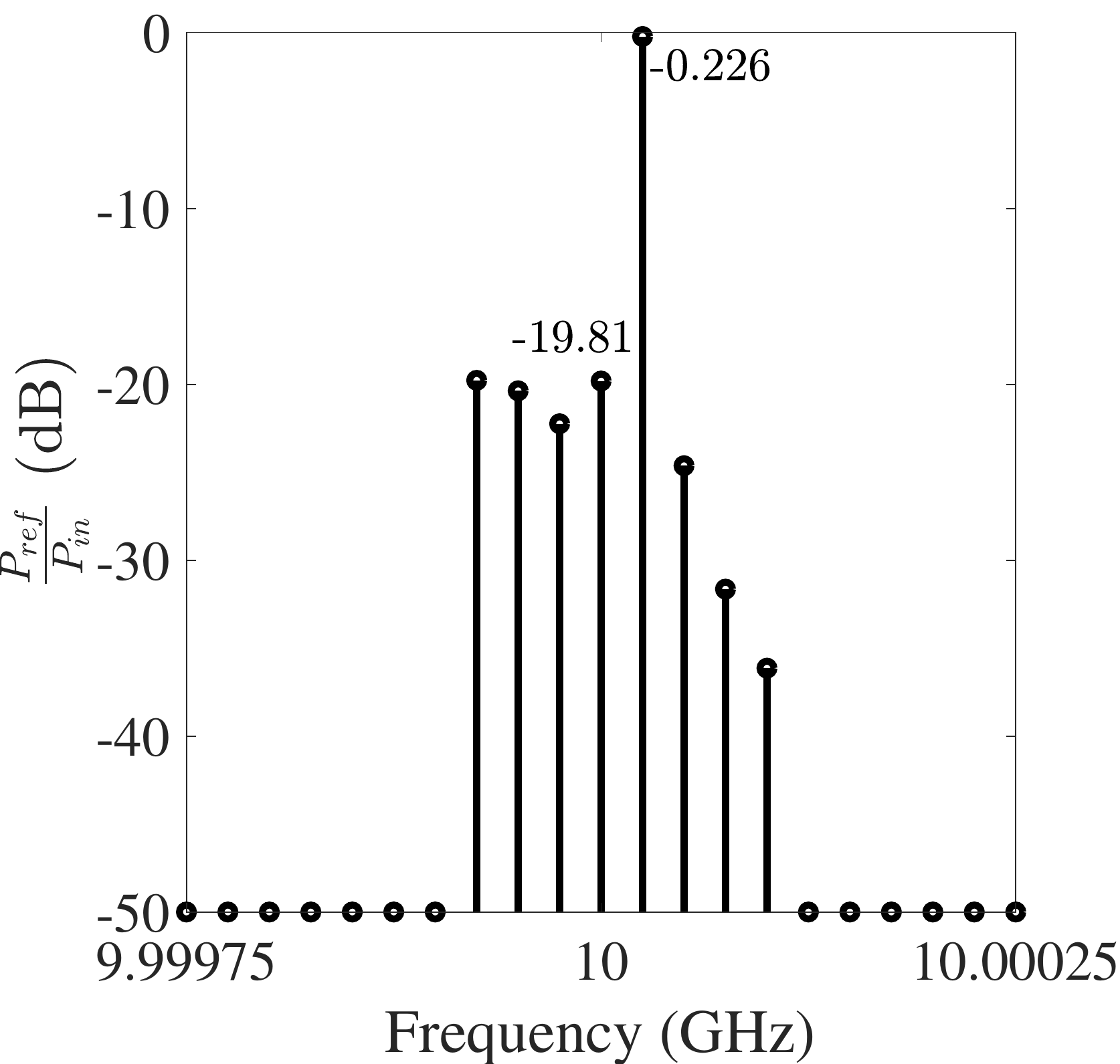}%
}\quad
\caption{Theoretical reflection spectrum of the homogenized, lossless, SD-TWM metasurface for an incident angle of $7.18^\circ$. (a) 20-path ($N=20$) modulation for TE polarization. (b) 20-path ($N=20$) modulation for TM polarization.}
\label{fig:Retro1A}
\end{figure}

Furthermore, in this regime of large spatial modulation period, the incident angle can be chosen to achieve retroreflection.
According to Eq. (\ref{Angle}), setting $\theta_{rq}=-\theta_i$ yields an expression for the incidence angles at which retroreflection occurs for the converted spatio-temporal harmonic. For this case, the modulation wavenumber $\beta_p=2k_x$, as shown in Fig. \ref{fig:Retro1}. The reflected wave propagates back to the source with an upconverted frequency. The retroreflection angle $\theta_i$ can be calculated by solving $\theta_{0,1}=-\theta_i$ in Eq. (\ref{Angle}),
\begin{equation}\label{RetroAngle1}
\theta_i=-\arcsin \frac{\beta_p}{2k_0}= - \arcsin \frac{\lambda _0}{2Nd_0}.
\end{equation} 
Here, the number of paths is chosen to be $N=20$, and the retroreflection angle is calculated to be $\theta_i=-7.18^\circ$.
The calculated reflection spectra for both polarizations are shown in Fig. \ref{fig:Retro1A}. The spectra clearly show a Doppler shift to frequency $f_0+f_p$.
The conversion loss and sideband suppression for both polarizations are provided in example 5 of Table \ref{tab:sim}.
Note that in Fig. \ref{fig:Retro1A}, only the harmonic pair $(r=0,q=1)$ (at frequency $f_0+f_p$) is retroreflective. The reflection angle of other harmonics can be calculated based on Eq. (\ref{Angle}). 

\subsection{Wavelength-scale spatial modulation period ($|k_x|+|\beta_p|>k_0,\ \&\  ||k_x|-|\beta_p||<k_0 $)} \label{sec:WavePeriod}

\begin{figure}[t]
\subfloat[\label{sfig:RetroCone2}]{%
  \includegraphics[clip,width=0.5\columnwidth]{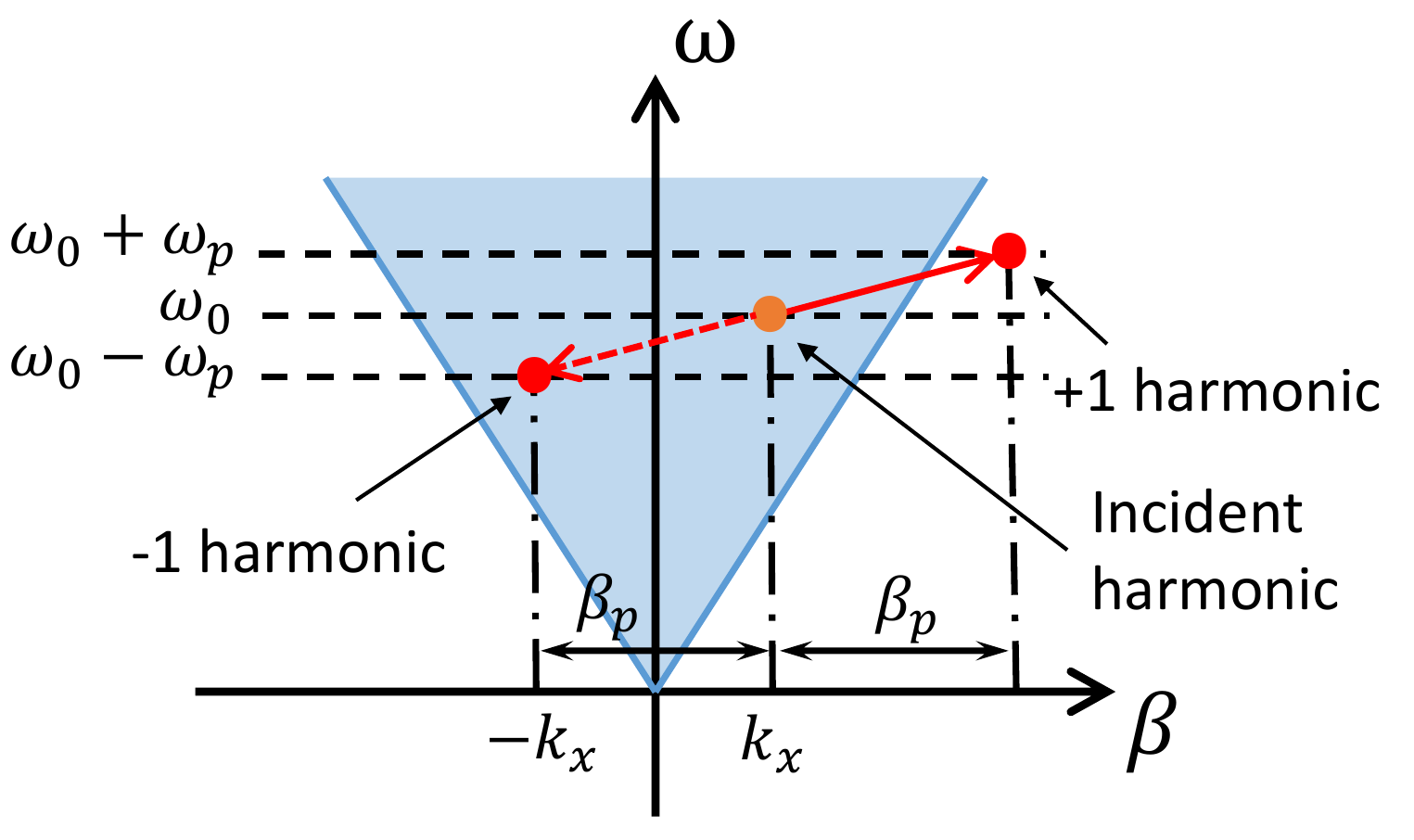}%
}\hfill
\subfloat[\label{sfig:RetroCone3}]{%
  \includegraphics[clip,width=0.5\columnwidth]{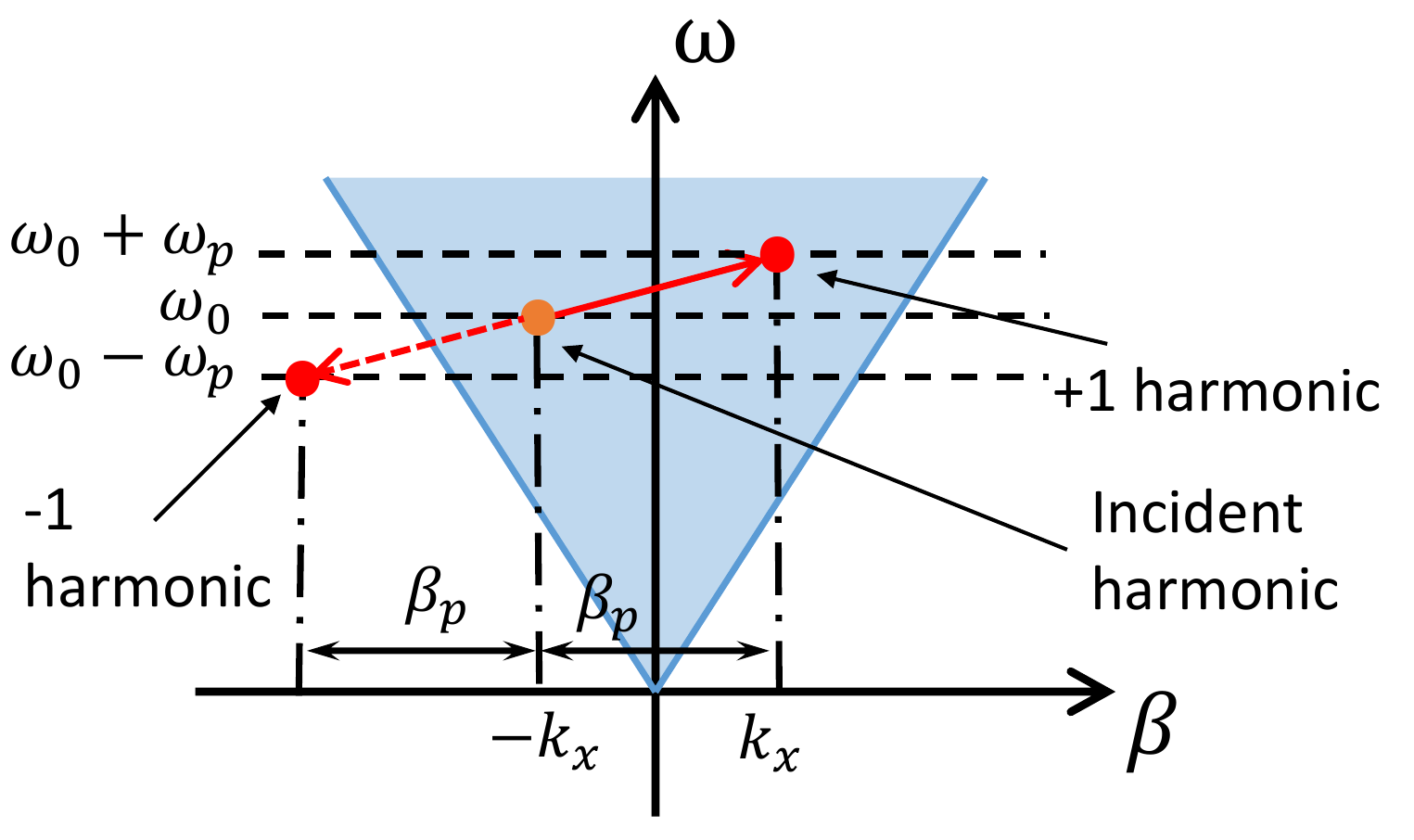}%
}\quad
\caption{Graphic representation of the spatial and temporal frequency shifts for a path number $N=4$. (a) The incident tangential wavenumber $k_x$ is positive. (b) The incident tangential wavenumber $k_x$ is negative.}
\label{fig:Retro2LightCone}
\end{figure}

In this section, we consider a spatial modulation period that is on the order of the wavelength of radiation ($|k_x|+|\beta_p|>k_0\ \&\  ||k_x|-|\beta_p||<k_0 $). In this regime, either the $+1$ ($k_x+\beta_p$) or the $-1$ ($k_x-\beta_p$) spatial harmonic is inside the light cone, as shown in Fig. \ref{fig:Retro2LightCone}. For the fixed stixel size of $d_0=\lambda_0/5$, 4-path modulation ($N=4$) is chosen to satisfy the wavelength-scale period condition. In this regime, the SD-TWM metasurface allows both small and large period electromagnetic effects. That is, both subharmonic mixing and angular deflection can be simultaneously achieved. In this section, the deflective and retroreflective behavior of the metasurface is showcased for various scenarios. In the first case, the metasurface exhibits simultaneous subharmonic frequency translation and deflection. The incident angle is specifically chosen to achieve subharmonic frequency translation in retroreflection. In the second case, we show that the retroreflective frequency can be switched by changing the temporal phase modulation waveform to sinusoidal.

\subsubsection{Deflective/retroreflective subharmonic frequency translation}\label{sec:RetroSub}

\begin{figure}[t]
\subfloat[\label{sfig:Retro2TEA}]{%
  \includegraphics[clip,width=0.5\columnwidth]{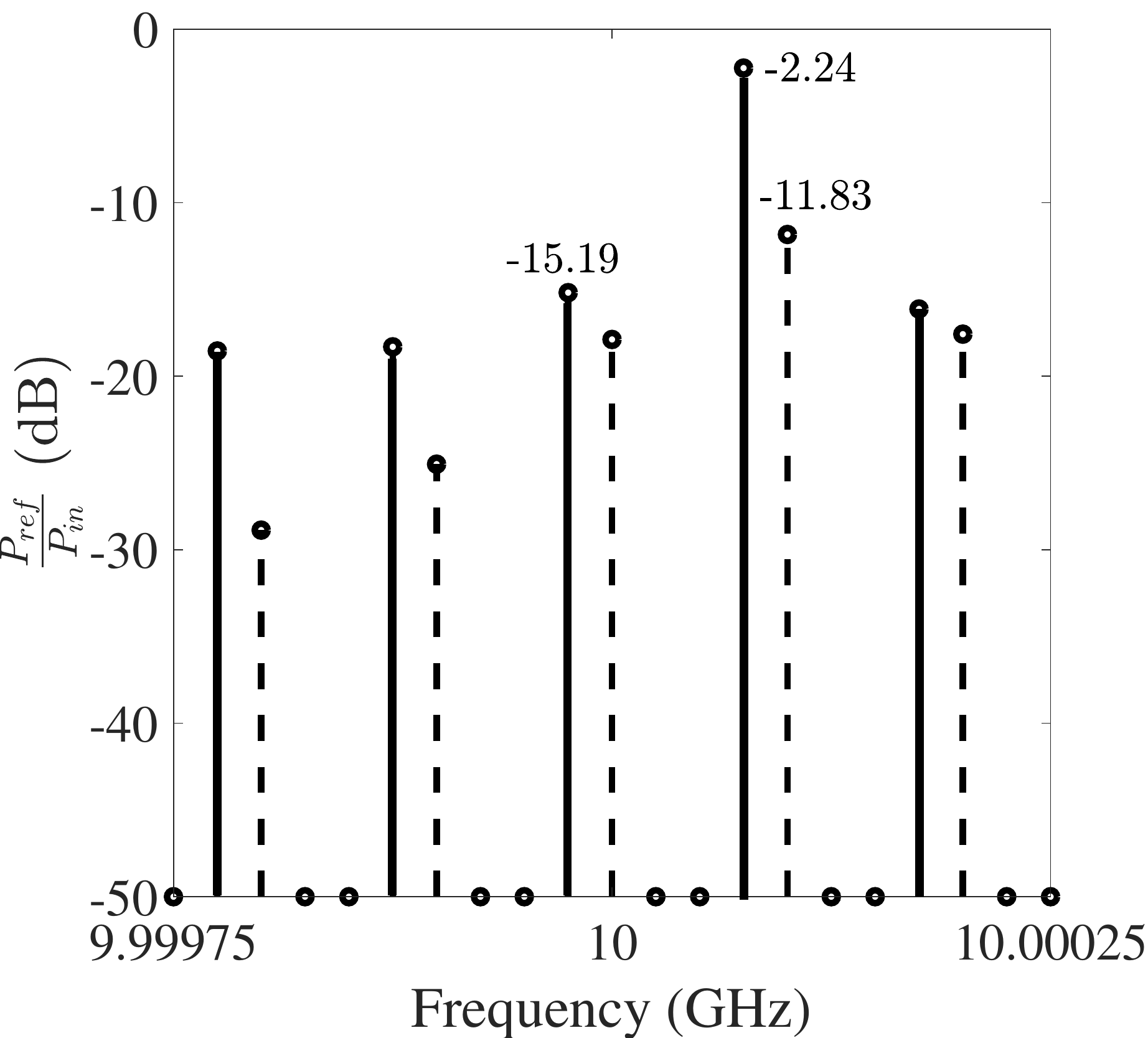}%
}\hfill
\subfloat[\label{sfig:Retro2NTEA}]{%
  \includegraphics[clip,width=0.5\columnwidth]{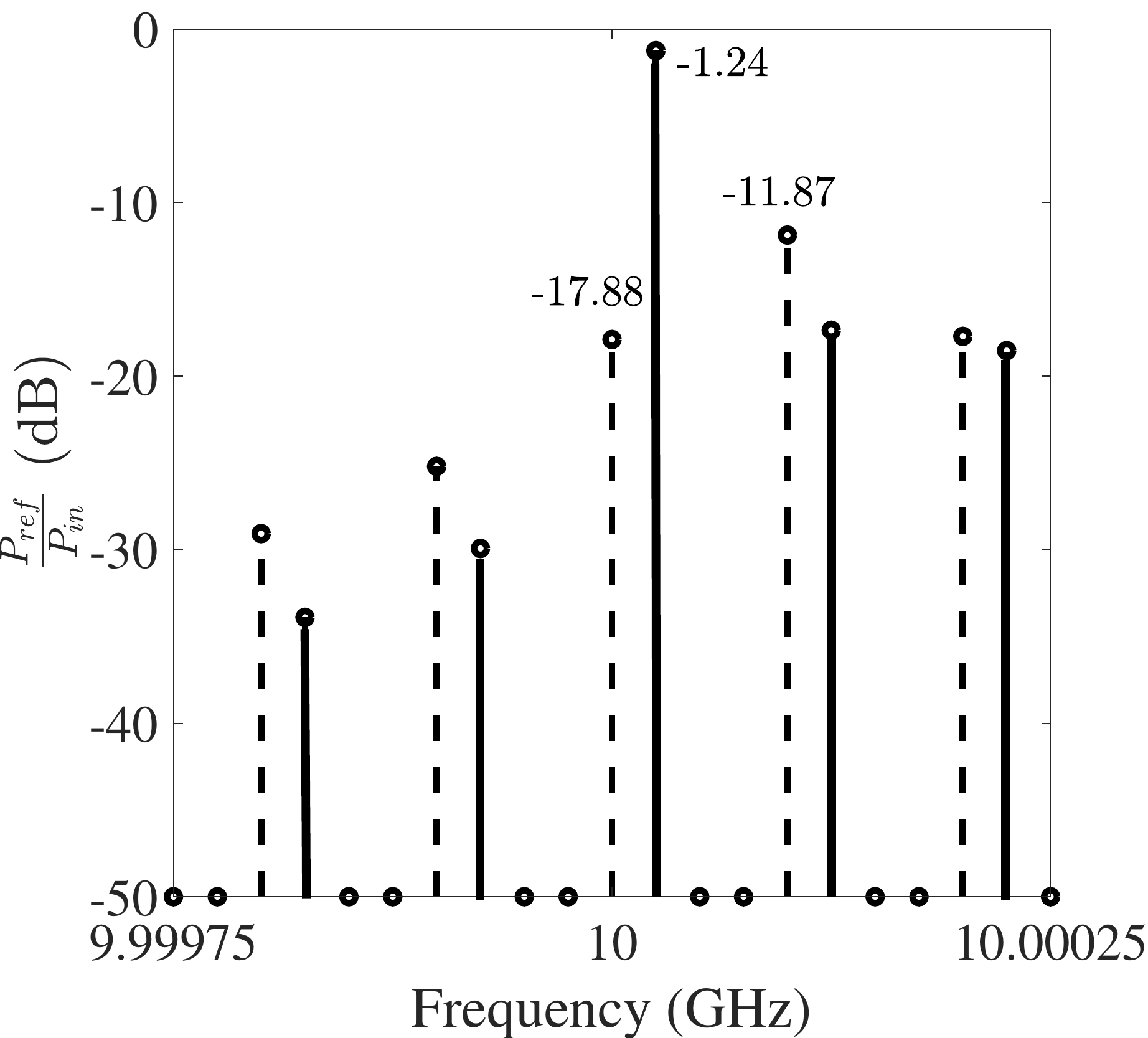}%
}\quad
\subfloat[\label{sfig:Retro2TMA}]{%
  \includegraphics[clip,width=0.5\columnwidth]{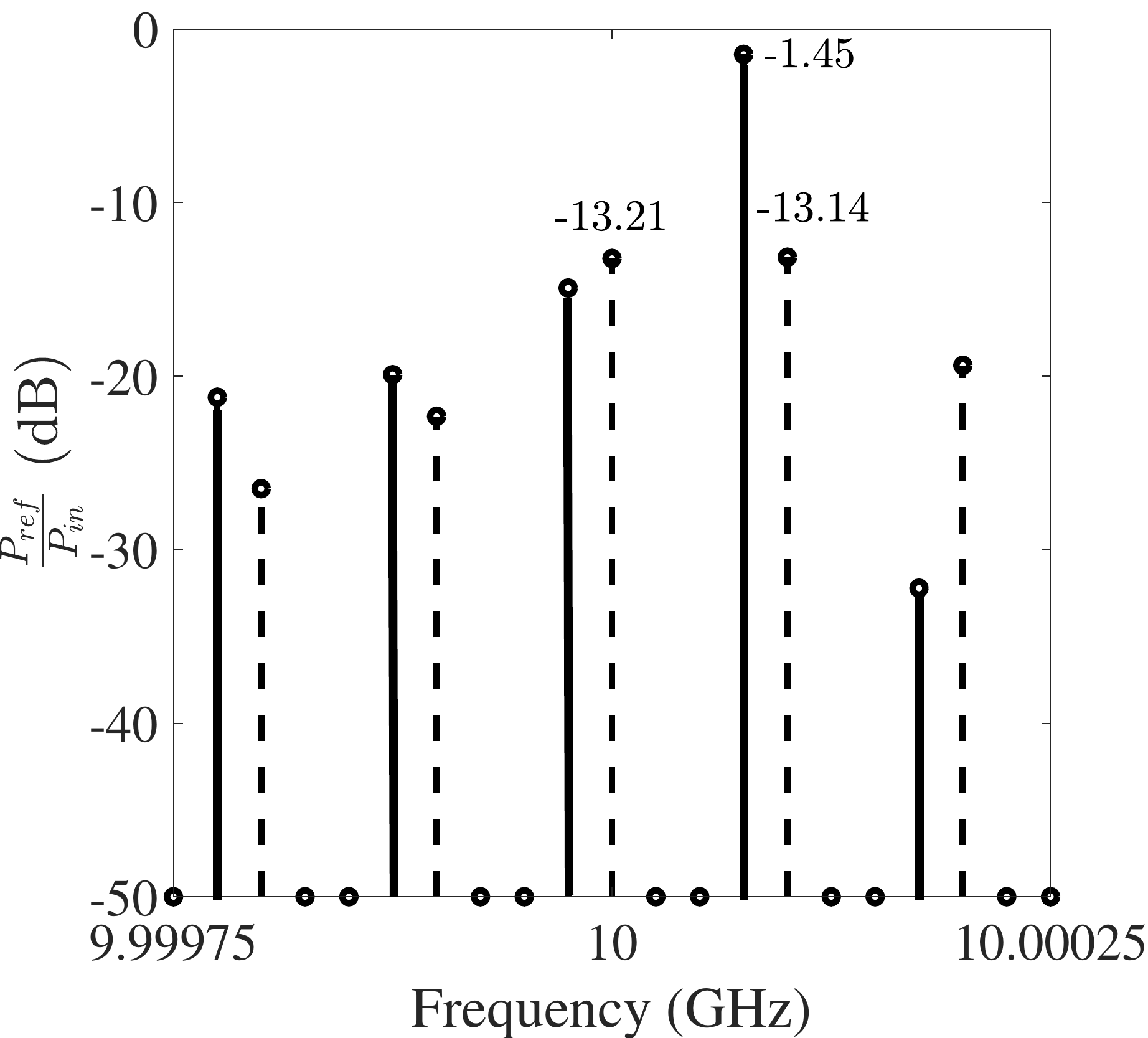}%
}\hfill
\subfloat[\label{sfig:Retro2NTMA}]{%
  \includegraphics[clip,width=0.5\columnwidth]{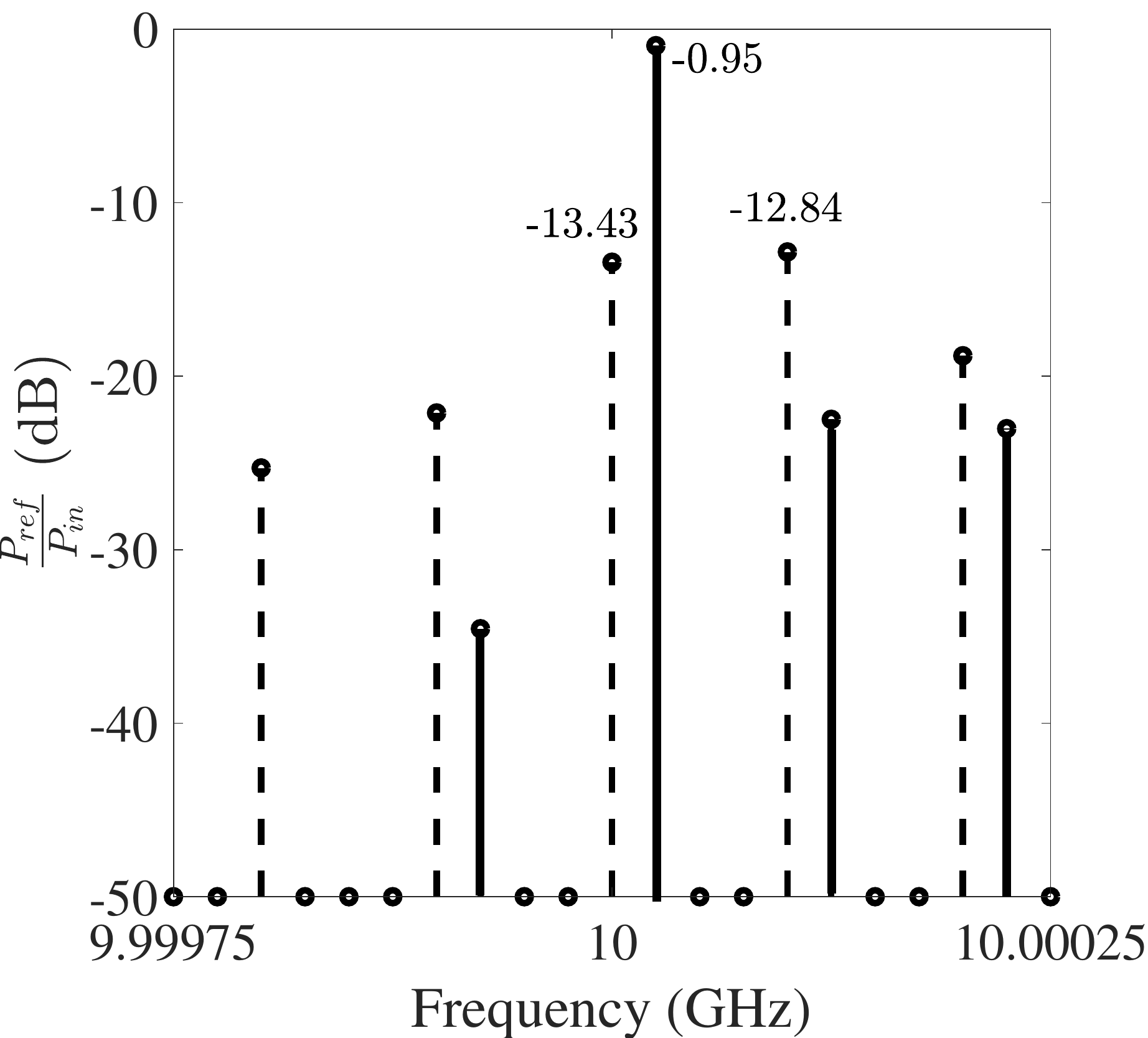}%
}
\caption{Theoretical retroreflection spectrum of the homogenized, lossless, SD-TWM metasurface. The harmonics denoted by solid lines retroreflect. The harmonics denoted by the dashed lines reflect in the specular direction. (a) 4-path ($N=4$) modulation for TE polarization for an incident angle of $39^\circ$. (b) 4-path ($N=4$) modulation for TE polarization for an incident angle of $-39^\circ$. (c) 4-path ($N=4$) modulation for TM polarization for an incident angle of $39^\circ$. (d) 4-path ($N=4$) modulation for TM polarization for an incident angle of $-39^\circ$.}
\label{fig:Retro2A}
\end{figure}

First, let us consider the example shown in Fig. \ref{sfig:RetroCone2}, where a wave is incident on the metasurface with a positive $k_x$ value. According to Eq. (\ref{Kx}), the radiated harmonics are those with:
\begin{align}\label{NpathHarmonicsRetro}
q+rN=0\ \text{or}\ q+rN=-1.
\end{align}
Eq. (\ref{NpathHarmonicsRetro}) implies that the radiated reflected wave contains frequency harmonics at $f_0+rNf_p$ and $f_0+(rN-1)f_p$, where $r\in \mathbb{Z}$. Under a capacitance variation that generates sawtooth reflection phase, the reflected wave is upconverted to the first radiated frequency harmonic, which in this case is the harmonic pair ($r=-1,q=3$). Therefore, the reflected wave is Doppler shifted to a frequency $f_0+3f_p$. In addition, an incident angle is chosen such that the wave is retroreflected: $\beta_p=2k_x$ (see Fig. \ref{sfig:RetroCone2}). The retroreflection angle can be calculated by setting $\theta_{-1,3}=-\theta_i$ in Eq. (\ref{Angle}),
which is $39^\circ$ for a path number of $N=4$.

The calculated retroreflection spectra are shown in Fig. \ref{sfig:Retro2TEA} and \ref{sfig:Retro2TMA}. Doppler-like frequency translation to frequency $f_0+3f_p$ occurs for the incident angle of $39^\circ$, for both polarizations. In addition, the signal at $f_0+3f_p$ is retroreflected. The conversion loss and sideband suppression for both polarizations are provided in example 6 of Table \ref{tab:sim}. Note that in Fig. \ref{sfig:Retro2TEA} and \ref{sfig:Retro2TMA}, only the harmonics represented by a solid line are retroreflected. The harmonics represented by dashed lines are reflected in the specular direction.

\begin{figure}[t]
\subfloat[\label{sfig:RetroReflect2}]{%
  \includegraphics[clip,width=0.5\columnwidth]{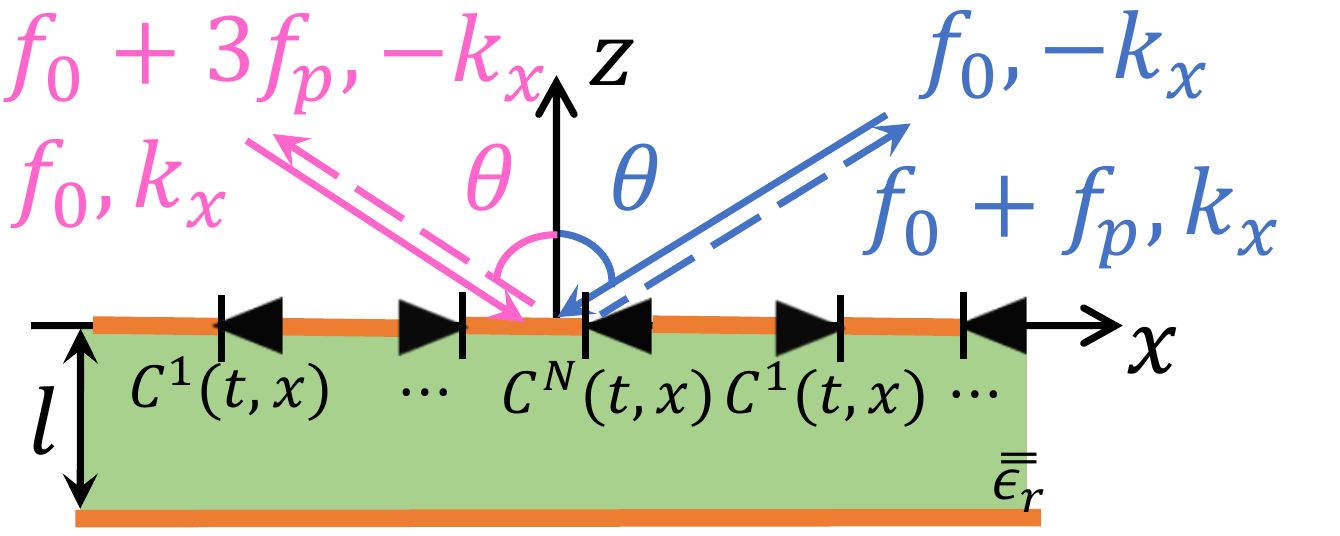}%
}\hfill
\subfloat[\label{sfig:RetroReflect3}]{%
  \includegraphics[clip,width=0.5\columnwidth]{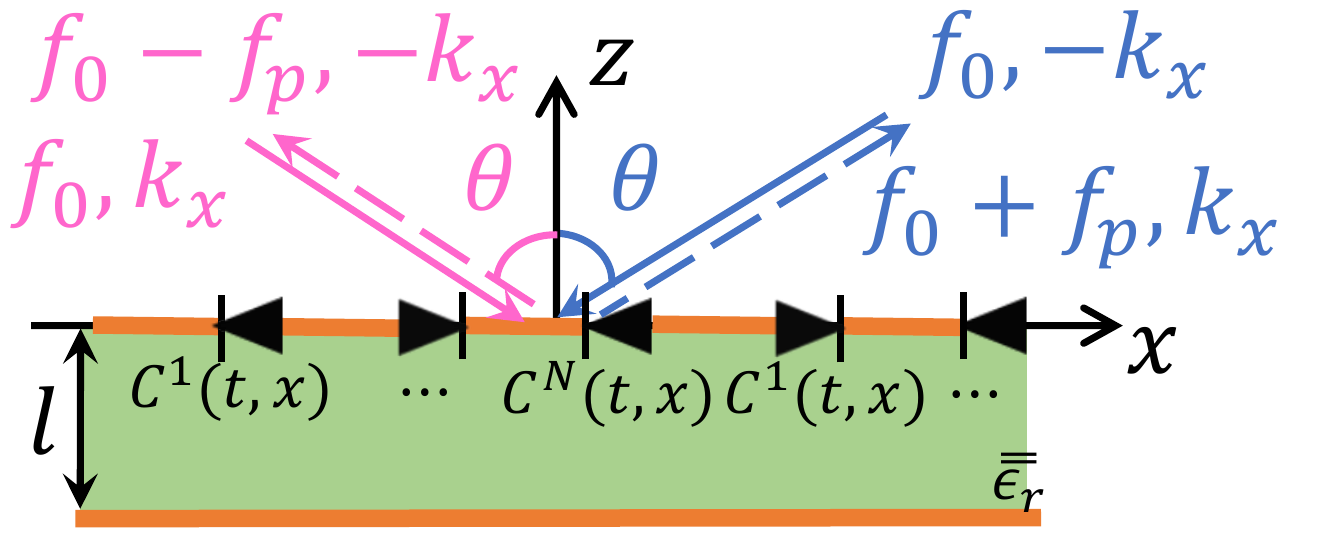}%
}\quad
\caption{Retroreflective performance of the SD-TWM metasurface. The path number is $N=4$, and the retroreflection angle is $\theta=39^\circ$. (a) The sheet capacitance generates a reflection phase on each column that is a sawtooth with respect to time. (b) The capacitance modulation on each column generates sinusoidal reflection phase with respect to time.}
\label{fig:Retro2Reflection}
\end{figure} 

With a wavelength-scale spatial modulation period, the performance of the metasurface is direction-dependent. When the incident angle is $-39^\circ$, as shown in Fig. \ref{sfig:RetroCone3}, it is clear that the radiated harmonics are those with:
\begin{align}\label{NpathHarmonicsRetro2}
q+rN=0\ \text{or}\ q+rN=1.
\end{align} 
In this case, the harmonic pair ($q=1,r=0$) is inside the light cone. Therefore, the metasurface performs serrodyne frequency translation: upconversion to a a frequency $f_0+f_p$. Since $\beta_p=2k_x$, the frequency of interest $f_0+f_p$ is also retroreflected. The calculated reflection spectra are shown in Fig. \ref{sfig:Retro2NTEA} and \ref{sfig:Retro2NTMA}.  Doppler-like frequency translation to frequency $f_0+f_p$ is observed for both polarizations. The conversion loss and sideband suppression for both polarizations are provided in example 7 of Table \ref{tab:sim}. An illustration of the direction-dependent retroreflective behavior of the metasurface, with a wavelength-scale spatial modulation period, is depicted in Fig. \ref{sfig:RetroReflect2}.

\subsubsection{Retroreflective frequency translation with a staggered sinusoidal reflection phase}
 
\begin{figure}[t]
\subfloat[\label{sfig:Retro3WaveformTE}]{%
  \includegraphics[clip,width=0.5\columnwidth]{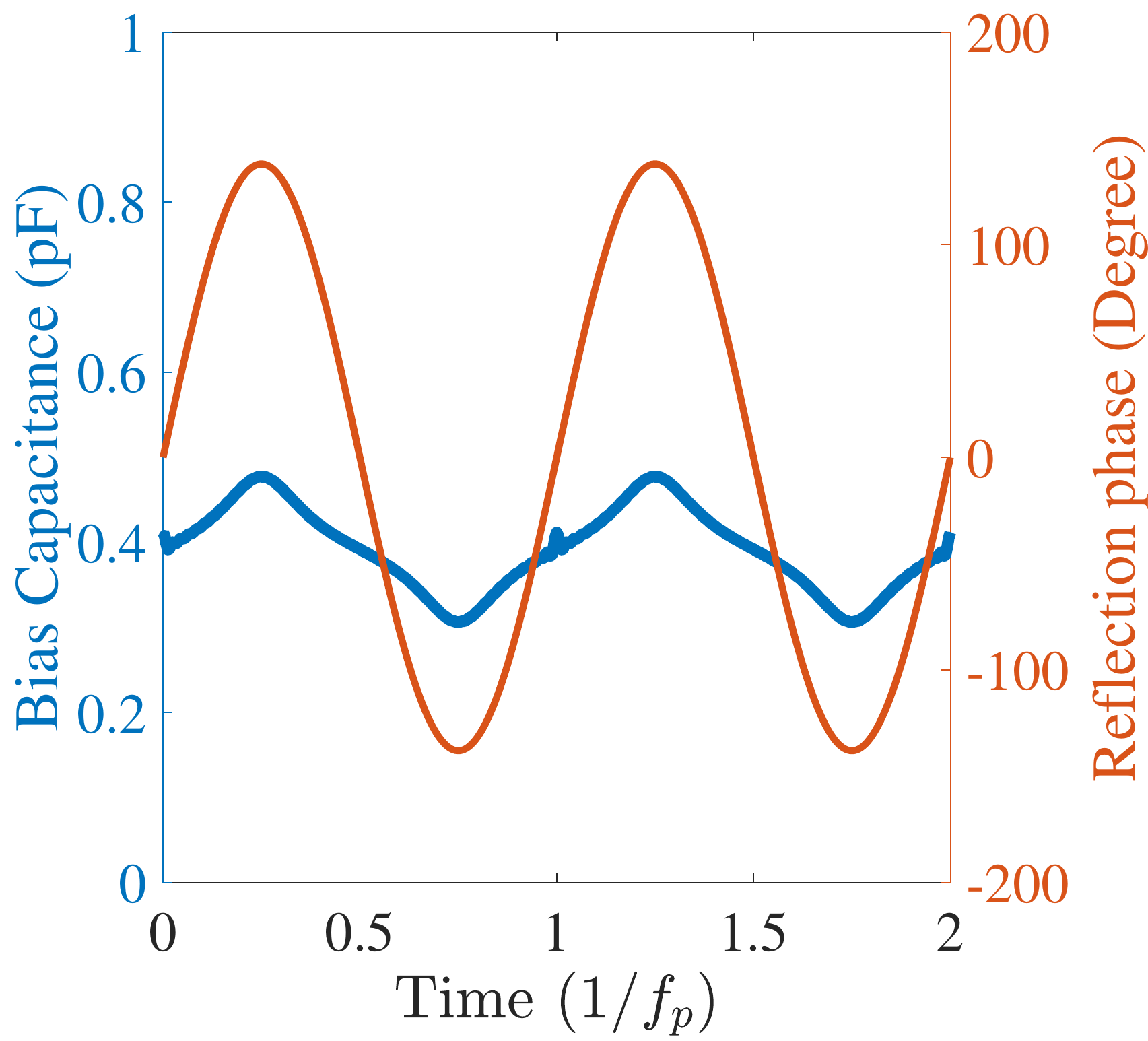}%
}\hfill
\subfloat[\label{sfig:SinSpectrumTE}]{%
  \includegraphics[clip,width=0.5\columnwidth]{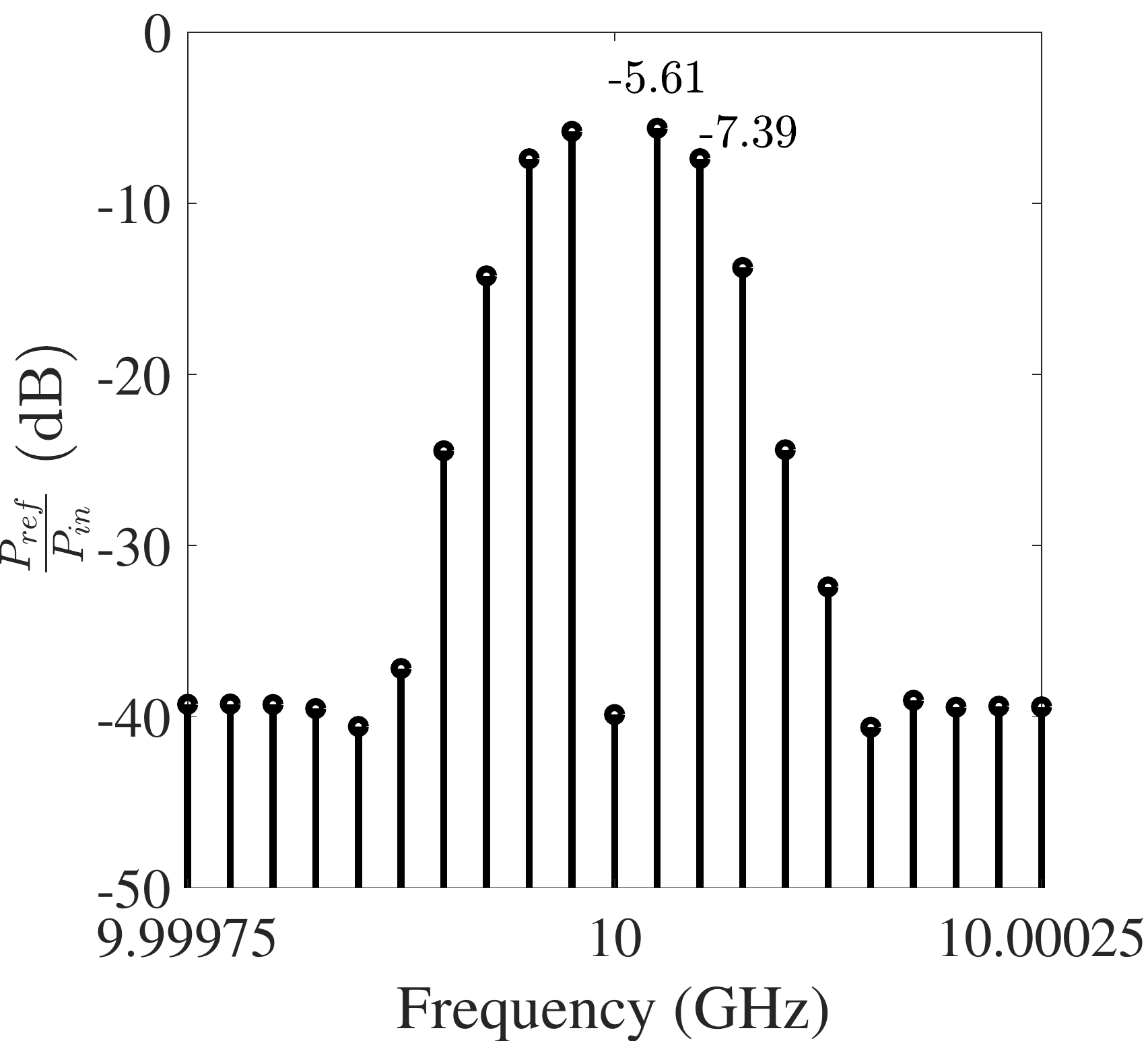}%
}\quad
\subfloat[\label{sfig:Retro3WaveformTM}]{%
  \includegraphics[clip,width=0.5\columnwidth]{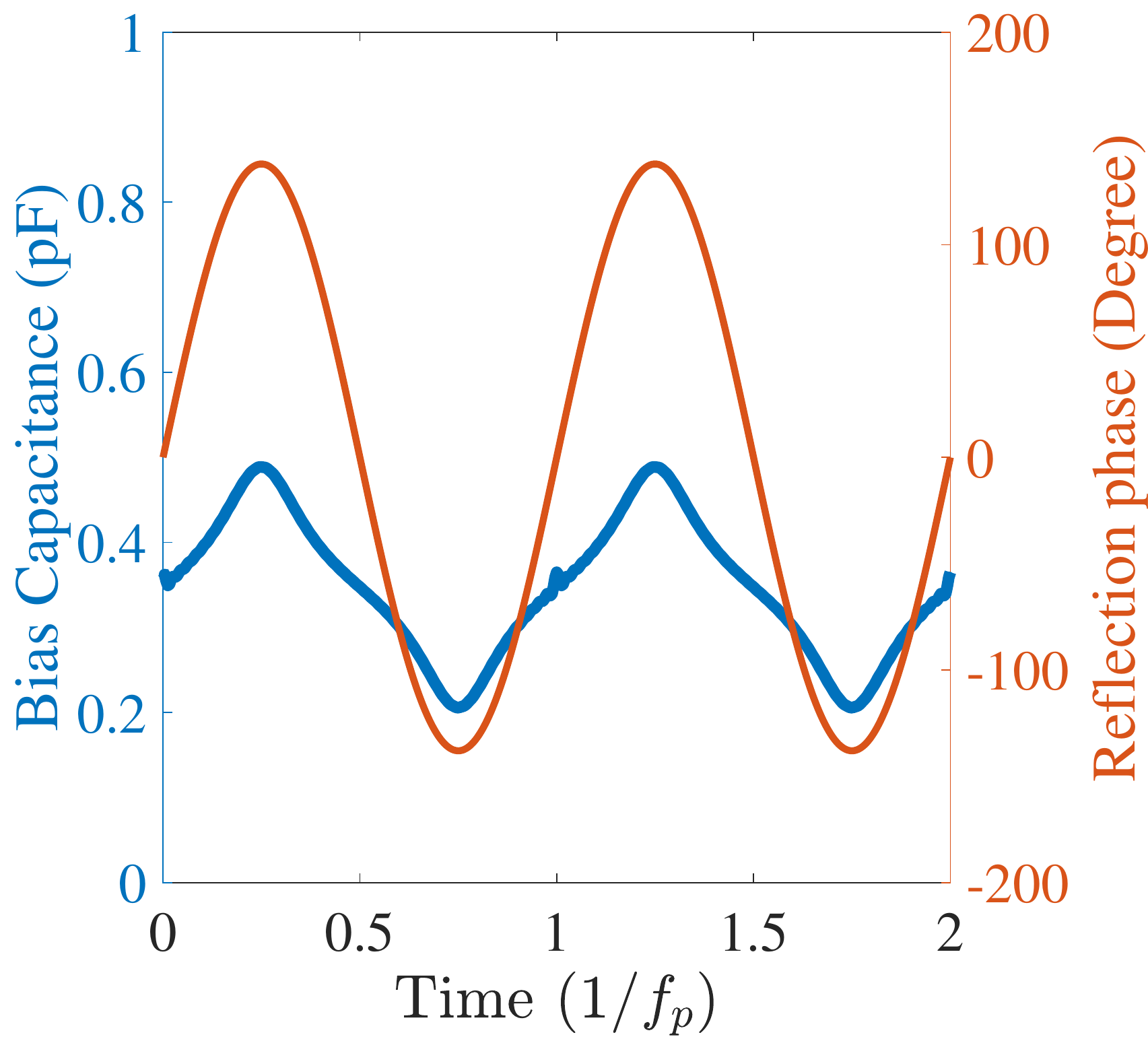}%
}\hfill
\subfloat[\label{sfig:SinSpectrumTM}]{%
  \includegraphics[clip,width=0.5\columnwidth]{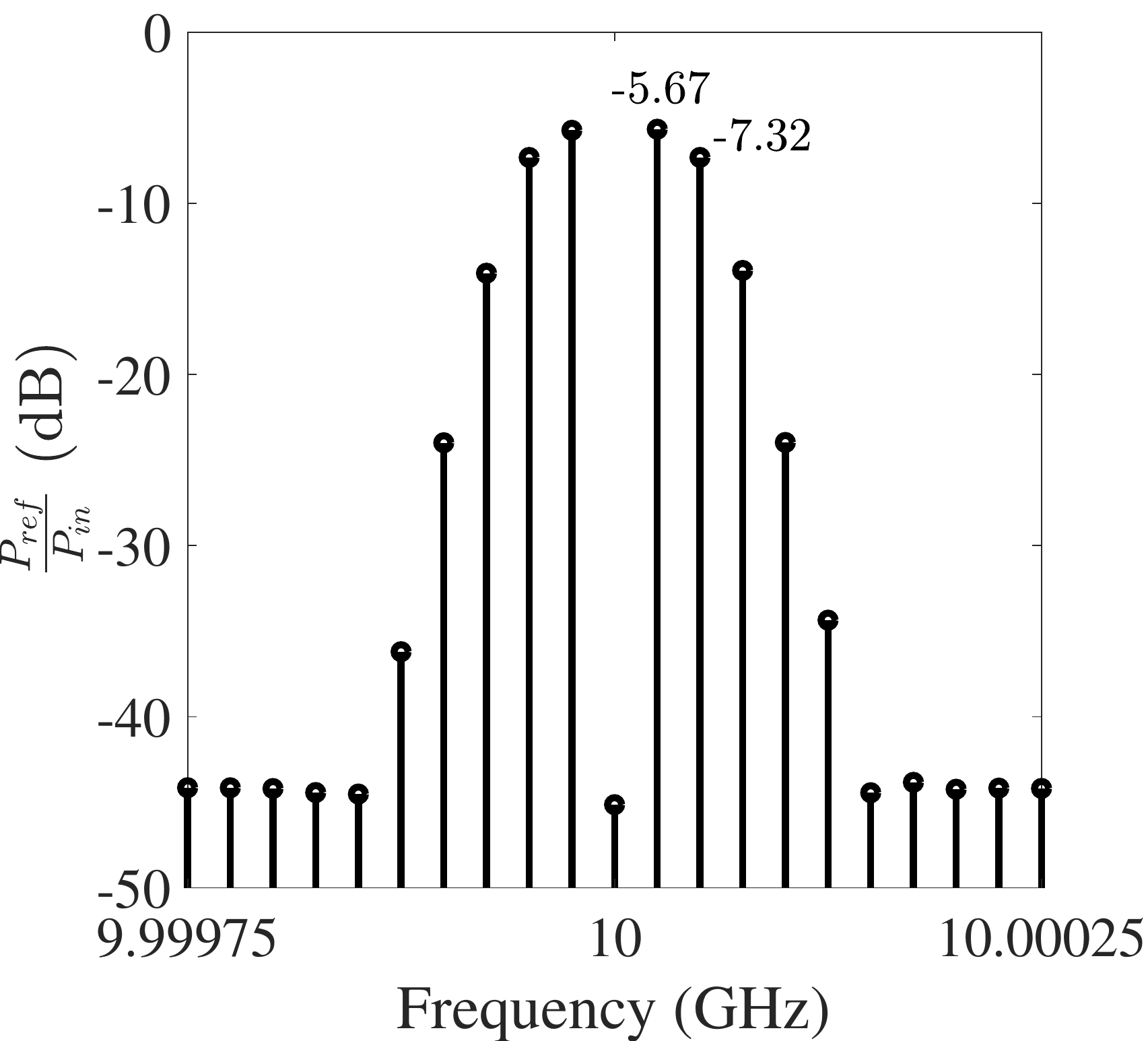}%
}
\caption{(a) Calculated capacitance modulation for a sinusoidal reflection phase versus time for TE polarization. (b)  Analytical reflection spectrum of the homogenized, lossless time-modulated metasurface for TE polarization. (c) Calculated capacitance modulation for a sinusoidal reflection phase versus time for TM polarization. (d) Analytical reflection spectrum of the homogenized lossless time-modulated metasurface for TM polarization.}
\label{fig:Retro3Waveform}
\end{figure}

Here, retroreflection is achieved using stixels that generate staggered sinusoidal reflection phase with respect to time. The capacitance modulation waveform is shown in Fig. \ref{sfig:Retro3WaveformTE} and Fig. \ref{sfig:Retro3WaveformTM} for each polarization. When all the columns of the metasurface are homogeneously biased with the same waveform, the reflection spectra take the form of a Bessel function, as shown in Fig. \ref{sfig:SinSpectrumTE} and Fig. \ref{sfig:SinSpectrumTM}. The reflection phase range is chosen to be $276^\circ$ to suppress the zeroth harmonic in reflection \cite{doerr2014silicon}. Unlike the sawtooth modulation, the sinusoidal reflection phase excites both $+1$ ($r=0,q=1$) and $-1$ ($r=0,q=-1$) frequency harmonics.

\begin{figure}[t]
\subfloat[\label{sfig:Retro2TEASin}]{%
  \includegraphics[clip,width=0.5\columnwidth]{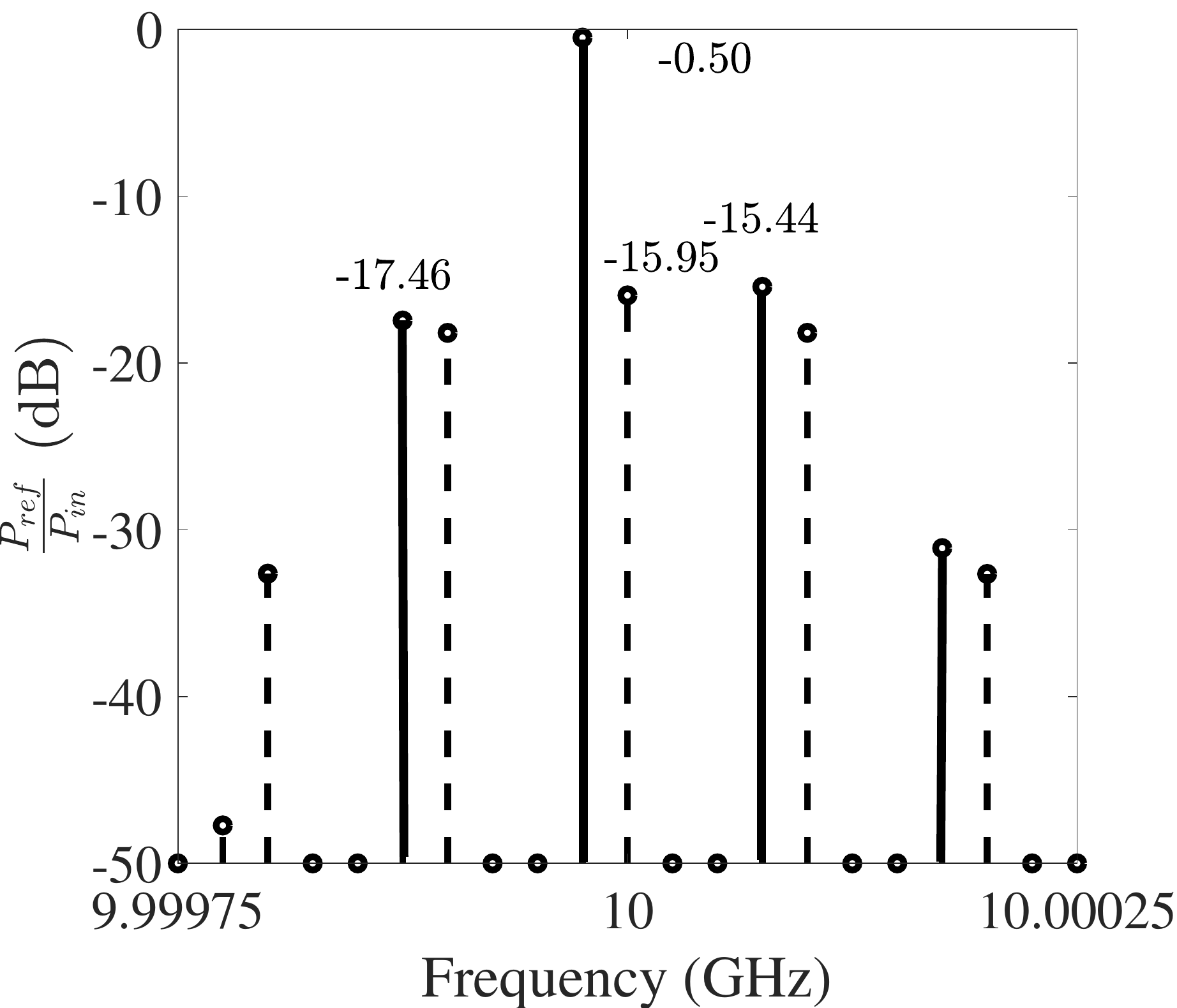}%
}\hfill
\subfloat[\label{sfig:Retro2NTEASin}]{%
  \includegraphics[clip,width=0.5\columnwidth]{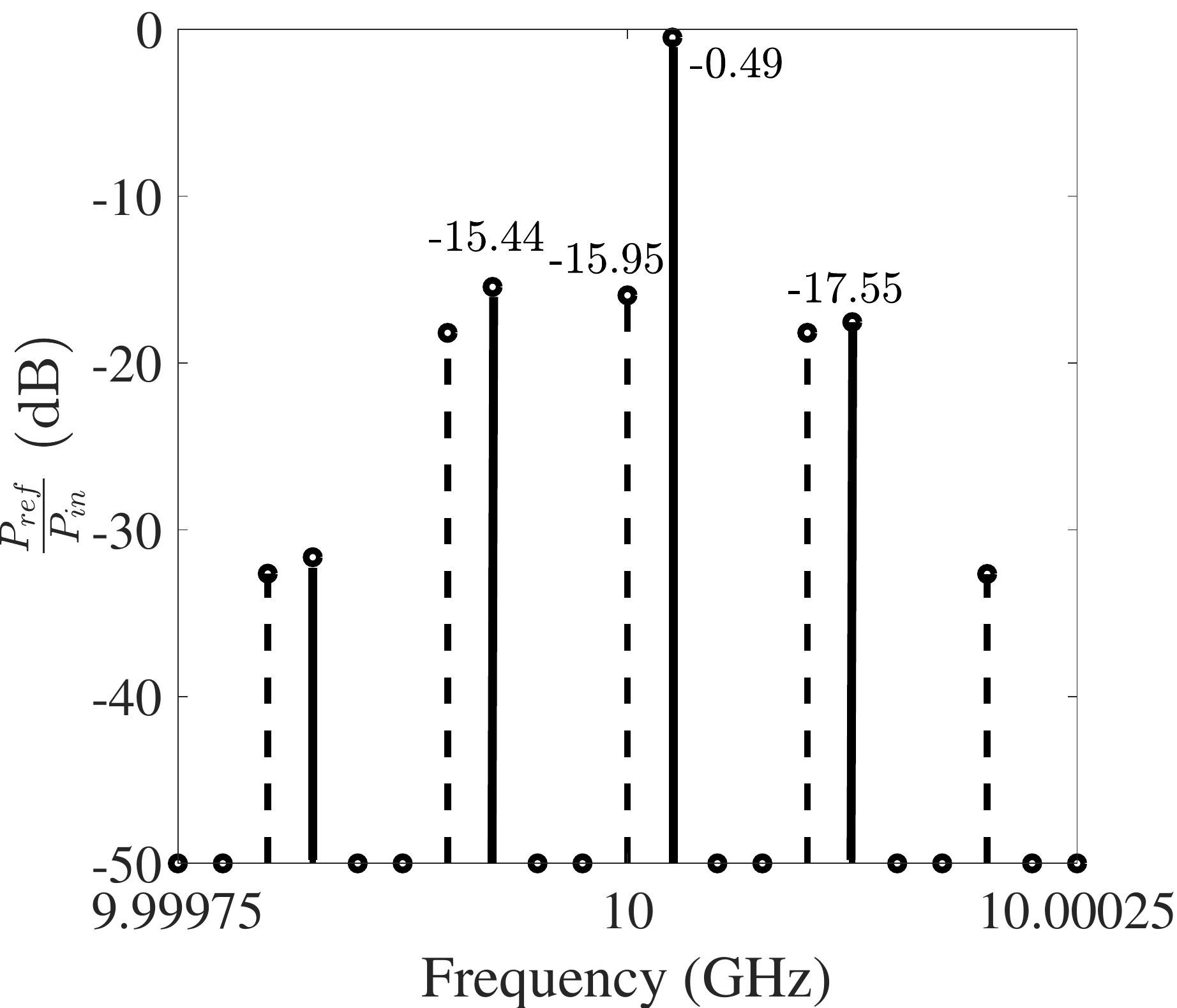}%
}\quad
\subfloat[\label{sfig:Retro2TMASin}]{%
  \includegraphics[clip,width=0.5\columnwidth]{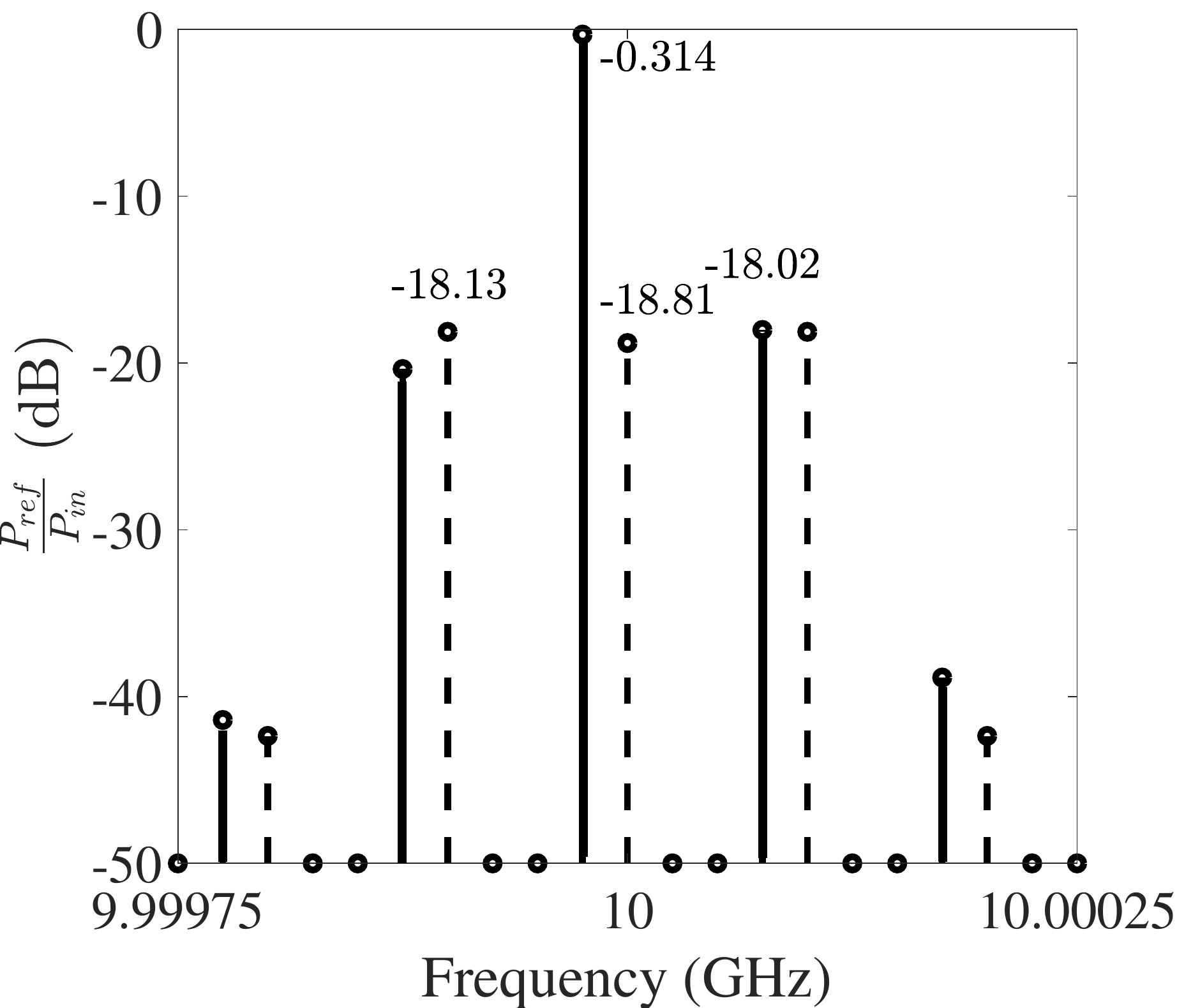}%
}\hfill
\subfloat[\label{sfig:Retro2NTMASin}]{%
  \includegraphics[clip,width=0.5\columnwidth]{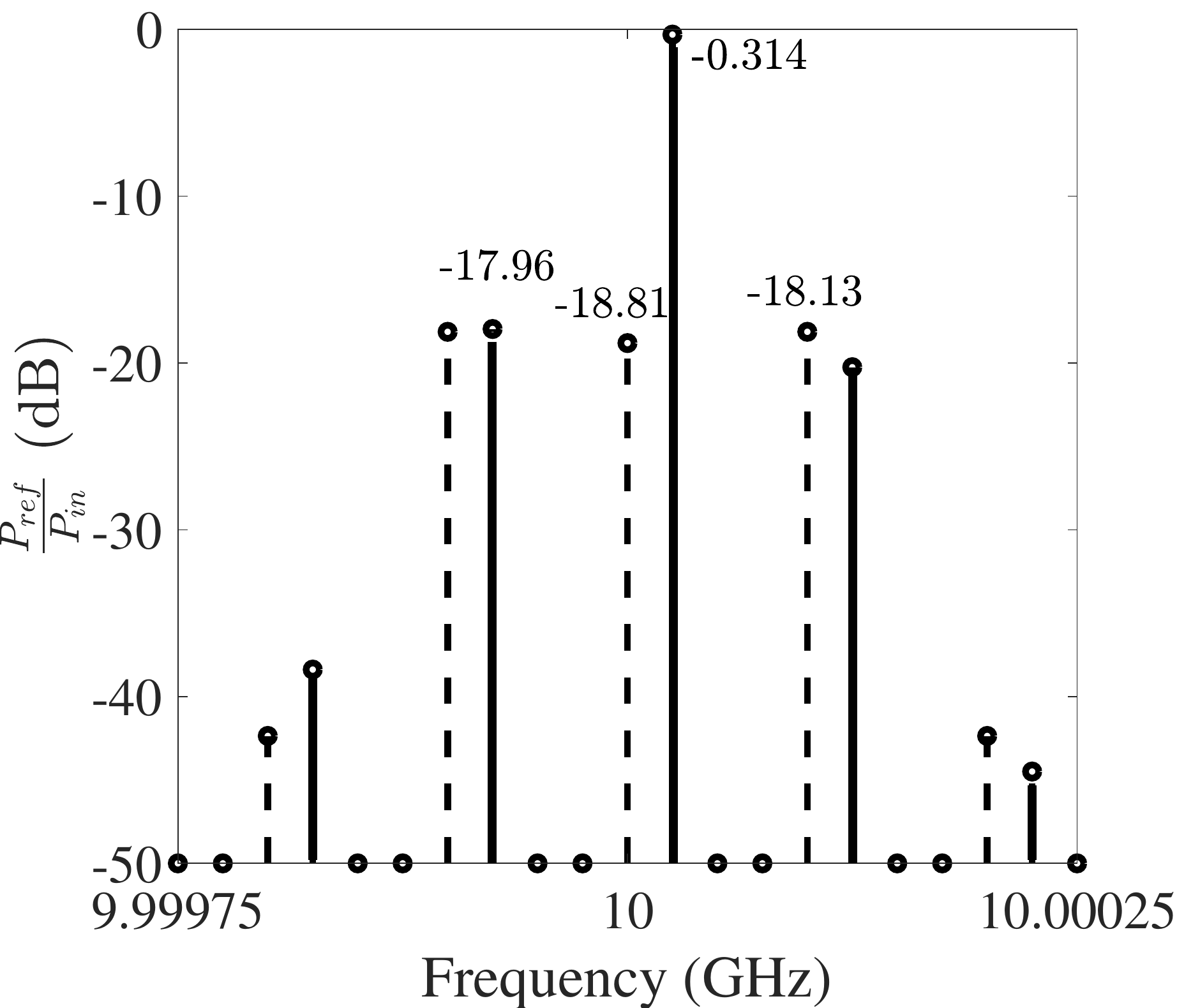}%
}
\caption{Analytical retroreflection spectrum of the homogenized, lossless, SD-TWM metasurface. The capacitance modulation generates a sinusoidal reflection phase on each path. The harmonics denoted by the solid lines are propagating in the retroreflective direction. The harmonics denoted by the dashed lines are propagating in specular direction. (a) 4-path ($N=4$) modulation for TE polarization for an incident angle of $39^\circ$. (b) 4-path ($N=4$) modulation for TE polarization for an incident angle of $-39^\circ$. (c) 4-path ($N=4$) modulation for TM polarization for an incident angle of $39^\circ$. (d) 4-path ($N=4$) modulation for TM polarization for an incident angle of $-39^\circ$.}
\label{fig:Retro2ASin}
\end{figure} 

Fig. \ref{fig:Retro2LightCone} shows that for a positive $k_x$ incident wavenumber, the reflected +1 frequency harmonic is outside of the light cone ($|k_x+\beta_p|>k_0$), and the refleted -1 frequency harmonic is inside the light cone ($|k_x-\beta_p|<k_0$). Since the +1 frequency harmonic does not radiate and is not supported by the metasurface as a surface wave, the power is reflected from the metasurface with a frequency of $f_0-f_p$. In other words, for a wavelength-scale spatial modulation period, the metasurface supports single-sideband frequency translation with the sinusoidal modulation. 

When the incident angle is $39^\circ$, the frequency of interest $f_0-f_p$ is retroreflective. Again, the retroreflective behavior of the metasurface is directionally dependent. When the incident angle is $-39^\circ$, the $+1$ frequency harmonic is inside the light cone. The retroreflected wave is radiated at a frequency $f_0+f_p$. For this case, the direction-dependent retroreflective behavior of the metasurface is depicted in Fig. \ref{sfig:RetroReflect3}. The calculated reflection spectra are shown in Fig. \ref{fig:Retro2ASin}. As expected, Doppler-like frequency translation to $f_0-f_p$ is observed for an incident angle of $39^\circ$, and $f_0+f_p$ for incident angle of $-39^\circ$.
The conversion loss and sideband suppression for both polarizations are provided in examples 8 and 9 of Table \ref{tab:sim}.

Note that the retroreflection angle for both of the two cases (with sawtooth and sinusoidal reflection phase) was $\pm39^\circ$ for a path number of $N=4$. By simply changing the temporal modulation waveform, the retroreflection frequency was changed from $f_0-f_p$ to $f_0+3f_p$, for the same incident angle of $39^\circ$.

%%%%%%%%%%%%%%%%%%%%%%%%%%%%%%%%%%%%%%%%%%%%%%%%%%
%%%%%%%%%%%%%%%%%%%%%%%%%%%%%%%%%%%%%%%%%%%%%%%%%%
%%%%%%%%%% Section 4
%%%%%%%%%%%%%%%%%%%%%%%%%%%%%%%%%%%%%%%%%%%%%%%%%%
%%%%%%%%%%%%%%%%%%%%%%%%%%%%%%%%%%%%%%%%%%%%%%%%%%

\section{\label{sec:Fab} Experimental Realization of an SD-TWM metasurface}

In this section, an experimental prototype of the SD-TWM metasurface is developed for experimental verification. Details of the metasurface realization, as well as the measurement setup used to characterize its performance, are given in Section \ref{sec:fab_setup}. The static performance of the metasurface under various DC bias conditions is presented in Section \ref{sec:fab_dc}. Based on this static (DC) characterization, the required bias waveform to attain a particular desired time-dependent reflection phase is discussed in Section \ref{sec:fab_time}. This section also includes the measured reflection spectra for time-variation alone. 

\subsection{Metasurface design and measurement setup} \label{sec:fab_setup}

\begin{figure}[t]
\subfloat[\label{sfig:BiasTop}]{%
  \includegraphics[clip,width=0.3\columnwidth]{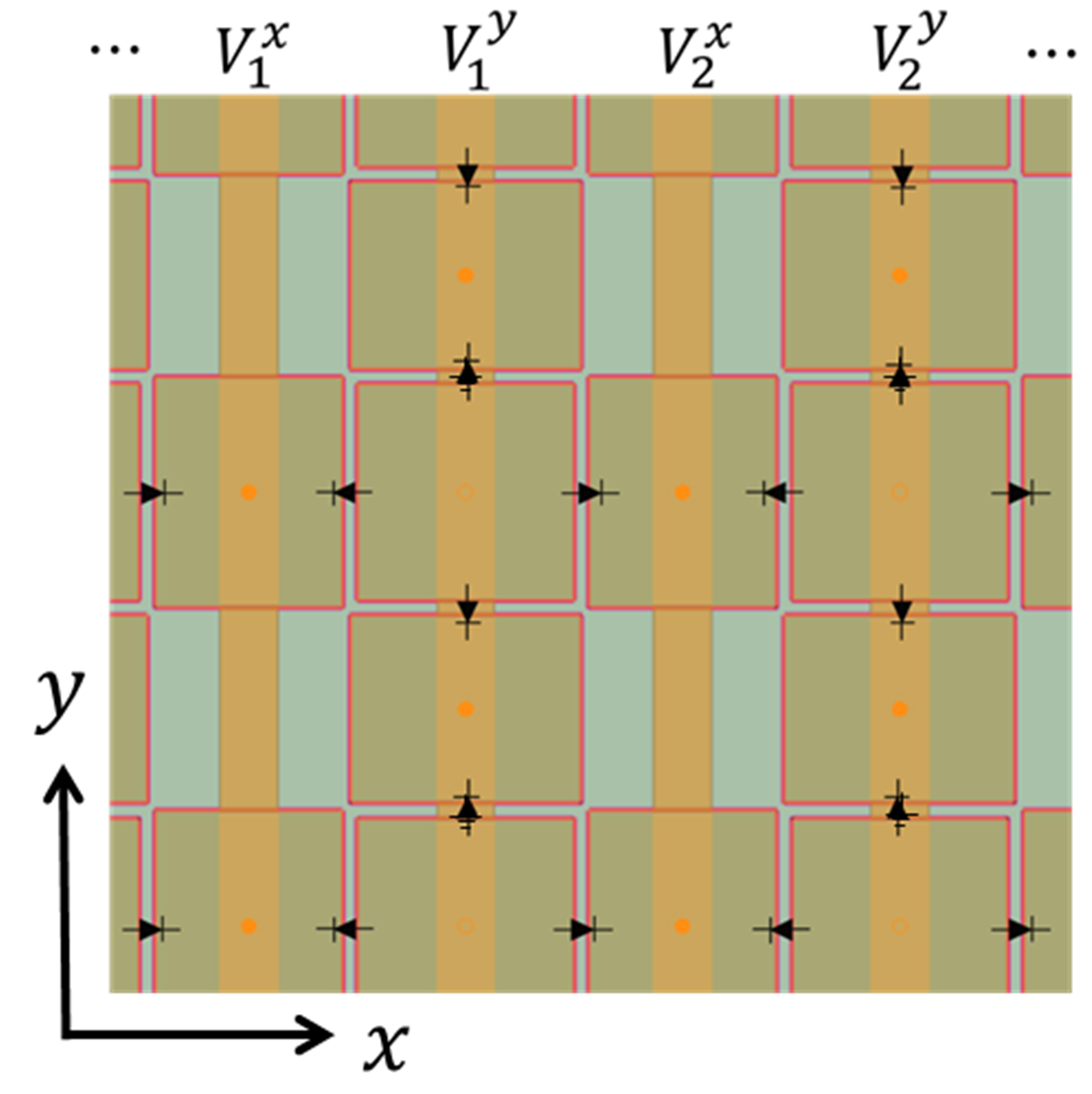}%
}\hfill
\subfloat[\label{sfig:fab}]{%
  \includegraphics[clip,width=0.53\columnwidth]{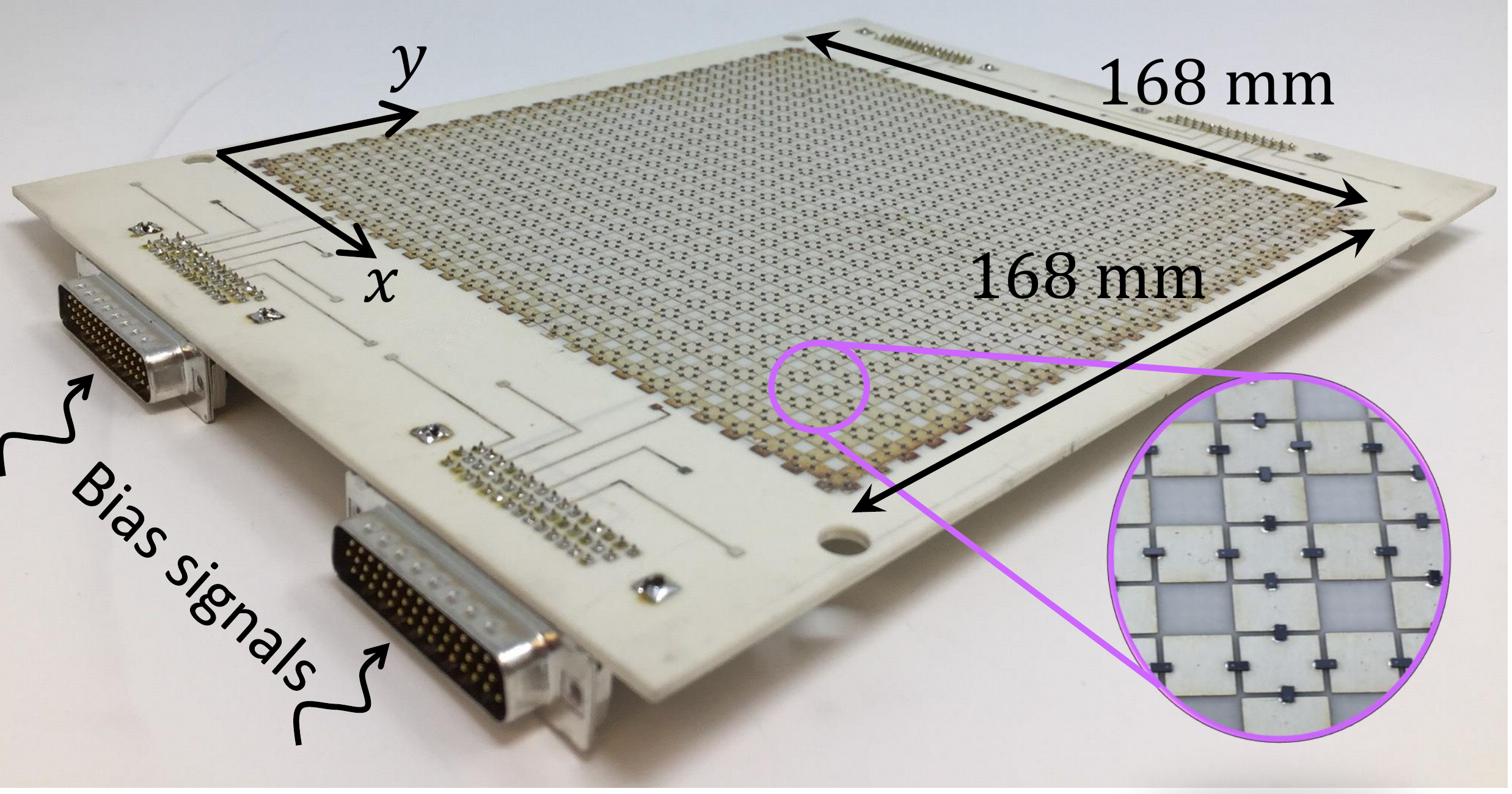}%
}\quad
\subfloat[\label{sfig:BiasSide}]{%
  \includegraphics[clip,width=0.7\columnwidth]{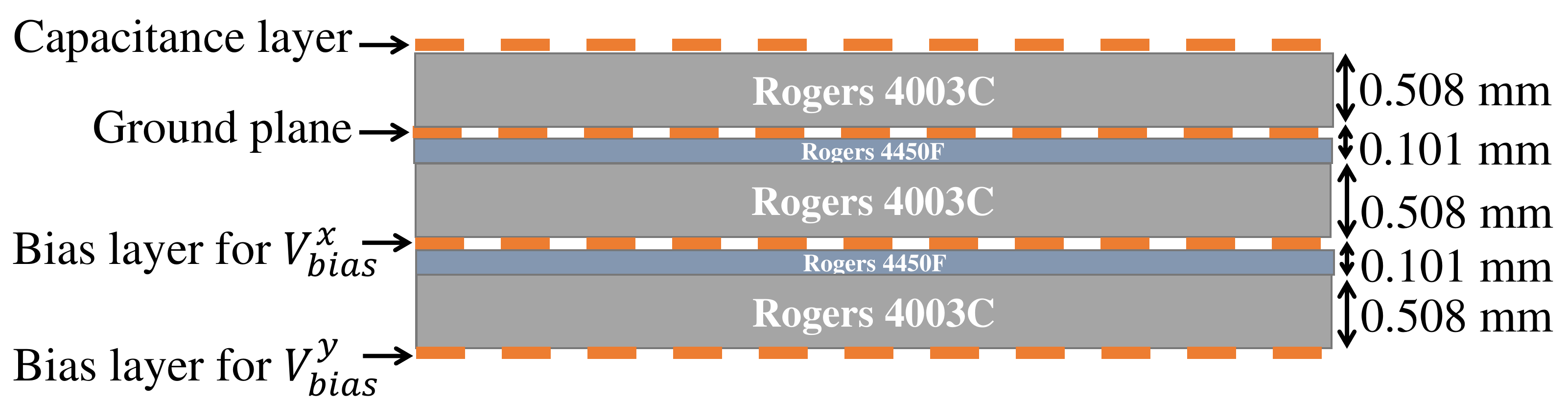}%
}\quad
\caption{(a) Transparent view of the metasurface prototype with two bias layers independently controlling each polarization. (b) Photograph of the fabricated metasurface. (c) Cross section of the fabricated metasurface.}
\label{fig:BiasScheme}
\end{figure} 

% metasurface description
A stixel (unit cell) of the dual-polarized metasurface is depicted in Fig. \ref{sfig:UnitCell}. Varactor diodes (MAVR-000120-1411 from MACOM \cite{macom}) are integrated onto the metasurface to act as tunable capacitances for two orthogonal polarizations.
The biasing networks for both polarizations were printed behind the ground plane, as shown in Fig. \ref{sfig:BiasTop} and Fig. \ref{sfig:BiasSide}.
Each bias layer consists of 28 metallic lines that can independently modulate all 28 columns of the metasurface. A total of $3136$ MAVR-000120-1411 varactor diodes were mounted onto the metasurface. The varactor diodes are biased through vias located in the center of the metallic patches. A photo of the fabricated metasurface is shown in Fig. \ref{sfig:fab}. The metallic traces of the bias layers are routed to four D-SUB connectors edge mounted to the metasurface. Rogers 4003C ($\epsilon_r = 3.55$ and $\tan \delta = 0.0027$) substrate with a thickness of $0.508$ mm was chosen for each layer. Rogers 4450F ($\epsilon_r=3.52$ and $\tan \delta=0.004$) bondply, with a thickness of $0.101$ mm was used as an adhesive layer.  A cross section of the material layers used to fabricate the metasurface is shown in Fig. \ref{sfig:BiasSide}. The total thickness of the fabricated metasurface is $1.726$ mm ($0.06\lambda$).

\begin{figure}[t]
\centering
\includegraphics[width=0.75\columnwidth]{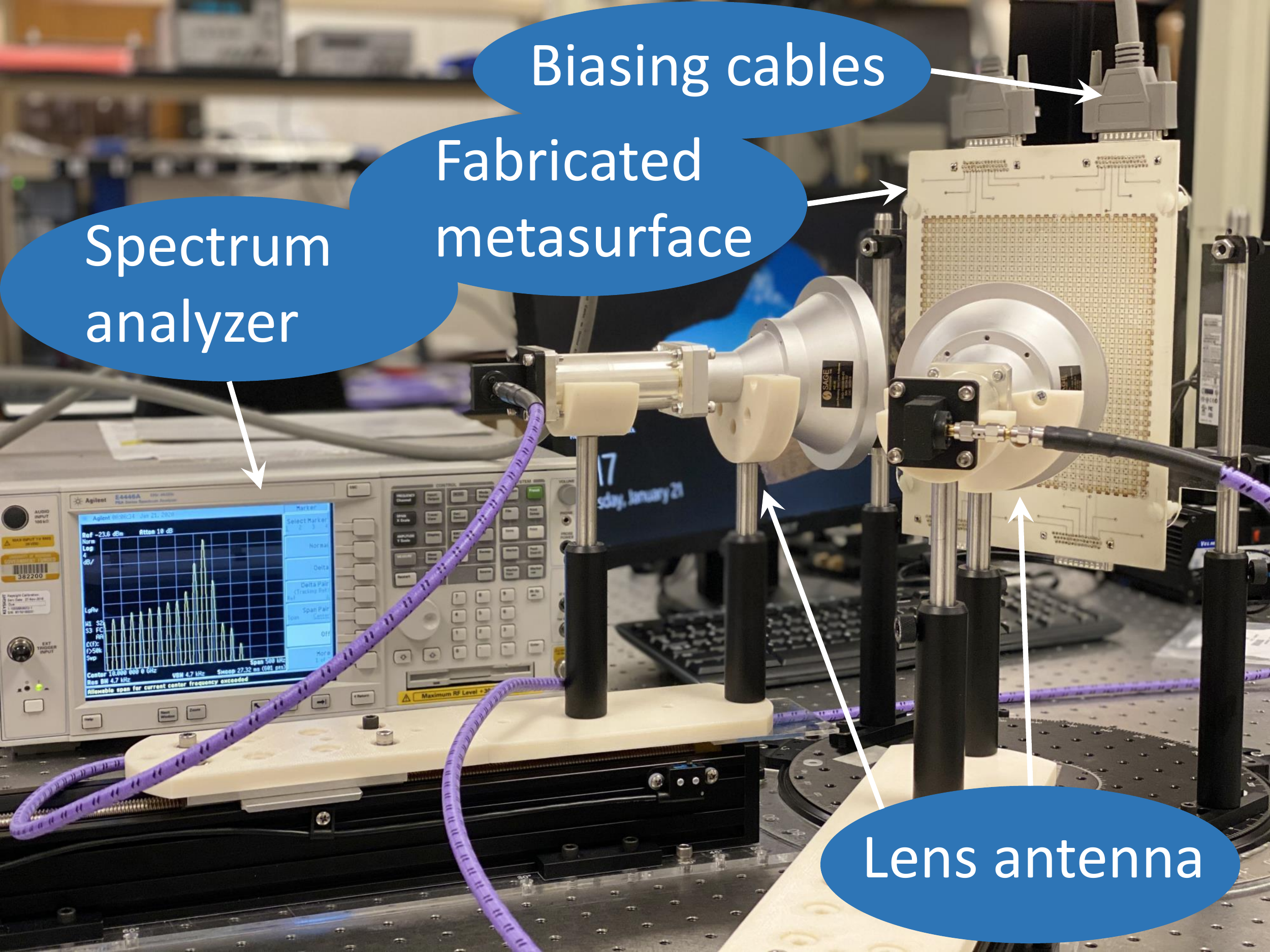}
\caption{Photograph of the quasi-optical, free-space measurement system.}
\label{fig:MeasurementSetup}
\end{figure}

% experimental setup
The metasurface was experimentally characterized using the quasi-optical Gaussian beam system shown in Fig. \ref{fig:MeasurementSetup}. In the experimental setup, the fabricated metasurface is illuminated by a spot-focusing lens antenna (SAQ-103039-90-S1). The antenna excites a Gaussian beam with a beamwidth of $50$ mm at a focal length of $10$ cm. The width of the fabricated metasurface is larger than $1.5$ times the beamwidth to limit edge diffraction. A continuous wave signal provided by Anritsu MS4644B vector network analyzer at $f_0=10$ GHz was used as the incident signal. The amplitude of the incident signal impinging on the metasurface was measured to be $-20$ dBm. An Agilent E4446A spectrum analyzer was used to measure the reflected spectrum. The path loss of the system was measured and calibrated out of the measurements. The metasurface was modulated by four Keysight M9188A 16-channel D/A converters. Each channel of the D/A converter was synchronized and staggered in time. The D/A converter has an output voltage range from $0$ V to $30$ V, and a maximum modulation frequency of $f_p=25$ kHz.  

\subsection{Measurements of a DC biased metasurface: tunable reflection phase} \label{sec:fab_dc}

% DC simulation setup
We will first look at the simulated and measured DC performance of the metasurface prototype. The capacitance provided by the varactors ranges from $0.18$ pF to $1.2$ pF. Using the commercial electromagnetic solver ANSYS HFSS, a full-wave simulation of the stixel (unit cell) shown in Fig. \ref{sfig:UnitCell} is conducted in the absence of time-variation. In the simulations, each varactor diode is modeled as a lumped capacitance in series with a resistance. The capacitance and resistance values of the varactor diode were extracted as a function of bias voltage from its SPICE model \cite{macom}. 

\begin{figure}[t]
\subfloat[\label{sfig:TAmpHFSSTE}]{%
  \includegraphics[clip,width=0.5\columnwidth]{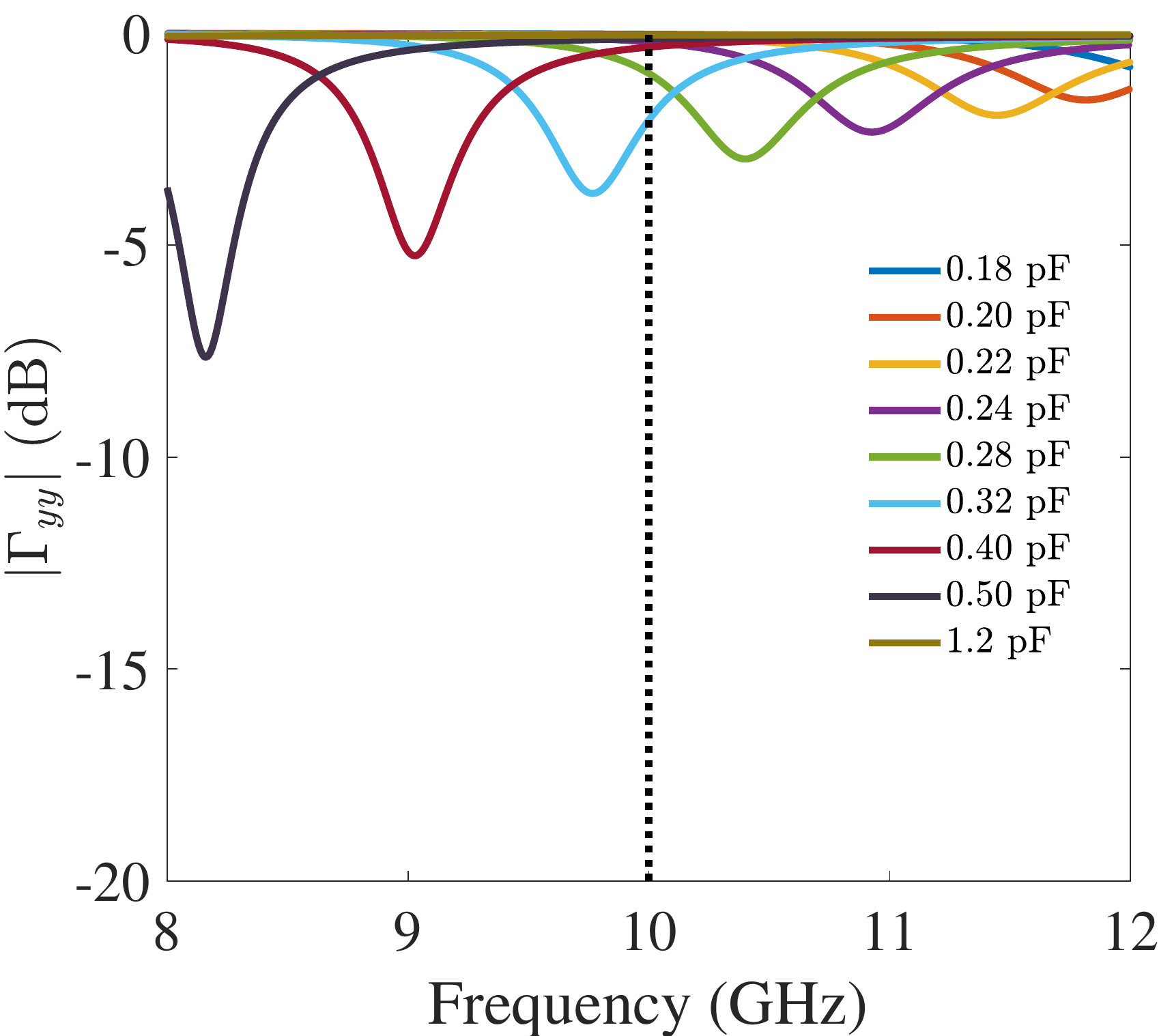}%
}\hfill
\subfloat[\label{sfig:TPhaseHFSSTE}]{%
  \includegraphics[clip,width=0.5\columnwidth]{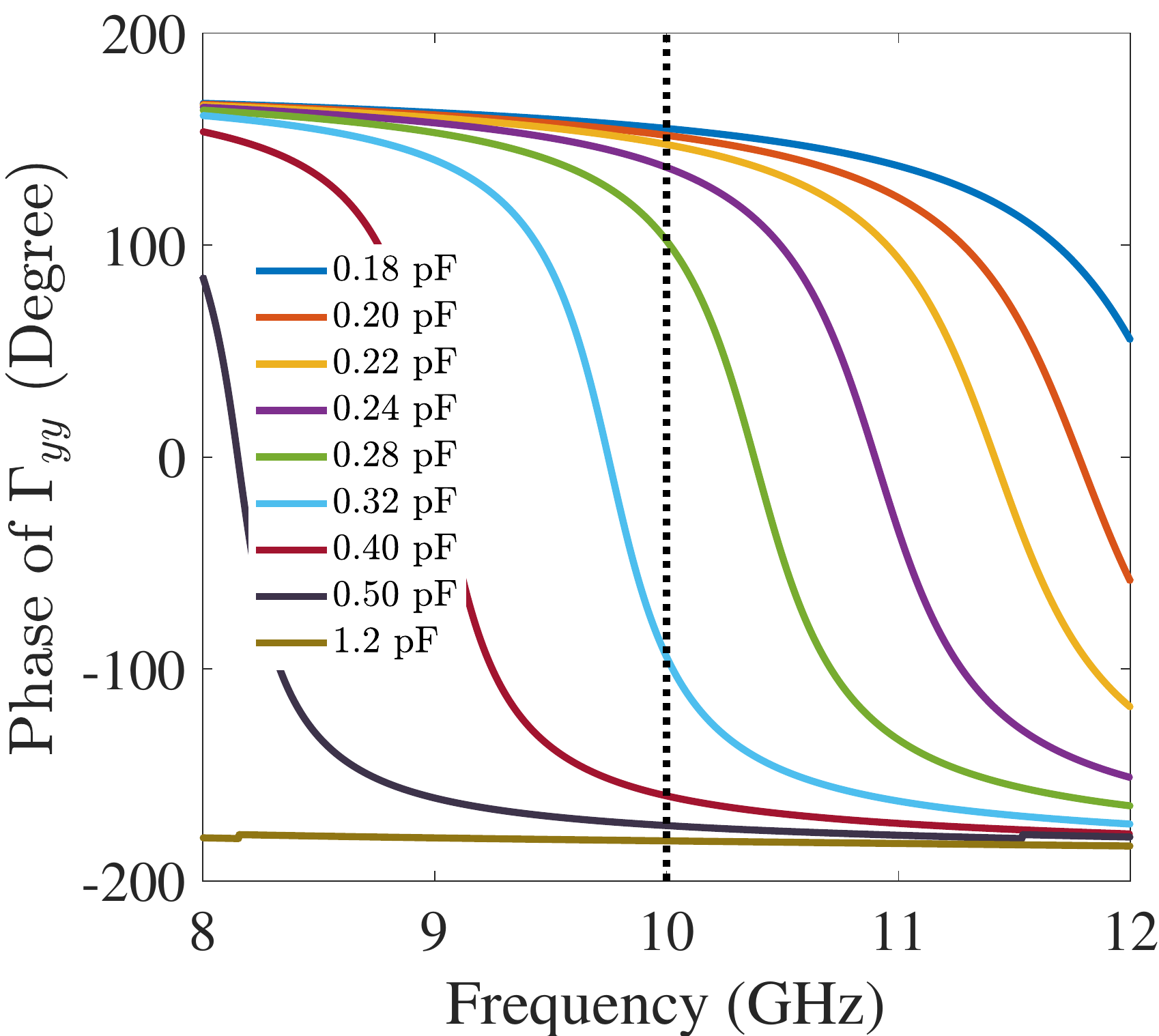}%
}\quad
\subfloat[\label{sfig:TAmpHFSSTM}]{%
  \includegraphics[clip,width=0.5\columnwidth]{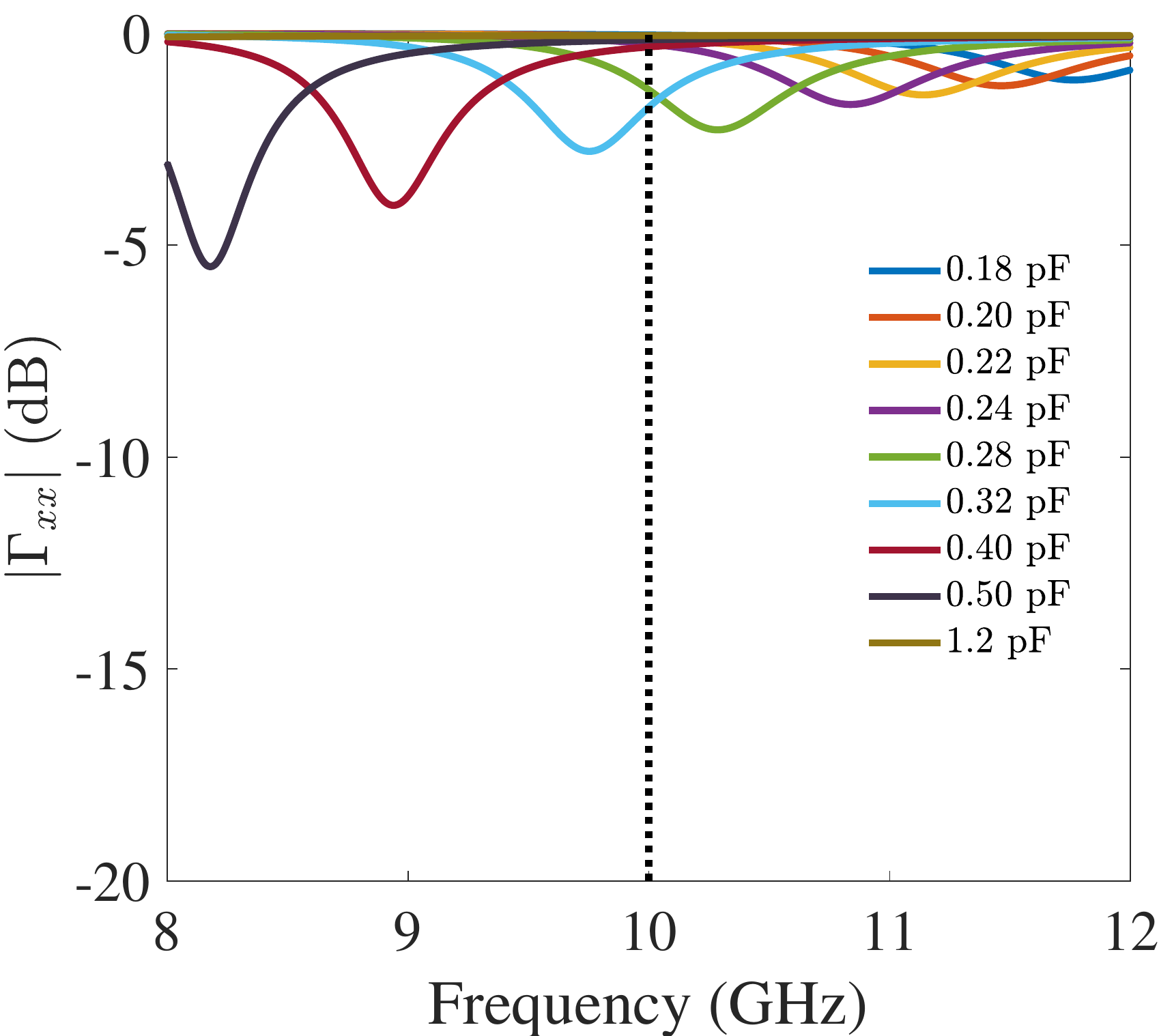}%
}\hfill
\subfloat[\label{sfig:TPhaseHFSSTM}]{%
  \includegraphics[clip,width=0.5\columnwidth]{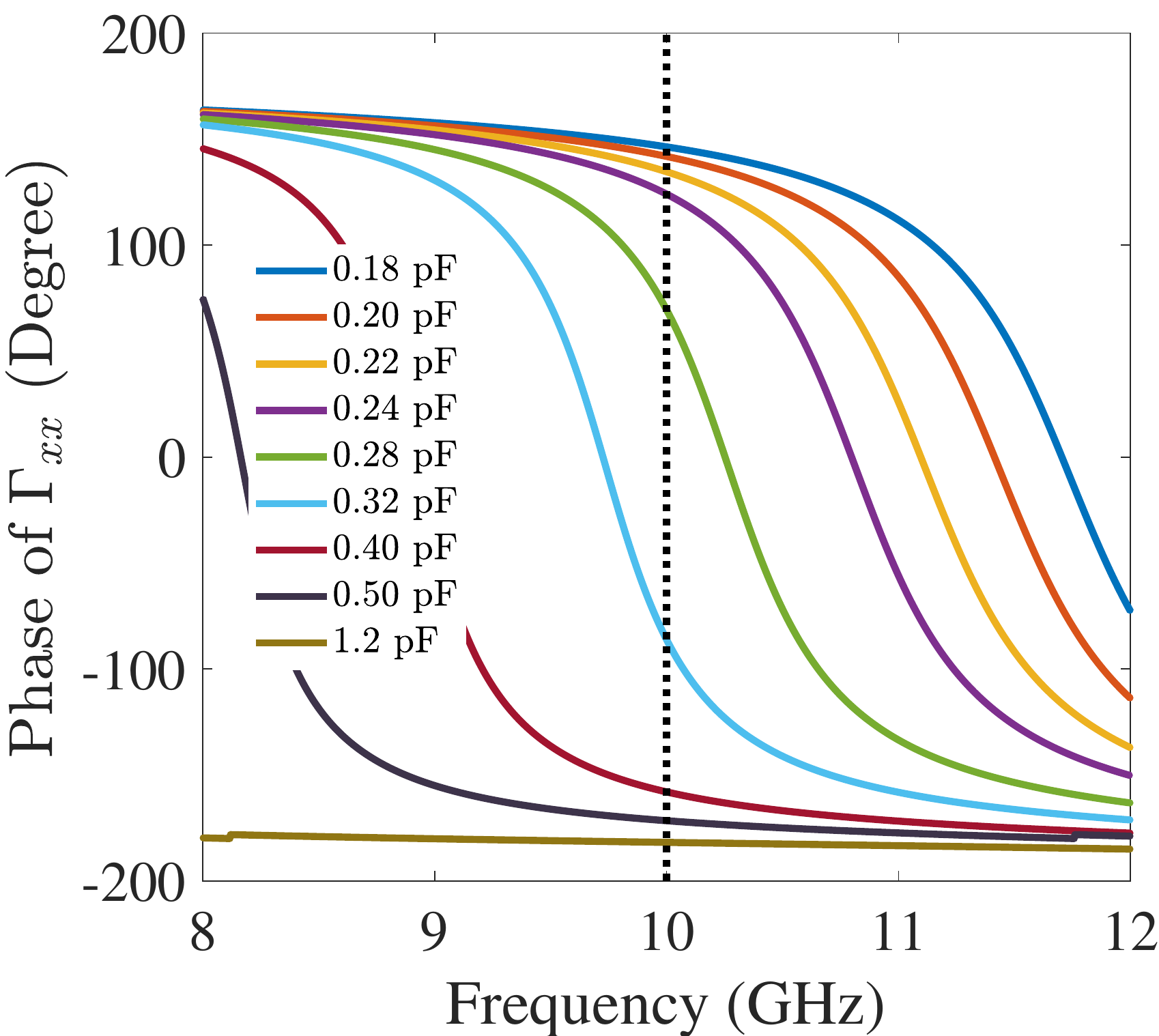}%
}\quad
\caption{Simulated reflection coefficient amplitude and phase of the realized metasurface for a range of varactor capacitances. The incident wave is obliquely incident at an angle of $25^\circ$. (a) Reflection amplitude for TE polarization. (b) Reflection phase for TE polarization. (c) Reflection amplitude for TM polarization. (d) Reflection phase for TM polarization.}
\label{fig:TunabilityHfSS}
\end{figure}

% reflection coefficient magnitude and phase dependence on capacitance
The simulated reflection coefficients of the metasurface for various varactor capacitance values are given in Fig. \ref{fig:TunabilityHfSS}. The incident angle is set to $25^\circ$. At the operating frequency of 10 GHz, the reflection phase of the metasurface can be varied from $-181.1^\circ$ to $155^\circ$ for TE polarization, providing a phase range of $336.1^\circ$. For TM polarization, the reflection phase of the metasurface can be varied from $-181.8^\circ$ to $146.3^\circ$, providing a phase range of $328.1^\circ$. At the operating frequency of 10 GHz, the simulated reflection amplitude for both polarizations remains greater than $-3$ dB across the entire phase range. Note that at the resonant frequency of the unit cell, the input susceptance goes to zero, and the surface admittance becomes purely resistive. The effective resistance seen by the incident wave is determined by the losses within the dielectric, the finite conductivity of the metallic patches and the losses of the varactor. As a result, the reflection coefficient magnitude dips at the resonance frequency, and the reflection phase becomes zero (a high impedance condition). The highest return loss at 10 GHz is $3.41$ dB for TE polarization, for a varactor capacitance of $0.313$ pF. For TM polarization, the highest return loss at 10 GHz is $2.47$ dB, for a varactor capacitance of $0.30$ pF. The metasurface incurs higher loss for TE polarization than TM polarization.
This is because, at the incident angle of $25^\circ$, the value of the free-space tangential wave impedance for TE polarization is closer (impedance matches better) to the purely resistive input impedance of the metasurface at resonance.
The simulated reflected cross-polarization of the metasurface is lower than $-50$ dB for all the orthogonal varactor capacitance variations. 

\begin{figure}[t]
\subfloat[\label{sfig:TAmpMeasTE}]{%
  \includegraphics[clip,width=0.5\columnwidth]{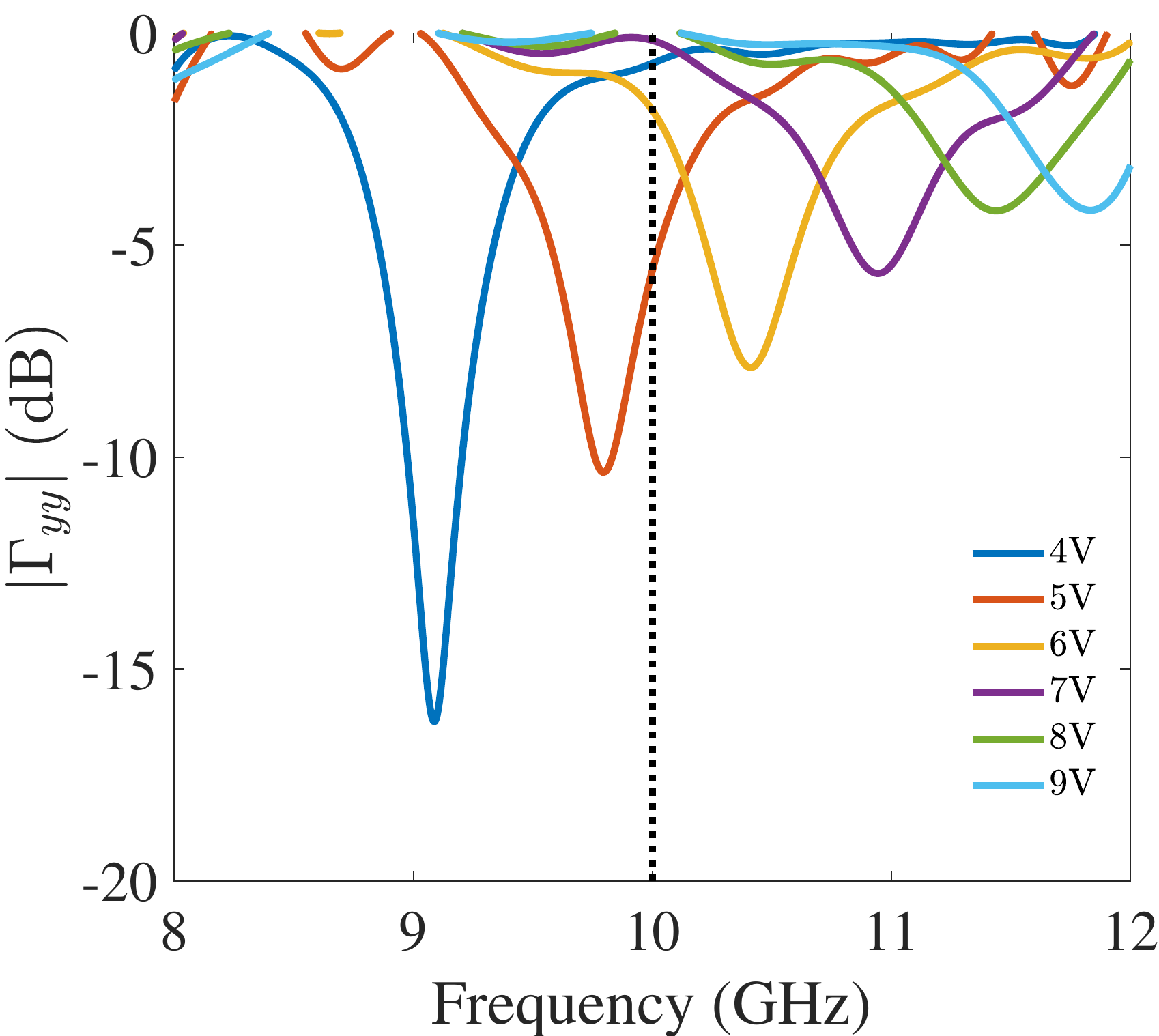}%
}\hfill
\subfloat[\label{sfig:TPhaseMeasTE}]{%
  \includegraphics[clip,width=0.5\columnwidth]{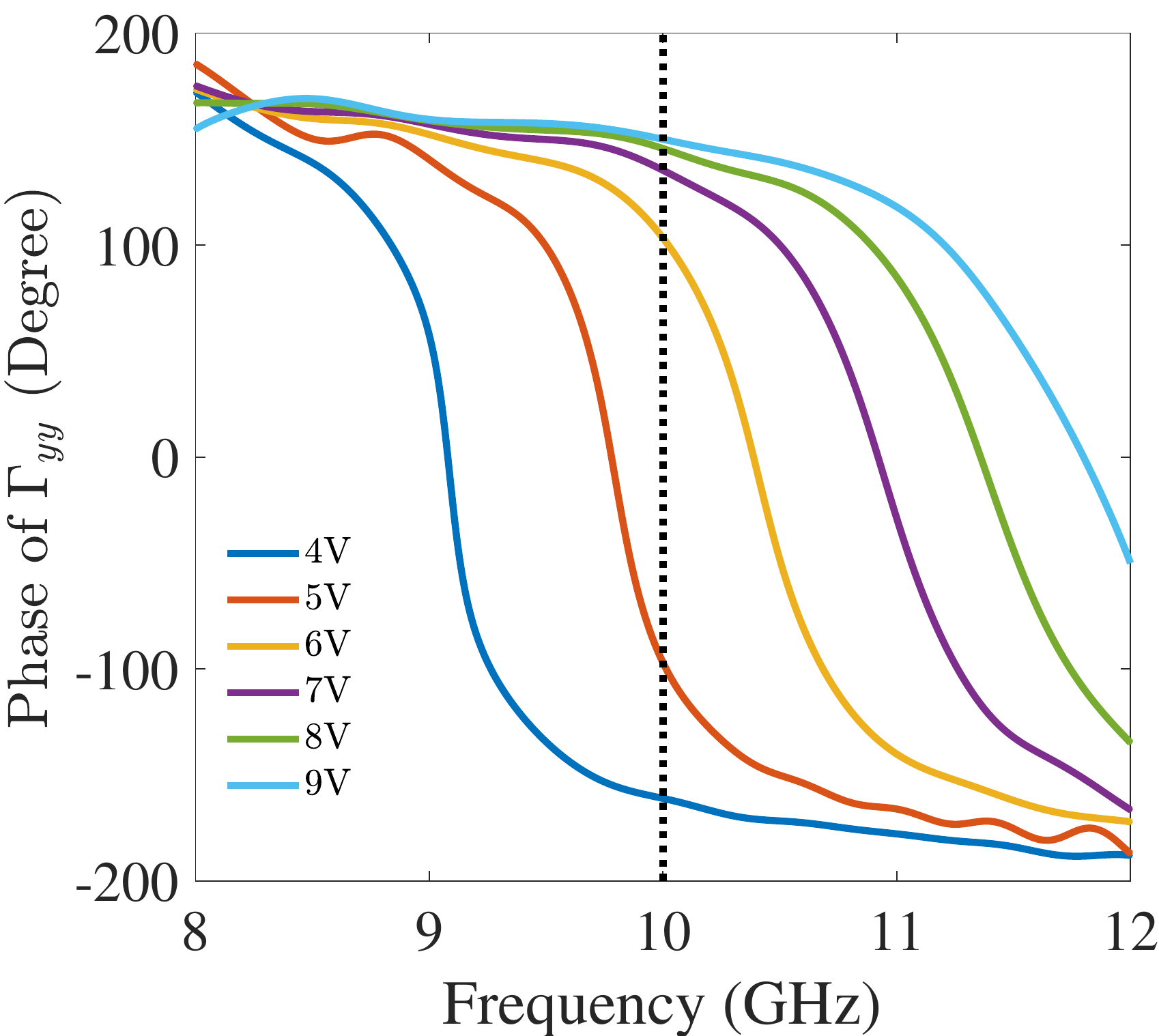}%
}\quad
\subfloat[\label{sfig:TAmpMeasTM}]{%
  \includegraphics[clip,width=0.5\columnwidth]{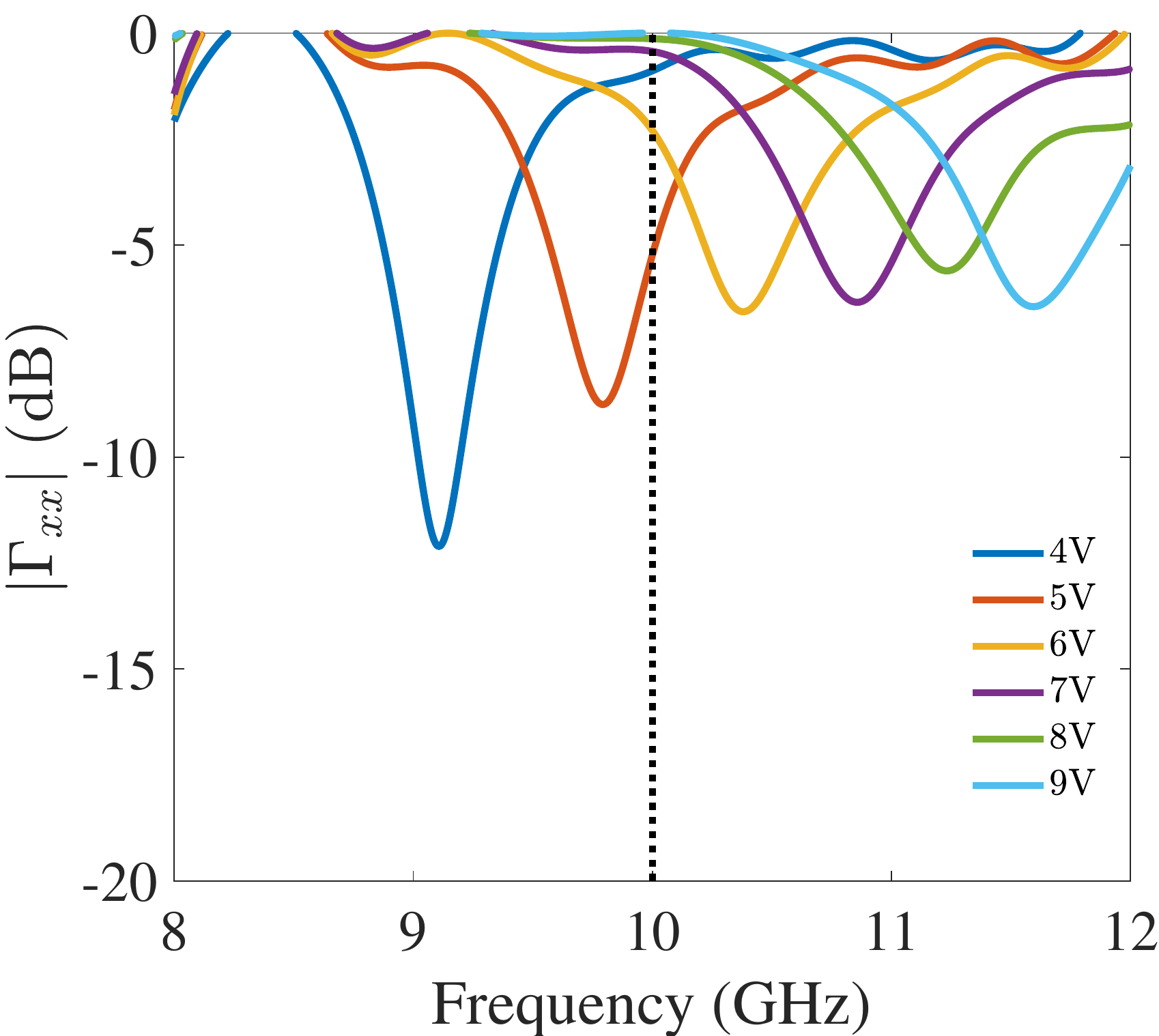}%
}\hfill
\subfloat[\label{sfig:TPhaseMeasTM}]{%
  \includegraphics[clip,width=0.5\columnwidth]{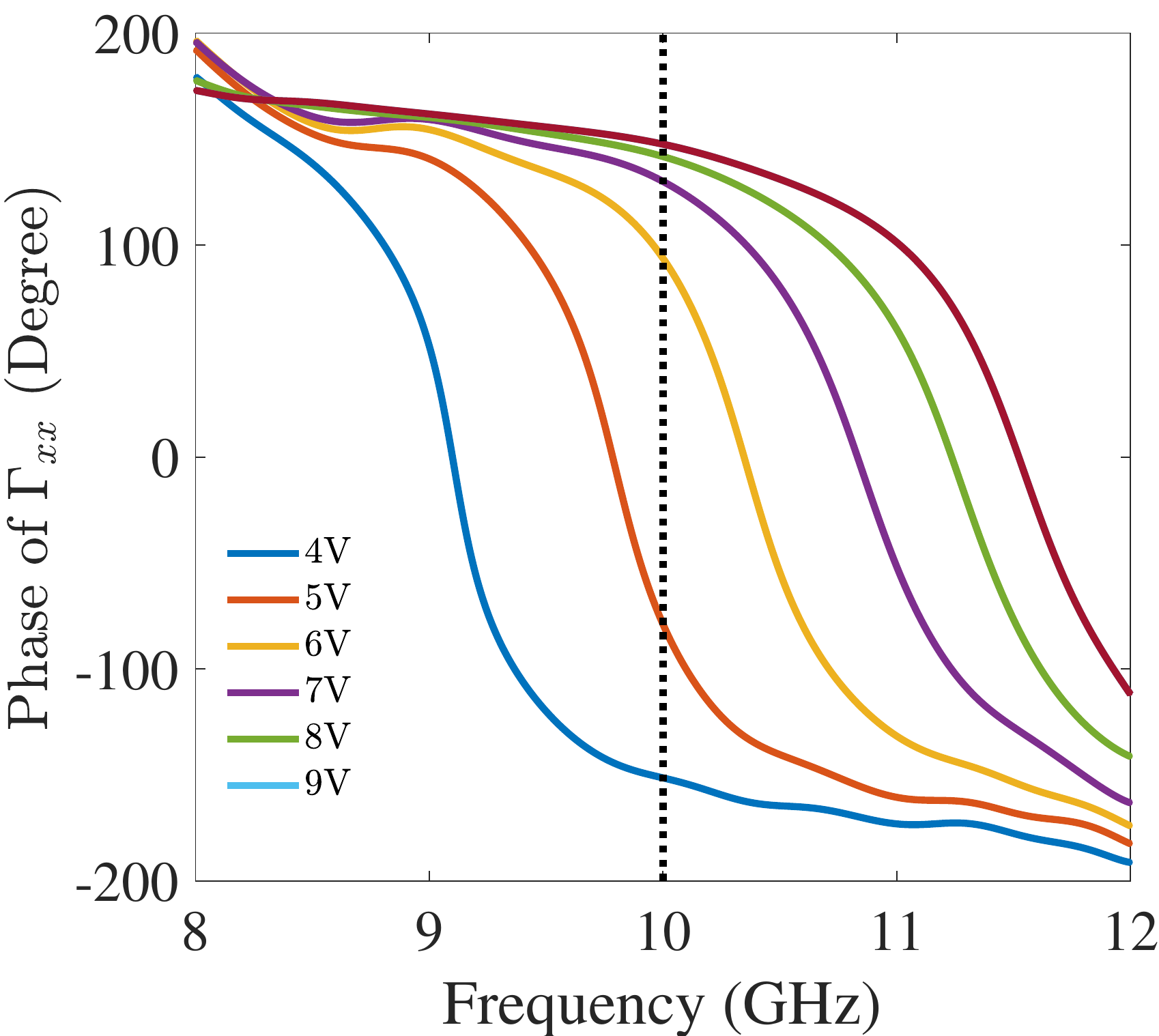}%
}\quad
\caption{Measured reflection coefficient amplitude and phase of the metasurface prototype for a range of bias voltages. The incident wave is oblique at an angle of $25^\circ$. (a) Reflection amplitude for TE polarization. (b) Reflection phase for TE polarization. (c) Reflection amplitude for TM polarization. (d) Reflection phase for TM polarization.}
\label{fig:TunabilityMeasure}
\end{figure}

% description of the DC measurement results
The static (DC biased) performance of the metasurface was measured under an oblique angle of $25^\circ$. The measured TE and TM reflection coefficients under various bias voltages are given in Fig \ref{fig:TunabilityMeasure}. The bias voltage used in measurement ranged from $0$ V to $15$ V, providing a varactor capacitance range of $0.18$ pF to $1.2$ pF. At the operating frequency of 10 GHz, the measured reflection phase of the metasurface could be varied from $-182.7^\circ$ to $149.9^\circ$ for TE polarization, corresponding to a phase range of $332.6^\circ$. For TM polarization, the measured reflection phase could be varied from $-176.9^\circ$ to $147.6^\circ$, corresponding to a phase range of $324.5^\circ$. At resonance, the measured reflection amplitude was found to be much lower than in simulation, indicating higher losses in the fabricated metasurface.
This could be attributed to additional ohmic loss within the diode as well as losses introduced by the tinning and soldering procedures used to mount the diodes. Nevertheless, the simulated and measured static performances of the metasurface are in good agreement. A detailed comparison between simulation and measurement for each bias voltage is given in Supplemental Materials V. 

A harmonic balance simulation using the Keysight ADS circuit solver was used to verify the theoretical analysis and to compute the reflection spectrum. However, to use the harmonic balance circuit solver, a circuit equivalent of the fabricated metasurface needed to be extracted for each polarization from full-wave scattering simulations. A voltage-dependent resistance is added to it to account for the added losses observed in measurement. The equivalent  circuits for the two polarizations under an oblique incident angle of $25^\circ$ are given in Supplemental Materials V. From the equivalent circuits, the capacitance modulation required to obtain a given reflection phase versus time dependence can be obtained, as detailed in Supplemental Material VI.

\subsection{Measurements of a time-modulated metasurface: serrodyne frequency translation} \label{sec:fab_time}

\begin{figure}[t]
\subfloat[\label{sfig:WaveTE_M}]{%
  \includegraphics[clip,width=0.49\columnwidth]{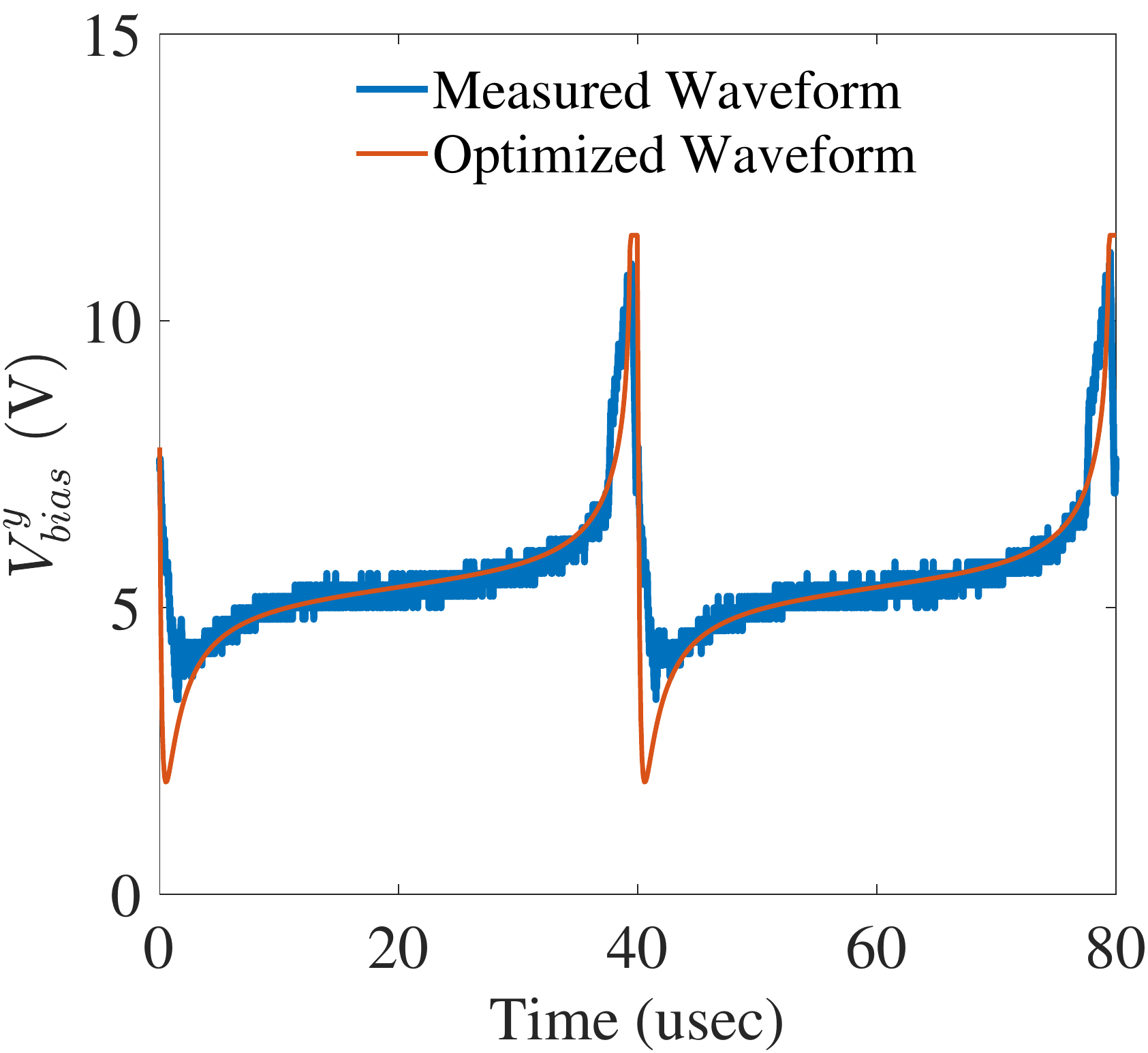}%
}\hfill
\subfloat[\label{sfig:1pathTE_M}]{%
  \includegraphics[clip,width=0.49\columnwidth]{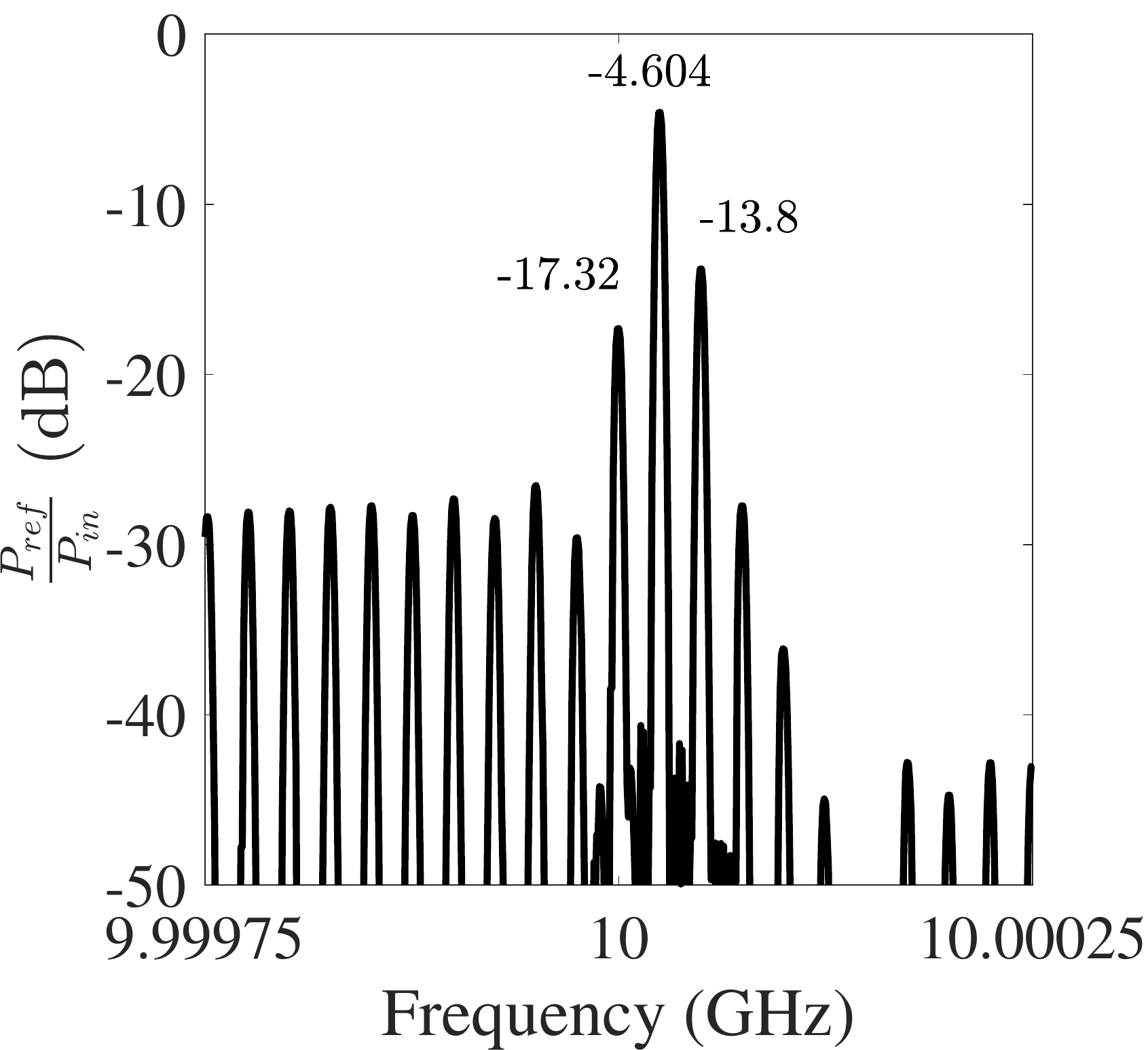}%
}\quad
\subfloat[\label{sfig:WaveTM_M}]{%
  \includegraphics[clip,width=0.5\columnwidth]{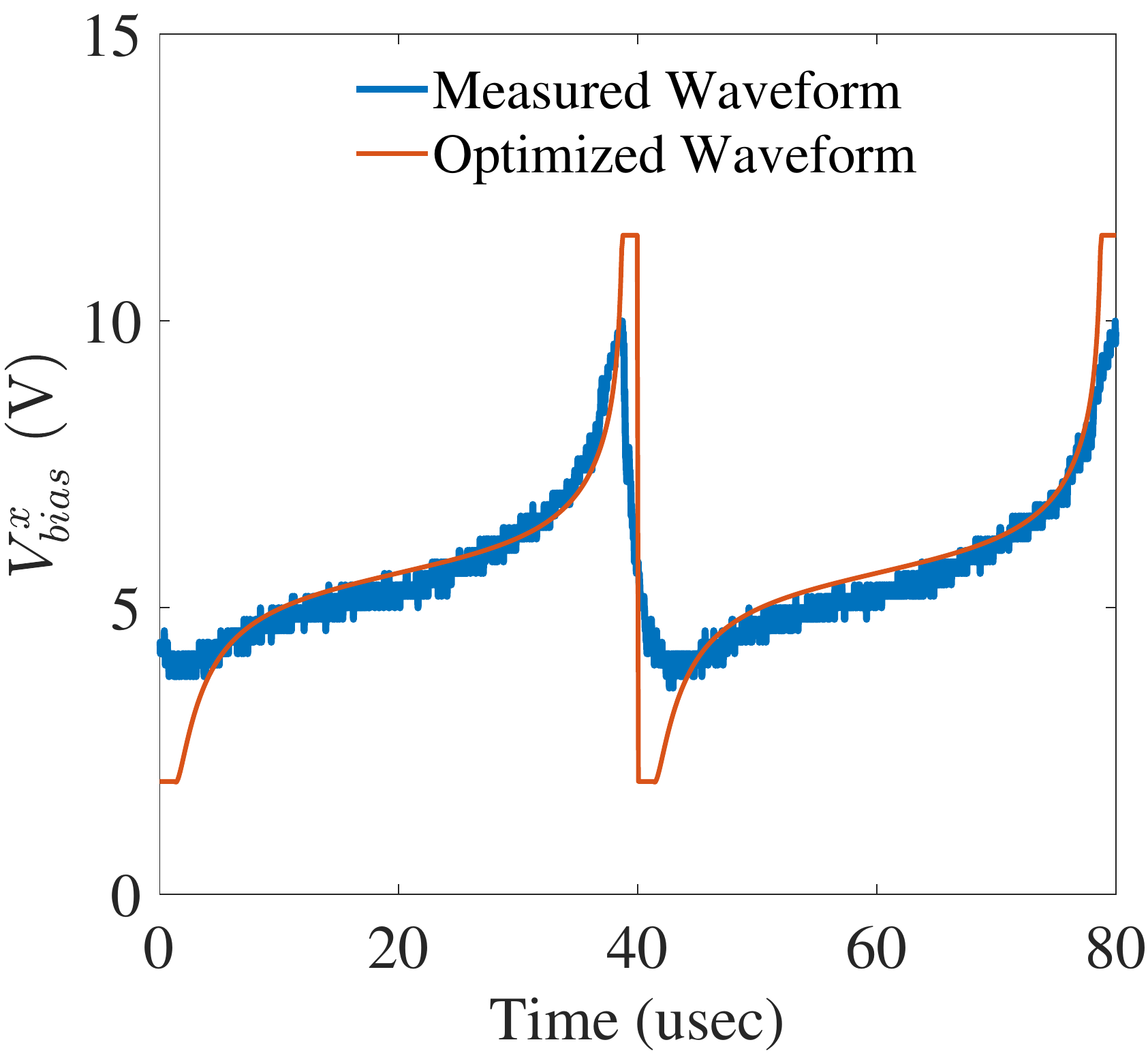}%
}\hfill
\subfloat[\label{sfig:1pathTM_M}]{%
  \includegraphics[clip,width=0.5\columnwidth]{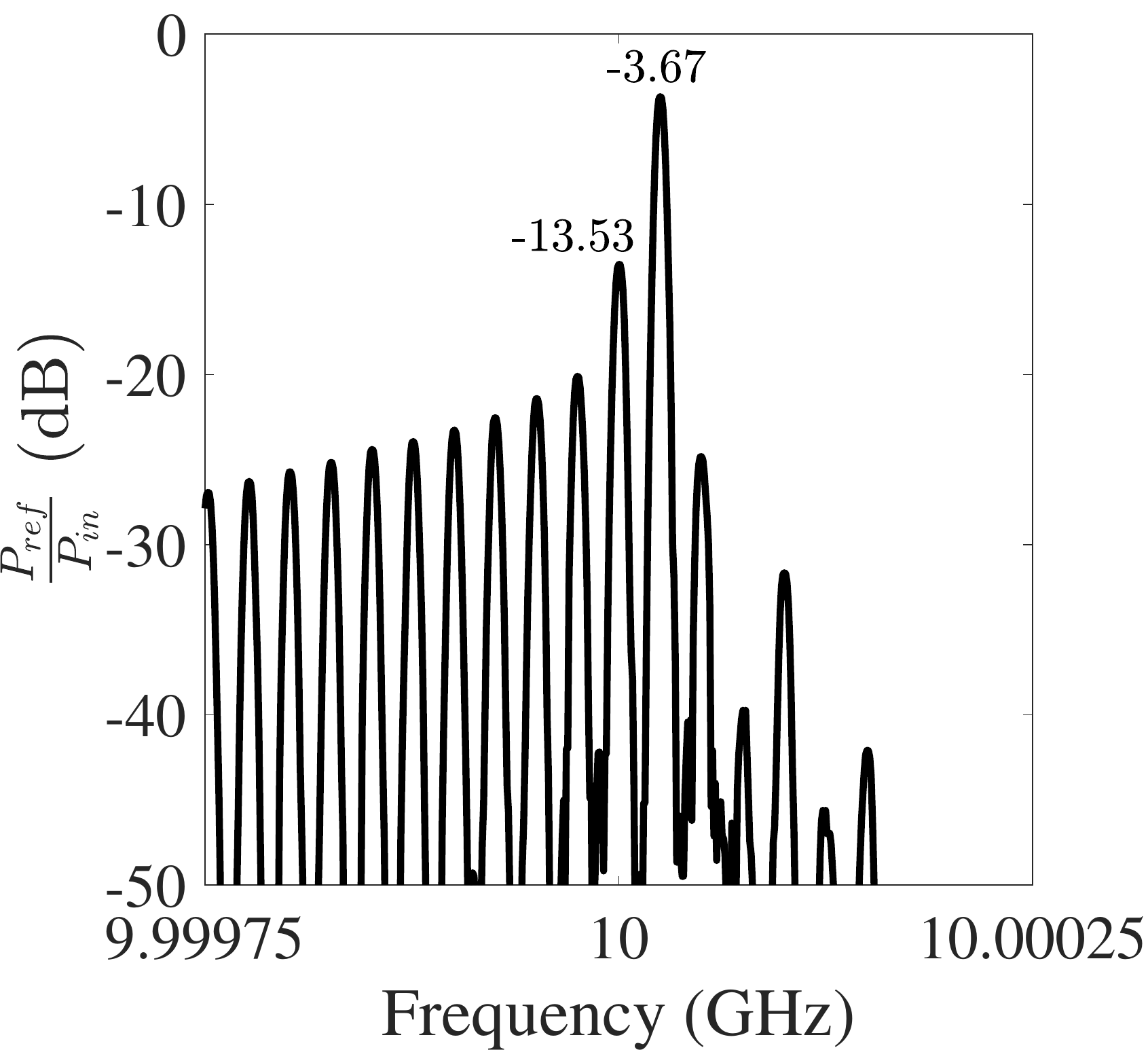}%
}
\caption{(a) Bias waveform of the time-modulated metasurface for TE polarization. (b) Measured reflection spectrum of the time-modulated metasurface prototype for TE polarization. (c) Bias waveform of the time-modulated metasurface prototype for TM polarization. (d) Measured reflection spectrum of the time-modulated metasurface for TM polarization.}
\label{fig:1path_Measure}
\end{figure} 

% Pure time-modulation results
As discussed in Section \ref{sec:TModOnly}, when the metasurface is uniformly biased with a sawtooth reflection phase waveform, serrodyne frequency translation is expected. As shown in Fig. \ref{fig:TunabilityHfSS} and Fig. \ref{fig:TunabilityMeasure}, the reflection amplitude is not unity due to the loss in the metasurface. Therefore, the capacitance modulation waveform had to be  numerically optimized.
The optimization process is detailed in Supplemental Materials VI.
In the experiment, the optimized waveform was sampled of $20$ data points per period ($T_p=40$ $\mu$sec), and entered into the D/A converter. All  channels of the D/A converter were synchronized with the same bias waveform. The bias waveform across several diodes was measured using a differential probe (Tektronix TMDP0200) and Tektronix oscilloscope MDO3024. The optimized and measured bias voltage waveforms are shown in Fig. \ref{sfig:WaveTE_M} for TE polarization and Fig. \ref{sfig:WaveTM_M} for TM polarization. The measured reflection spectrum for an oblique angle of $25^\circ$ is shown in Fig. \ref{sfig:1pathTE_M} for TE polarization and Fig. \ref{sfig:1pathTM_M} for TM polarization. Both polarizations show serrodyne frequency translation to $f=f_0+f_p$. For TE polarization, a $4.604$ dB conversion loss and $9.196$ dB of sideband suppression are achieved. For TM polarization, a $3.67$ dB conversion loss and $9.86$ dB sideband suppression are achieved. For each polarization, the measured reflection spectrum in Fig. \ref{sfig:1pathTE_M} and \ref{sfig:1pathTM_M} generally agrees with harmonic balance simulations of its extracted circuit models, as shown in Fig. S6 of Supplemental Materials VII.

\section{Experimental validation of an SD-STM metasurface} \label{sec:fab_stp}

In this section, measurements of the prototype metasurface are reported to verify the theory behind SD-TWM metasurfaces. Experimental scattering for the three regimes of metasurface operation introduced in Section \ref{sec:ThreeRegime} is reported. Specifically, effects such as specular subharmonic frequency translation, deflective/retroreflective serrodyne frequency translation, and deflective/retroreflective subharmonic frequency translation are experimentally examined. Again, the incident frequency $f_0$ and modulation frequency $f_p$ are set to 10 GHz and 25 kHz, respectively. Unless stated otherwise, the reflection phase of each column (path) is a sawtooth function in time. The measured conversion loss and sideband suppression for various examples introduced in Section \ref{sec:ThreeRegime} are provided in Table \ref{tab:meas}, and will be referred to throughout this section.

\begin{table}[b]
\renewcommand{\arraystretch}{1.3}
\caption{Measured conversion loss and and sideband suppression for desired reflected frequency harmonic $f^\prime$ given: $N$ - the number of paths, $\theta_i$ - the incident angle, $\theta_{\rm{obs}}$ - the observation angle, and the temporal phase modulation waveform (either a sawtooth or a sinusoid). Note that positive values of $\theta_i$ and $\theta_{\rm{obs}}$ correspond to waves traveling along the positive $x$ direction.  }
\label{tab:meas}
\centering
\begin{tabular}{|c|c|c|c|c|c||c|c|c|c|}
\hline
\multirow{2}{*}{Ex.} &
\multirow{2}{*}{N} &
\multirow{2}{*}{$\theta_i$} &
\multirow{2}{*}{$\theta_{\rm{obs}}$} &
\multirow{2}{*}{\begin{tabular}[c]{@{}c@{}}Wave-\\form\end{tabular}} &
\multirow{2}{*}{$f^\prime$} &
\multicolumn{2}{c|}{\begin{tabular}[c]{@{}c@{}}Conversion\\ Loss\\ (dB)\end{tabular}} &
\multicolumn{2}{c|}{\begin{tabular}[c]{@{}c@{}}Sideband\\ Supression\\ (dB)\end{tabular}} \\
\cline{7-10} 
& & & & & & TE & TM & TE & TM \\
\hhline{|=|=|=|=|=|=||=|=|=|=|}
0 & 1 & 25$^\circ$ & 25$^\circ$ & saw & $f_0+f_p$
& 4.694 & 3.67 & 9.196 & 9.86 \\
\hhline{|-|-|-|-|-|-||-|-|-|-|}
1 & 2 & 25$^\circ$ & 25$^\circ$ & saw & $f_0+2f_p$
& 6.86 & 5.038 & 12.71 & 8.80 \\
\hhline{|-|-|-|-|-|-||-|-|-|-|}
2 & 3 & 25$^\circ$ & 25$^\circ$ & saw & $f_0+3f_p$
& 10.69 & 8.47 & 6.55 & 5.53 \\
\hhline{|-|-|-|-|-|-||-|-|-|-|}
3 & 20 & 25$^\circ$ & 42$^\circ$ & saw & $f_0+f_p$
& 5.15 & 3.74 & 11.68 & 15.09 \\
\hhline{|-|-|-|-|-|-||-|-|-|-|}
4 & 20 & -42$^\circ$ & -25$^\circ$ & saw & $f_0+f_p$
& 5.06 & 3.79 & 11.46 & 15.20 \\
% \hhline{|-|-|-|-|-|-||-|-|-|-|}
% 5 & 6 & 25$^\circ$ & -25$^\circ$ & saw & $f_0+f_p$
% & 0.51 & 0.49 & 12.48 & 13.29 \\
\hhline{|-|-|-|-|-|-||-|-|-|-|}
5 & 4 & 39$^\circ$ & -39$^\circ$ & saw & $f_0+3f_p$
& 10.85 & 6.64 & 9.98 & 5.45 \\
\hhline{|-|-|-|-|-|-||-|-|-|-|}
6 & 4 & -39$^\circ$ & 39$^\circ$ & saw & $f_0+f_p$
& 4.02& 3.39 & 19.39 & 15.35 \\
% \hhline{|-|-|-|-|-|-||-|-|-|-|}
% 6 & 4 & -39$^\circ$ & 39$^\circ$ & saw & $f_0+f_p$
% & 1.24 & 0.95 & 10.63 & 11.89 \\
\hhline{|-|-|-|-|-|-||-|-|-|-|}
7 & 4 & 39$^\circ$ & -39$^\circ$ & sin & $f_0-f_p$
& 8.64 & 5.32 & 11.97 & 16.00 \\
\hhline{|-|-|-|-|-|-||-|-|-|-|}
 8 & 4 & -39$^\circ$ & 39$^\circ$ & sin & $f_0+f_p$
 & 7.76 & 5.30 & 9.45 & 15.87 \\
\hline
\end{tabular}
\end{table}

\subsection{Small spatial modulation period ($|k_x\pm\beta_p|>k_0$)} 
\label{sec:SmallPeriod_Fab}

\begin{figure}[t]
\subfloat[\label{sfig:2pathTEM}]{%
  \includegraphics[clip,width=0.5\columnwidth]{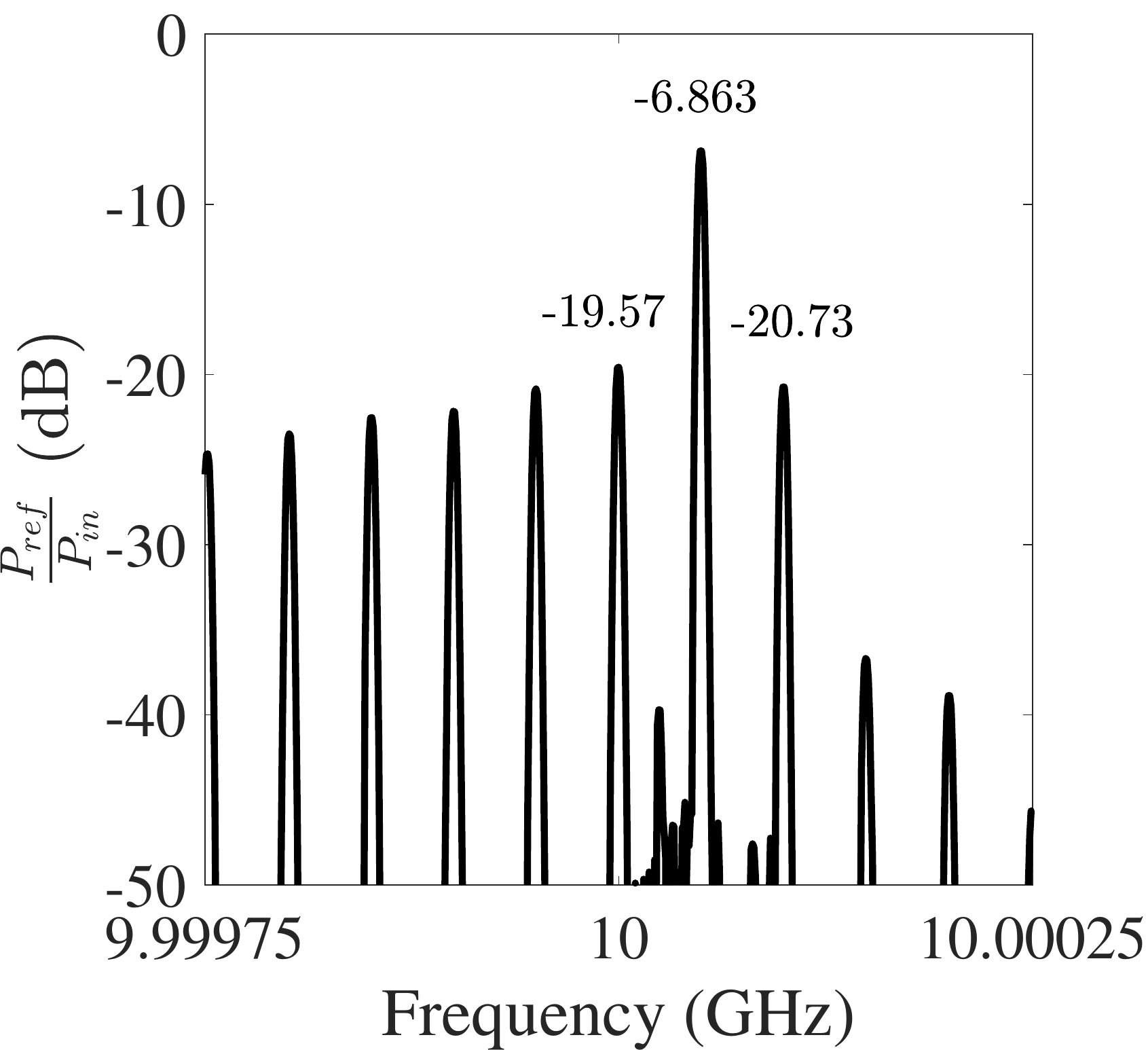}%
}\hfill
\subfloat[\label{sfig:3pathTEM}]{%
  \includegraphics[clip,width=0.5\columnwidth]{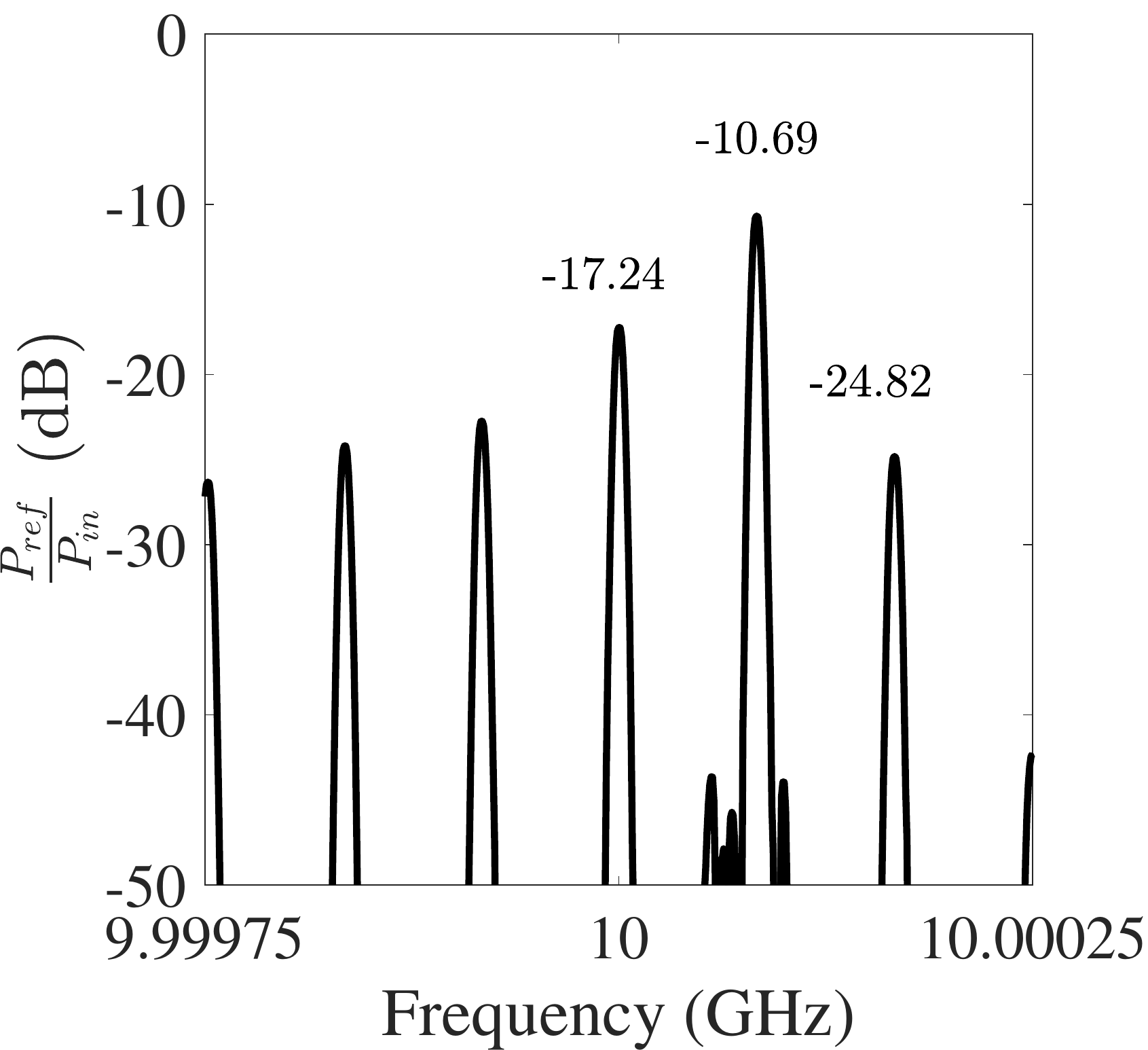}%
}\quad
\subfloat[\label{sfig:2pathTMM}]{%
  \includegraphics[clip,width=0.5\columnwidth]{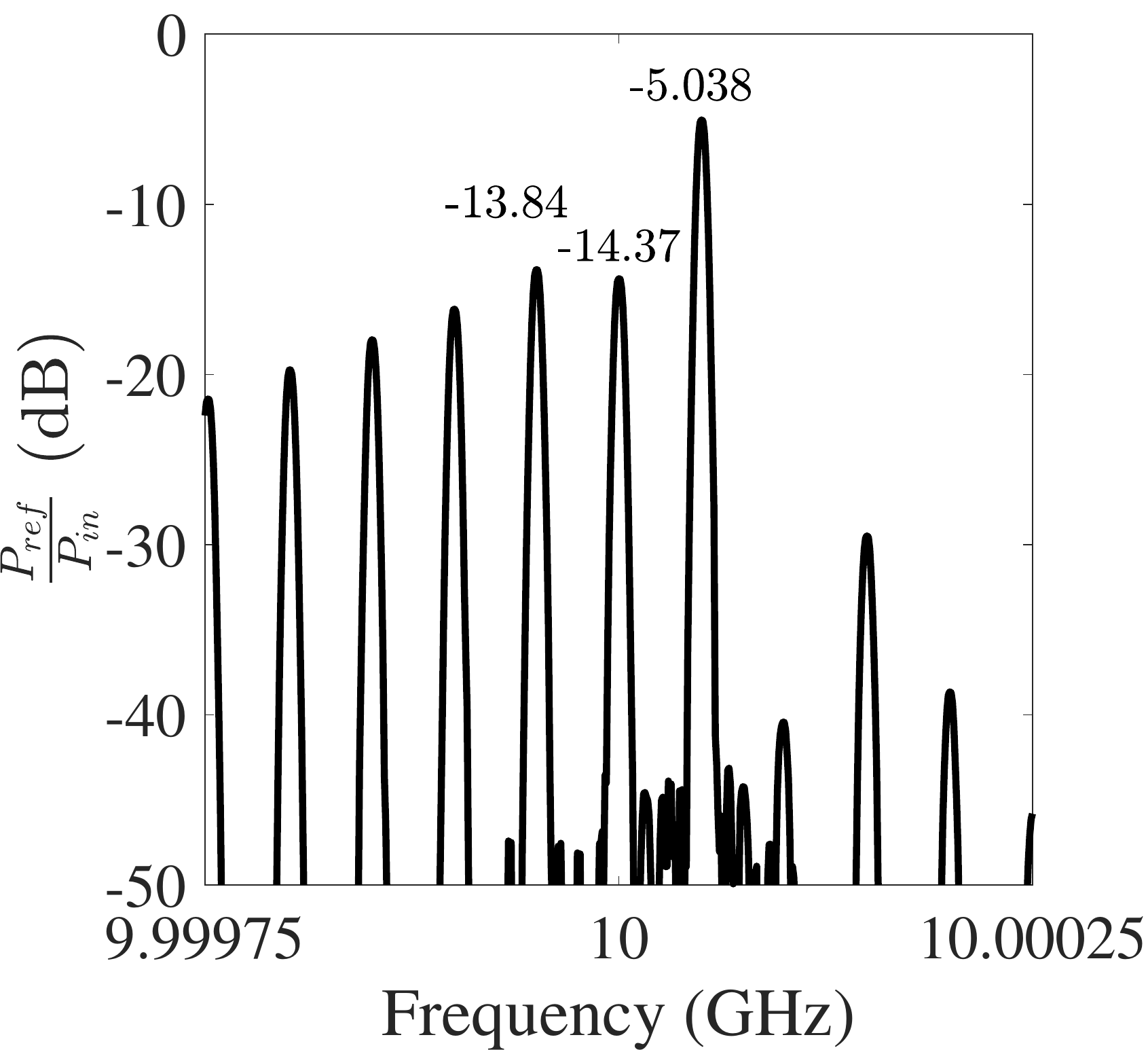}%
}\hfill
\subfloat[\label{sfig:3pathTMM}]{%
  \includegraphics[clip,width=0.5\columnwidth]{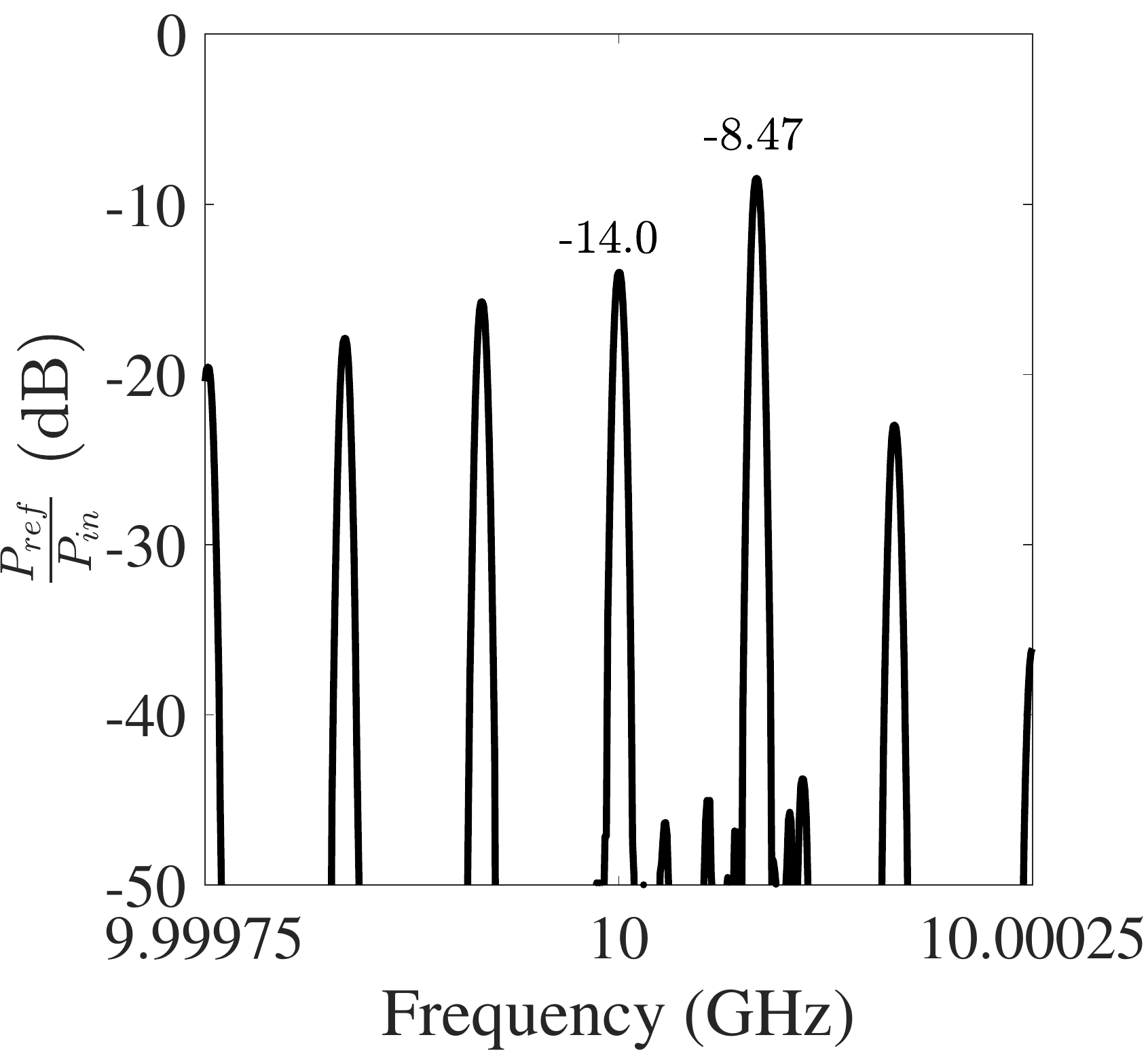}%
}
\caption{Measured reflection spectrum of the SD-TWM metasurface prototype. (a) 2-path ($N=2$) modulation for TE polarization. (b) 3-path ($N=3$) modulation for TE polarization. (c)  2-path ($N=2$) modulation for TM polarization. (d) 3-path ($N=3$) modulation for TM polarization.}
\label{fig:23pathMeas}
\end{figure}

In this section, electrically-small spatial modulation periods are considered ($|k_x\pm \beta_p|>k_0$). The incident wave is chosen to impinge on the metasurface with an oblique angle of $25^\circ$. The measured reflection spectra for 2-path ($N=2$) and 3-path ($N=3$) modulation schemes are given in Fig. \ref{fig:23pathMeas}. The reflection spectra are measured at a reflection angle of $\theta=25^\circ$ (see Fig. \ref{fig:NpathReflectionSub}). The measured spectra for both polarizations clearly demonstrate subharmonic frequency translation, where the only radiated harmonics are those at frequencies $f=f+rNf_s$ and $r\in \mathbb{Z}$. 
The measured conversion loss and sideband suppression for both polarizations are reported in examples 1 and 2 in Table \ref{tab:meas}.
Compared to the homogenized, lossless metasurface presented in Section \ref{sec:SmallPeriod}, the conversion loss and sideband suppression degrade more as the number of paths is increased. This is attributed to the evanescent harmonic pairs on the surface of the structure, which is discussed further in Supplemental Material VII. 

\subsection{Large spatial modulation period ($|k_x\pm\beta_p|<k_0$)} 
\label{sec:LargePeriod_Fab}

\begin{figure}[t]
\subfloat[\label{sfig:25steeringTEAM}]{%
  \includegraphics[clip,width=0.5\columnwidth]{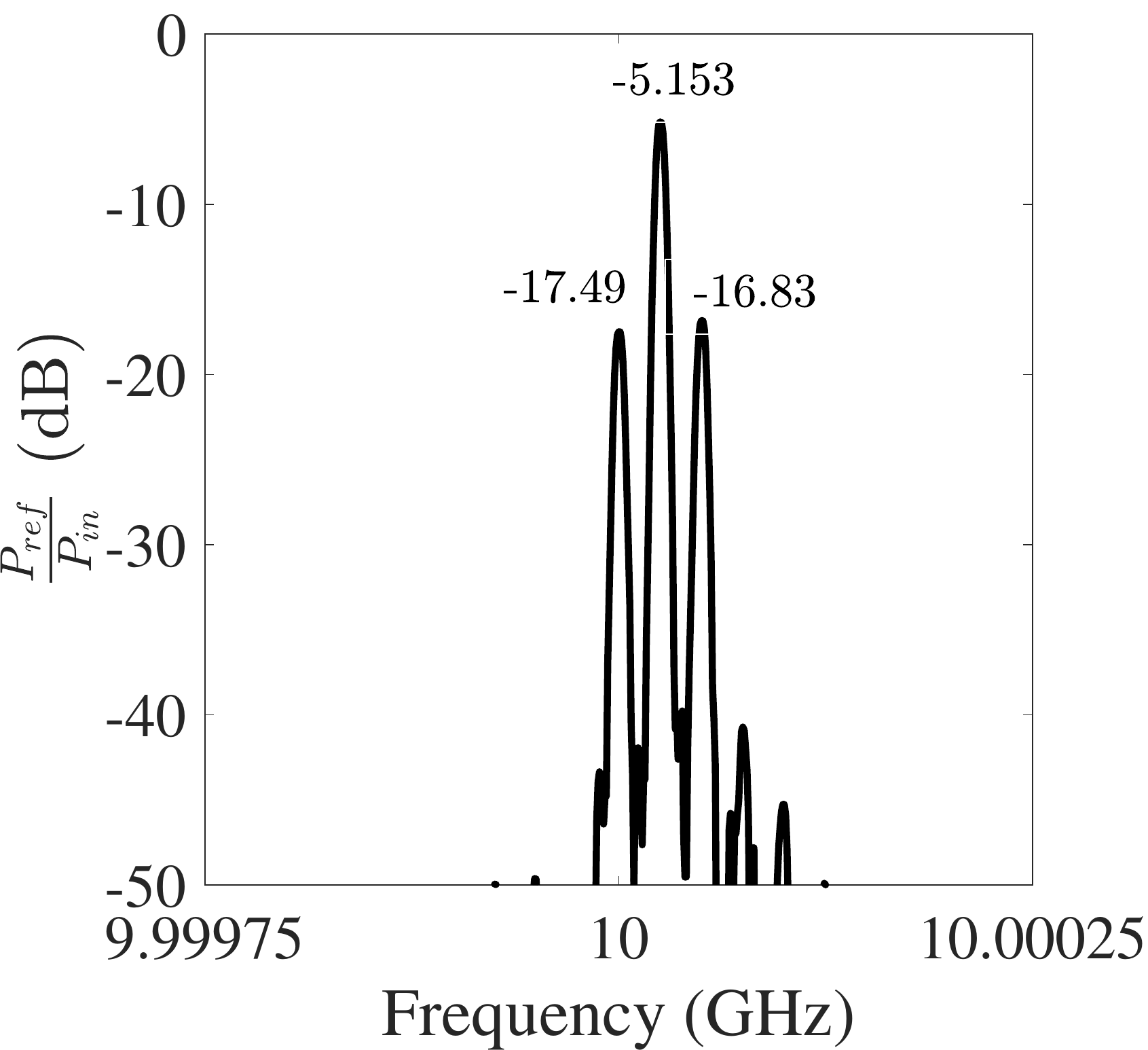}%
}\hfill
\subfloat[\label{sfig:25steeringTMM}]{%
  \includegraphics[clip,width=0.5\columnwidth]{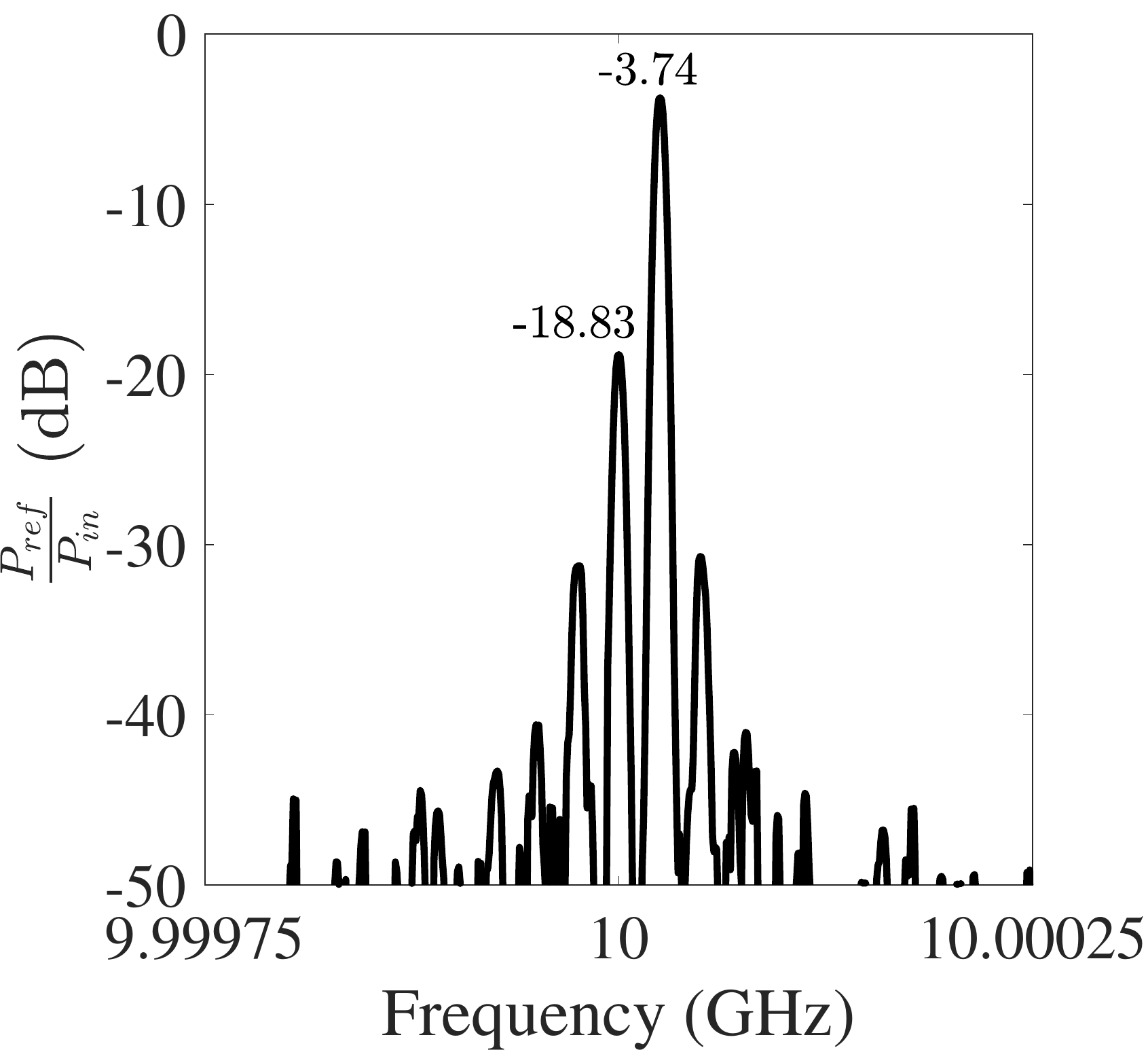}%
}\quad
\caption{Measured reflection spectrum of the SD-TWM metasurface prototype with an incident angle of $25^\circ$.(a) 20-path ($N=20$) modulation for TE polarization. (b) 20-path ($N=20$) modulation for TM polarization.}
\label{fig:25steeringM}
\end{figure}

In this section, the spatial modulation period is chosen to be electrically large ($|k_x\pm \beta_p|<k_0$). The path number is set to $N=20$. 
For the capacitance variation shown in Fig. \ref{sfig:WaveTE_M} and Fig. \ref{sfig:WaveTM_M}, the metasurface acts as a serrodyne frequency translator, that simultaneously deflects the wave to a different angle. When the incident angle is $\theta_1=25^\circ$, the measured reflection spectra at the reflection angle $\theta_2=42^\circ$ is shown in Fig. \ref{fig:25steeringM}. The spectra for both polarizations clearly show a Doppler shift to frequency $f_0+f_p$. The measured conversion loss and sideband suppression for both polarizations are provided in example 3 in Table \ref{tab:meas}.
Note that the reflection angles for harmonics with frequency $f_0$ and $f_0+2f_p$ are $25^\circ$ and $67^\circ$ respectively. We can see from Fig. \ref{fig:25steeringM} that a fraction of the reflected power from these two harmonics was still captured by the finite aperture of the receive antenna due to its relatively close proximity to the metasurface.

\begin{figure}[t]
\subfloat[\label{sfig:N42steeringTEM}]{%
  \includegraphics[clip,width=0.49\columnwidth]{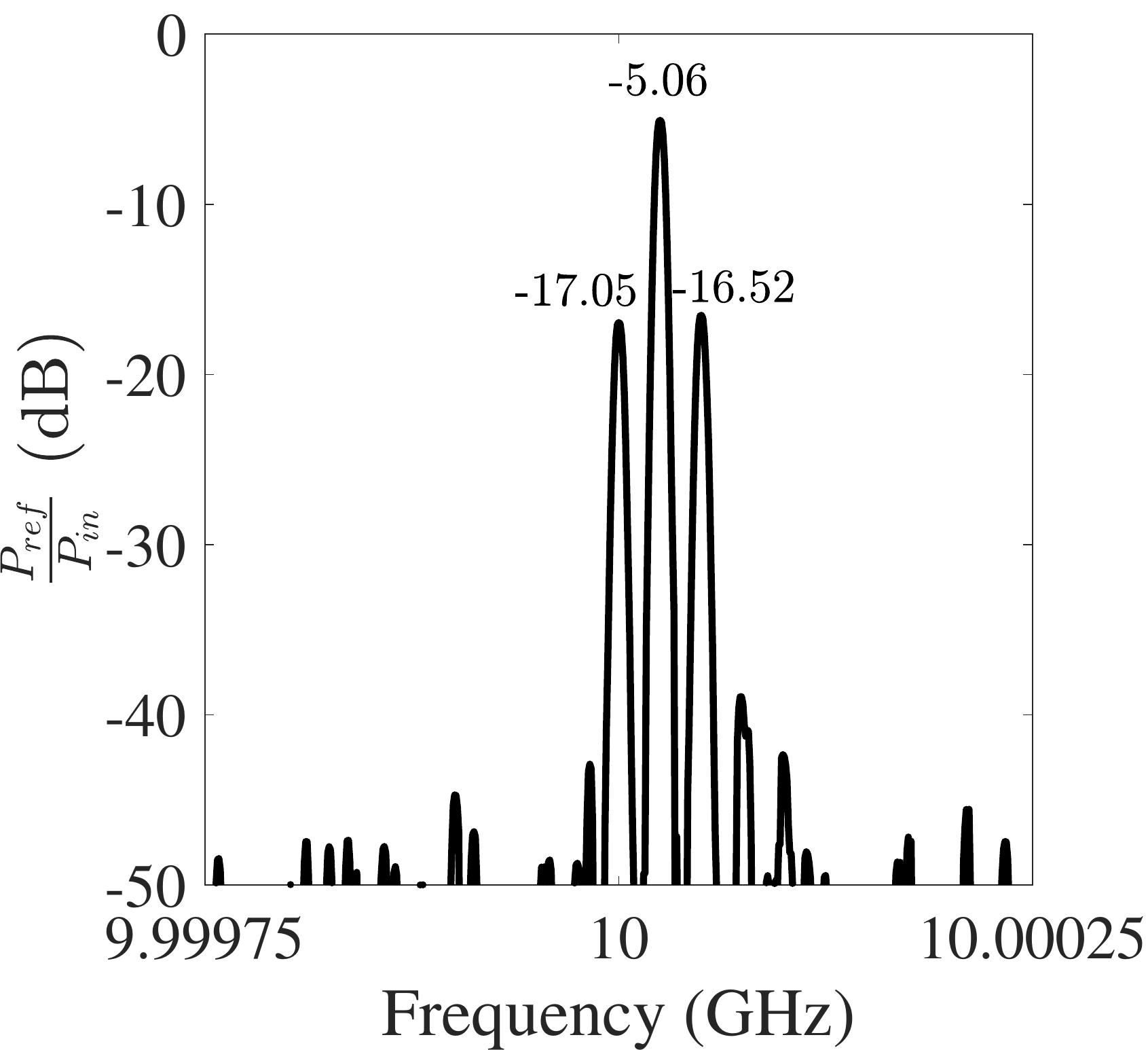}%
}\hfill
\subfloat[\label{sfig:N42steeringTMM}]{%
  \includegraphics[clip,width=0.49\columnwidth]{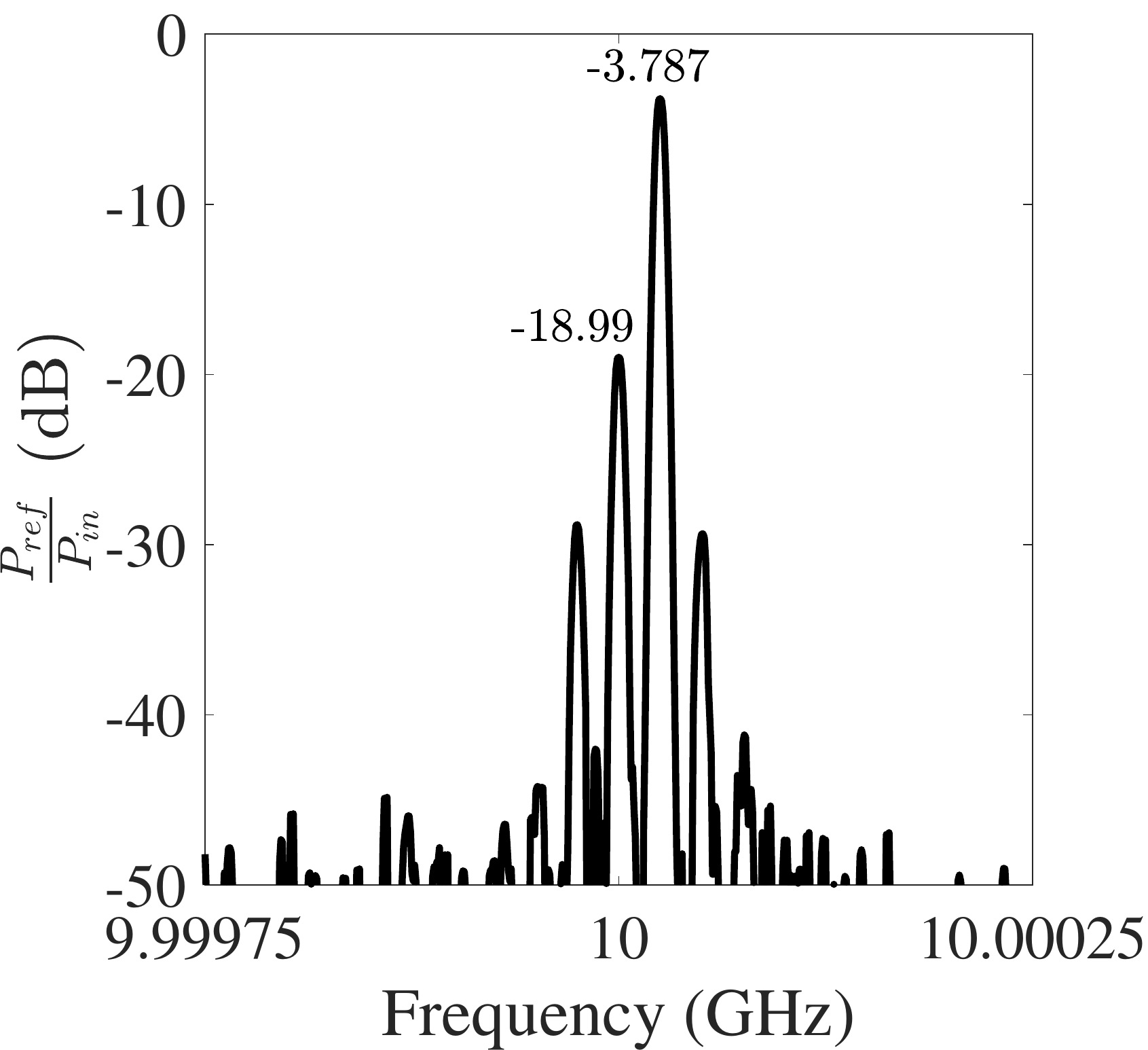}
}
\caption{Measured reflection spectrum of the SD-TWM metasurface prototype with an incident angle of $-42^\circ$. (a) 20-path ($N=20$) modulation for TE polarization. (b) 20-path ($N=20$) modulation for TM polarization.}
\label{fig:N42steeringM}
\end{figure}

By simply interchanging the transmitting and receiving antennas, we can measure the reflection spectra for the case where the incident angle is $\theta_2=-42^\circ$. The measured reflected spectra at the reflection angle of $\theta_1=-25^\circ$ are given in Fig. \ref{fig:N42steeringM}. Again, the spectra for both polarizations clearly show a Doppler shift to a frequency $f_0+f_p$.
The measured conversion loss and sideband suppression for both polarizations are reported in example 4 in Table \ref{tab:meas}.
The harmonics with frequency $f_0$ and $f_0+2f_p$ are reflected at $-42^\circ$ and $9.74^\circ$ respectively, which are also captured by the receiving antenna. Note that, the reflected spectra in Fig. \ref{fig:25steeringM} and Fig. \ref{fig:N42steeringM} are almost identical. This is due to the fact that the modulation frequency is far lower than the signal frequency. Otherwise, the reflection angle would differ from $\theta_1=-25^\circ$, as discussed in Supplemental Material V. 

\subsection{Wavelength-scale spatial modulation period ($|k_x|+|\beta_p|>k_0,\ \&\  ||k_x|-|\beta_p||<k_0 $)} \label{sec:WavePeriod_Fab}

In this section, the spatial modulation period is on the order of the wavelength of radiation ($|k_x|+|\beta_p|>k_0,\ \&\  ||k_x|-|\beta_p||<k_0 $). The path number is chosen to be $N=4$. As in Section \ref{sec:WavePeriod}, the retroreflective angle is chosen to be $\pm 39^\circ$. In experiment, a $3$ dB directional coupler (Omni-spectra 2030-6377-00) was attached to the antenna in order to measure the retroreflected spectra. Note that the modulation waveform on each column is optimized with the same procedure given in Supplemental material VI, for an incident angle of $39^\circ$. In this section, the retroreflective behavior of the metasurface is experimentally verified for the various scenarios introduced in Section \ref{sec:WavePeriod}.

\subsubsection{Measured retroreflective subharmonic frequency translation} \label{sec:meas_retro}

\begin{figure}[t]
\subfloat[\label{sfig:Retro2TEM}]{%
  \includegraphics[clip,width=0.5\columnwidth]{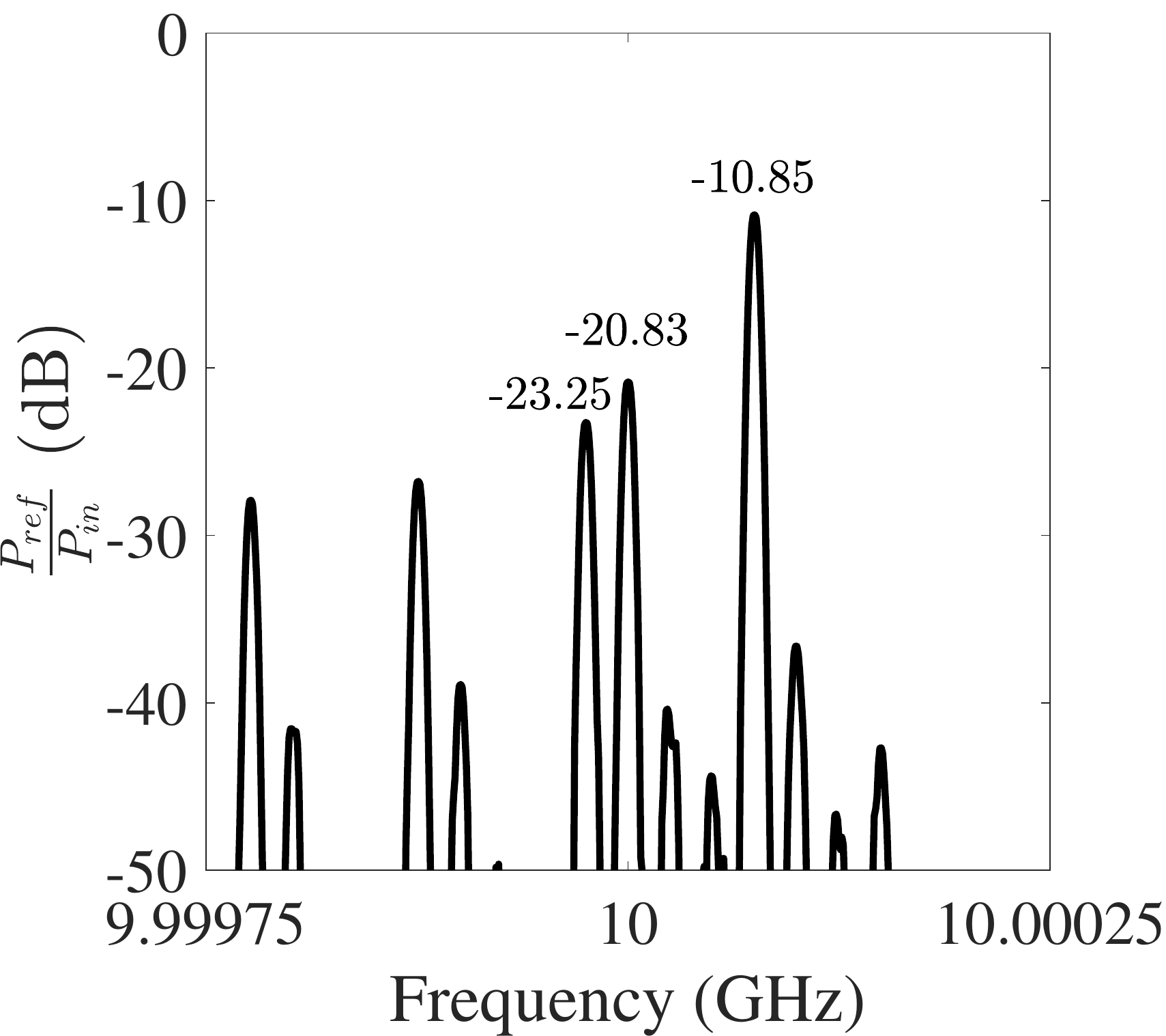}%
}\hfill
\subfloat[\label{sfig:Retro2TEM_N}]{%
  \includegraphics[clip,width=0.5\columnwidth]{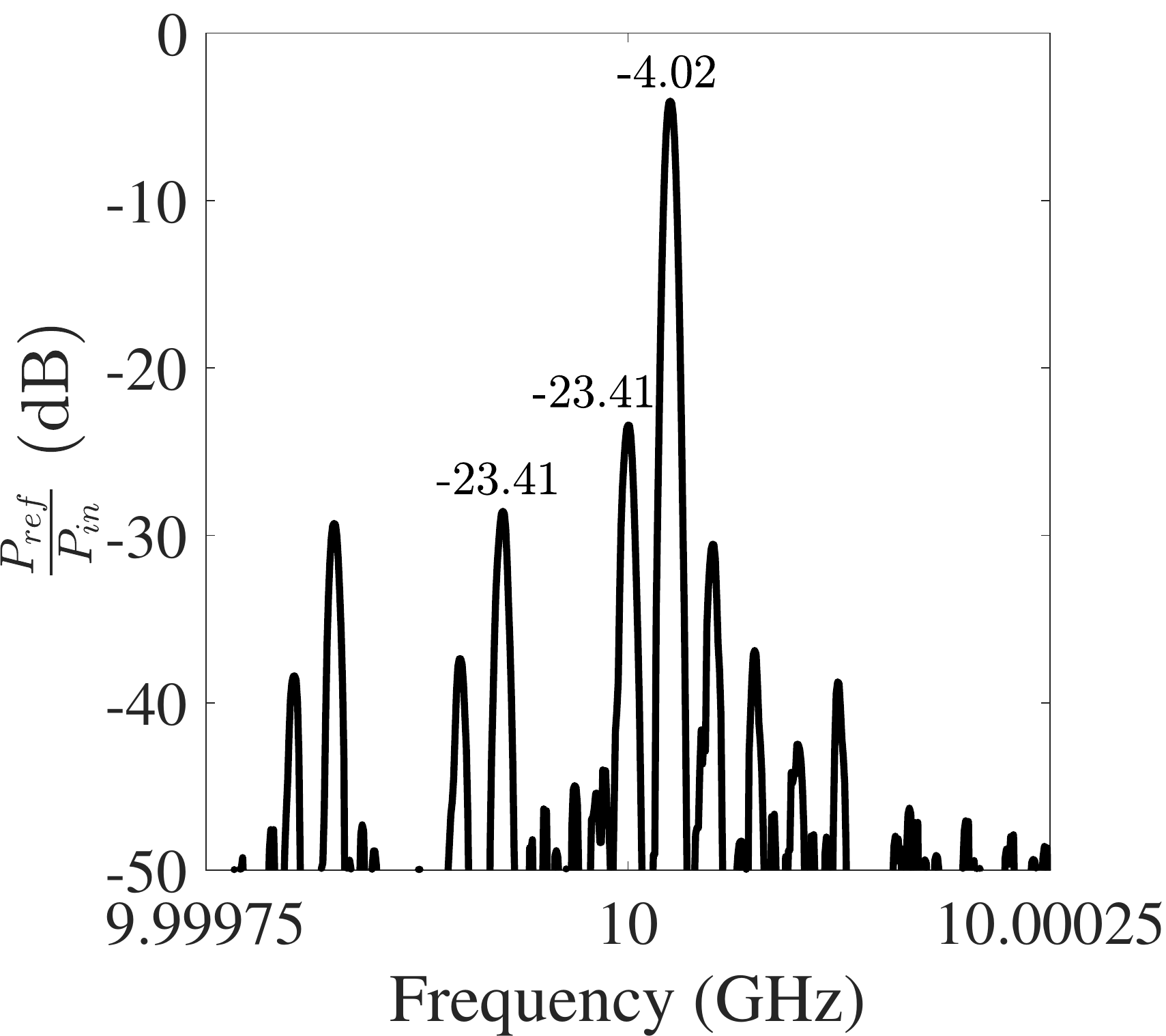}%
}\quad
\subfloat[\label{sfig:Retro2TMM}]{%
  \includegraphics[clip,width=0.5\columnwidth]{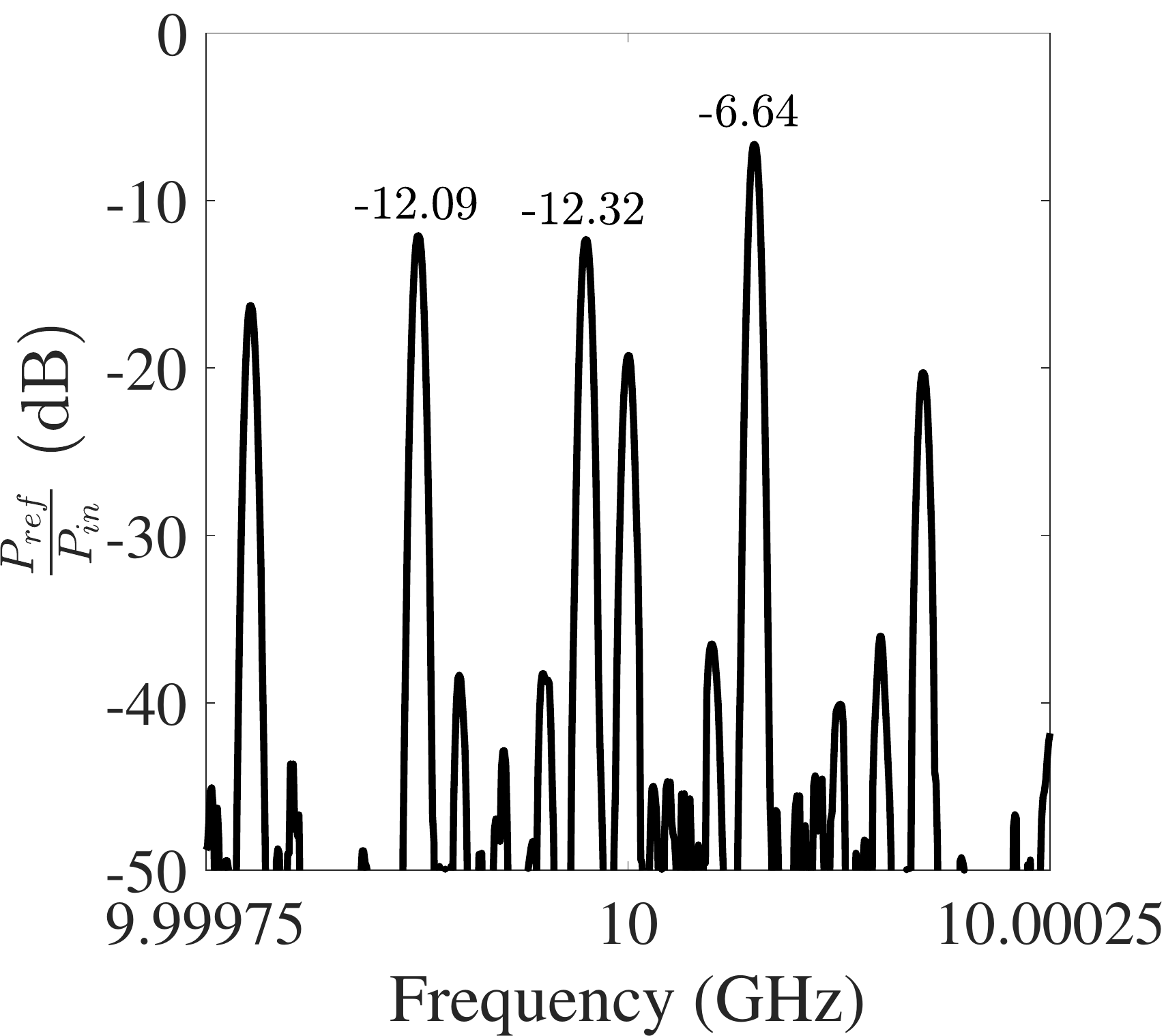}%
}\hfill
\subfloat[\label{sfig:Retro2TMM_N}]{%
  \includegraphics[clip,width=0.5\columnwidth]{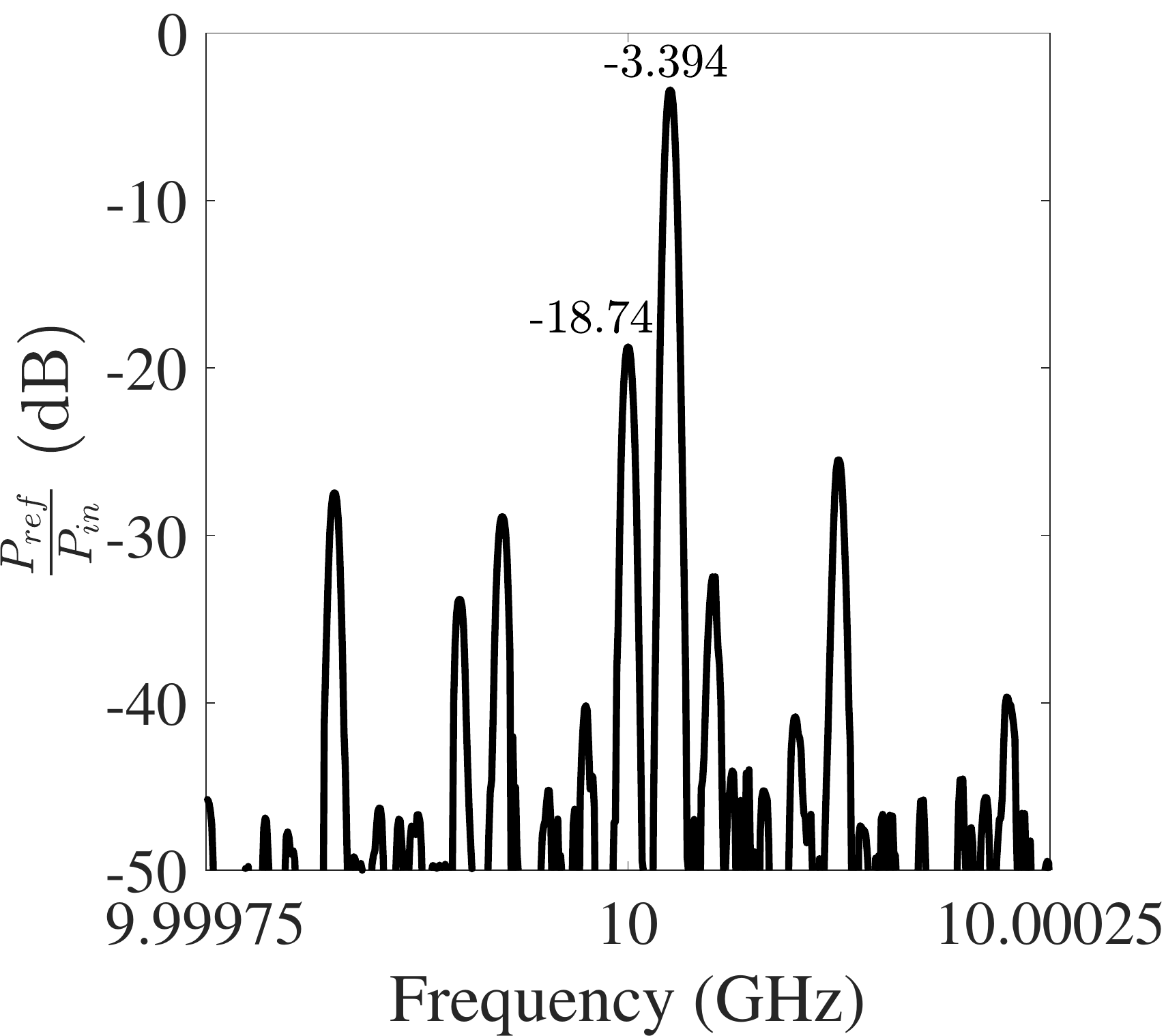}%
}\quad
\caption{Measured retroreflection spectrum of the SD-TWM metasurface prototype. (a) 4-path ($N=4$) modulation for TE polarization for an incident angle of $39^\circ$. (b) 4-path ($N=4$) modulation for TE polarization for an incident angle of $-39^\circ$. (c) 4-path ($N=4$) modulation for TM polarization for an incident angle of $39^\circ$. (d) 4-path ($N=4$) modulation for TM polarization for an incident angle of $-39^\circ$.}
\label{fig:Retro2M}
\end{figure}

The measured retroreflection spectra at an oblique angle of $39^\circ$ are given in Fig. \ref{sfig:Retro2TEM} and \ref{sfig:Retro2TMM} for TE and TM polarization, respectively. As predicted, frequency translation to $f_0+3f_p$ is observed for both polarizations.
The measured conversion loss and sideband suppression for both polarizations are reported in example 5 in Table \ref{tab:meas}. Note that, comparing Fig. \ref{sfig:Retro2TEM} and \ref{sfig:Retro2TMM} to Fig. \ref{sfig:Retro2TEA} and \ref{sfig:Retro2TMA}, only the harmonics in solid lines are captured by the antenna.

For an incident angle of $-39^\circ$, the measured retroreflection spectra are shown in Fig. \ref{sfig:Retro2TEM_N} and \ref{sfig:Retro2TMM_N} for TE and TM polarization, respectively. As predicted, the spectra for both polarizations show a Doppler shift to frequency $f_0+f_p$. The measured conversion loss and sideband suppression for both polarizations are reported in example 6 in Table \ref{tab:meas}. Again, comparing Fig. \ref{sfig:Retro2TEM_N} and \ref{sfig:Retro2TMM_N} to Fig. \ref{sfig:Retro2NTEA} and \ref{sfig:Retro2NTMA}, only the harmonics in solid lines are captured by the antenna.

\subsubsection{Measured retroreflective frequency translation with a staggered sinusoidal reflection phase} \label{sec:meas_retro_sin}

\begin{figure}[t]
\subfloat[\label{sfig:Retro3TEM}]{%
  \includegraphics[clip,width=0.5\columnwidth]{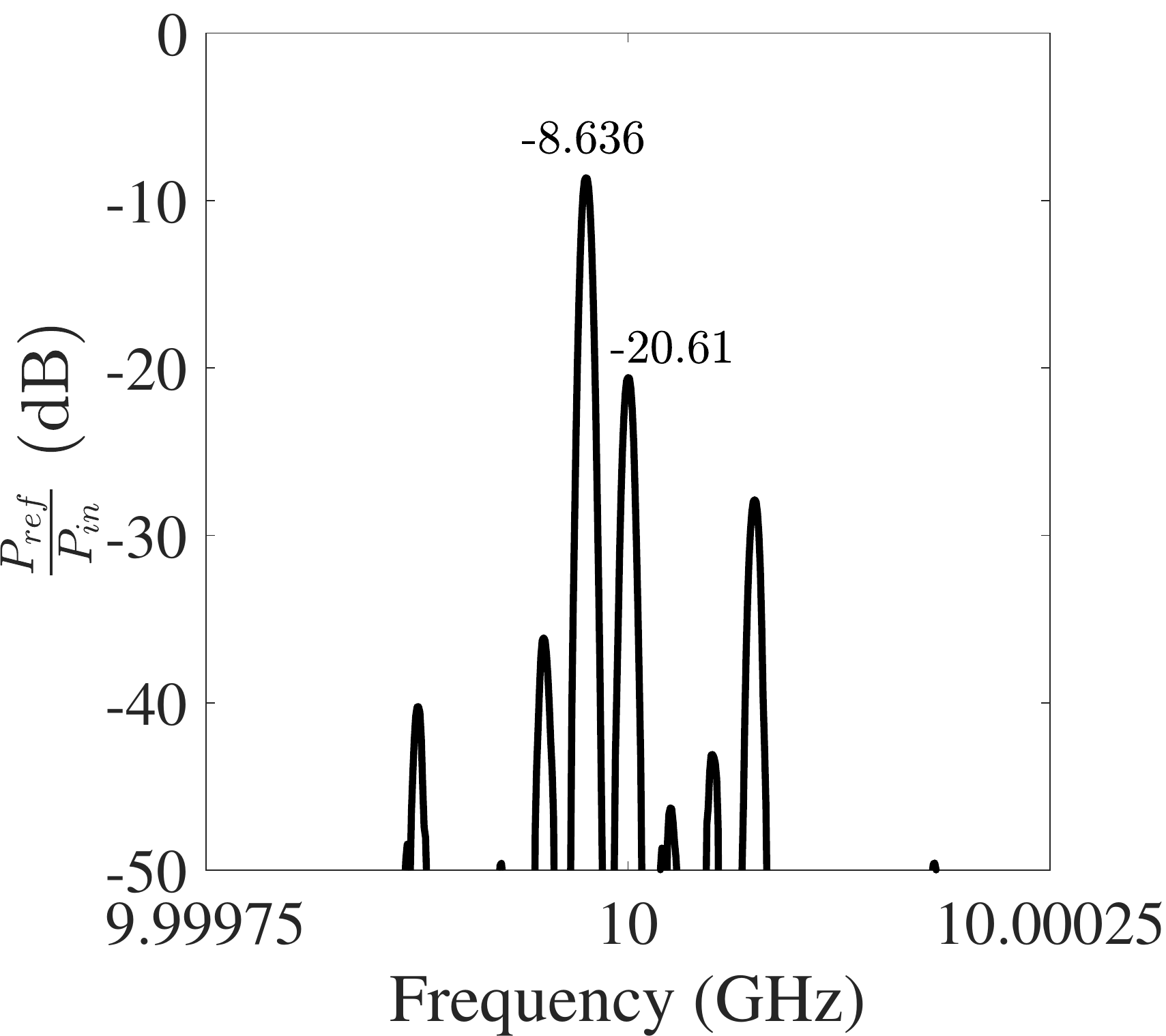}%
}\hfill
\subfloat[\label{sfig:Retro3TEM_N}]{%
  \includegraphics[clip,width=0.5\columnwidth]{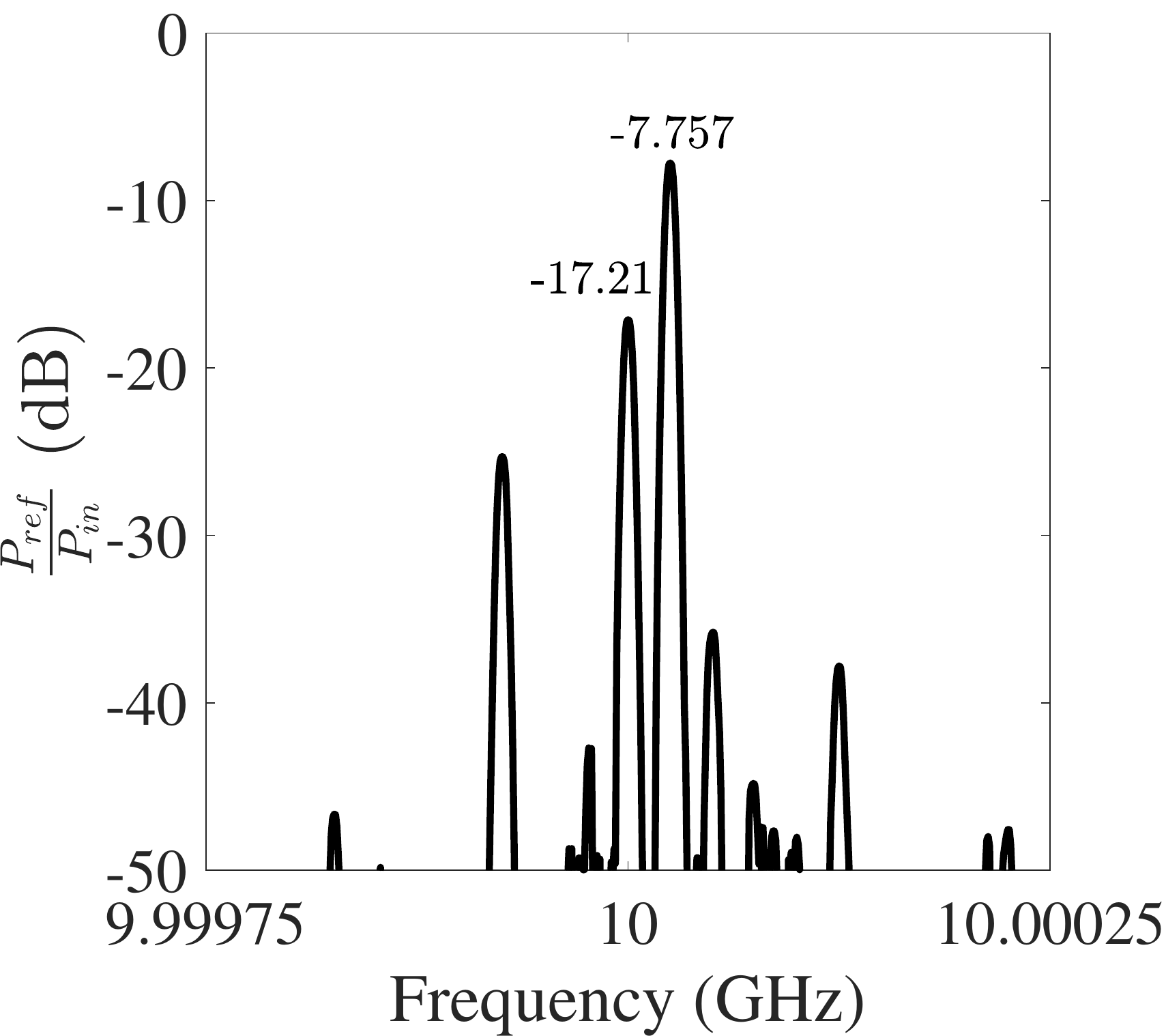}%
}\quad
\subfloat[\label{sfig:Retro3TMM}]{%
  \includegraphics[clip,width=0.5\columnwidth]{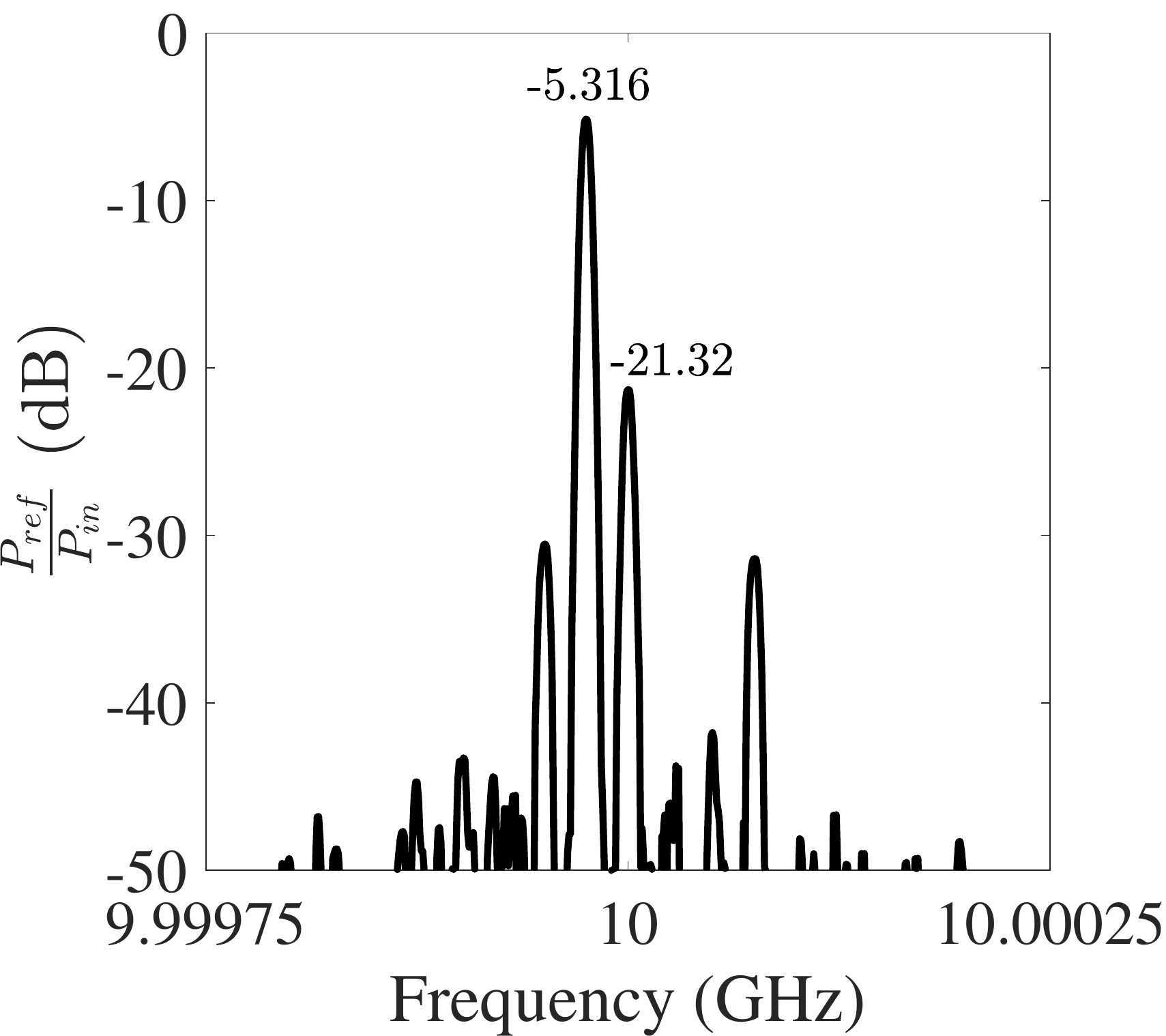}%
}\hfill
\subfloat[\label{sfig:Retro3TMM_N}]{%
  \includegraphics[clip,width=0.5\columnwidth]{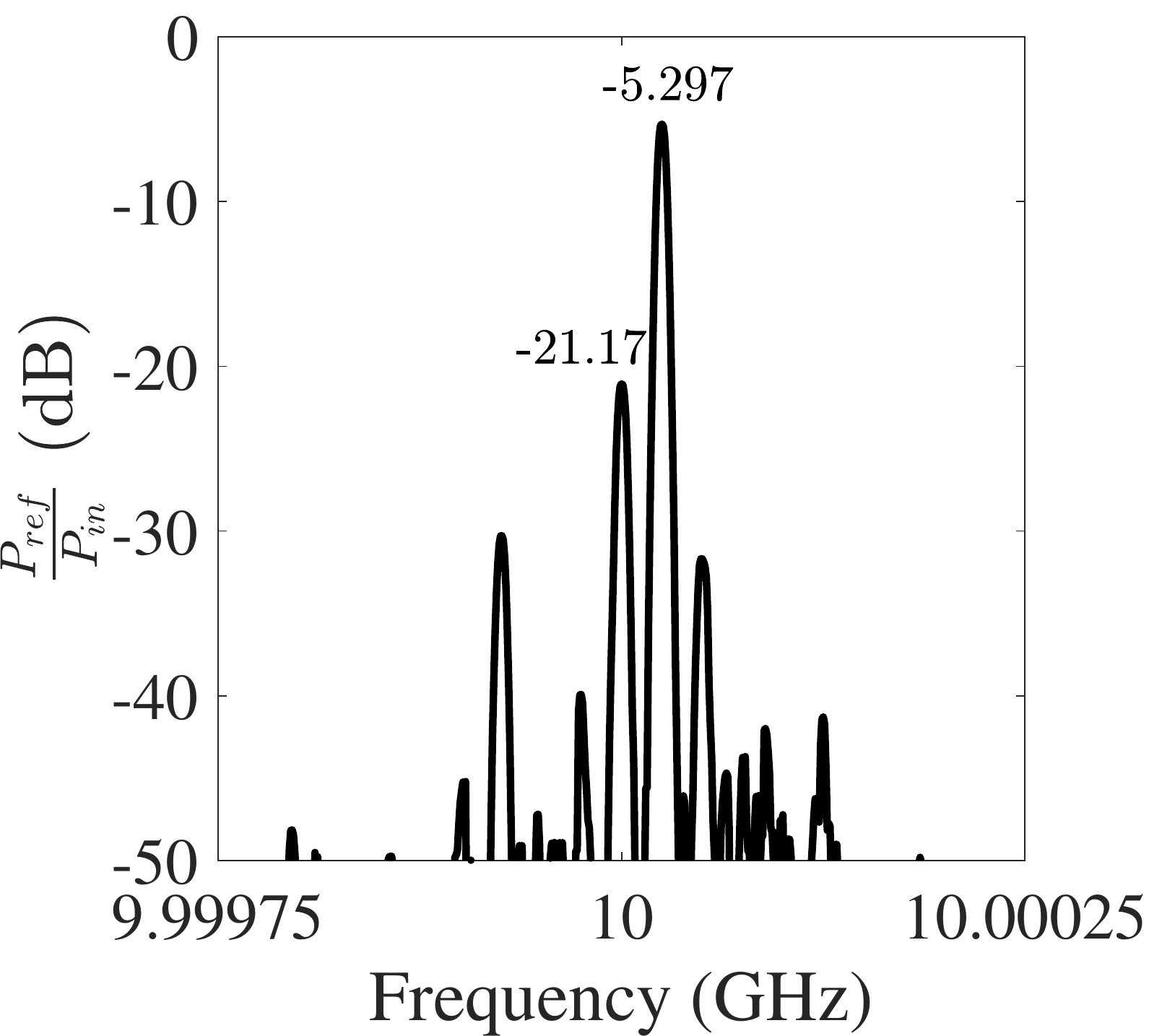}%
}\hfill
\caption{Measured retroreflection spectrum of the SD-TWM metasurface prototype. The bias waveforms generates a sinusoidal reflection phase. (a) 4-path ($N=4$) modulation for TE polarization for an incident angle of $39^\circ$. (b) 4-path ($N=4$) modulation for TE polarization for an incident angle of $-39^\circ$. (c) 4-path ($N=4$) modulation for TM polarization for an incident angle of $39^\circ$. (d) 4-path ($N=4$) modulation for TM polarization for an incident angle of $-39^\circ$.}
\label{fig:Retro3M}
\end{figure}

In this section, the bias waveforms on adjacent columns generate staggered sinusoidal reflection phases. The measured retroreflection spectra at an oblique angle of $39^\circ$ are given in Fig. \ref{sfig:Retro3TEM} and \ref{sfig:Retro3TMM}. As predicted, frequency translation to $f_0-f_p$ is observed for both polarizations.
The measured conversion loss and sideband suppression for both polarizations are provided in example 7 of Table \ref{tab:meas}. The measured retroreflection spectra at an oblique angle of $-39^\circ$ are given in Fig. \ref{sfig:Retro3TEM_N} and \ref{sfig:Retro3TMM_N}. Frequency translation to $f_0+f_p$ is observed for both polarizations.
The measured conversion loss and sideband suppression for both polarizations are provided in example 8 of Table \ref{tab:meas}.

Note that, both of the two retroreflection cases used $4$ paths per spatial modulation period for a retroreflection angle of $39^\circ$ (examples 5 and 7 of Table \ref{tab:meas}). The only difference between the two cases was the time-dependence of the reflection phase (sawtooth versus sinusoidal).
Thus, we have shown that simply changing the temporal modulation waveform of the stixels, the retroreflection frequency can be changed. In this case, it changed from $f_0-f_p$ to $f_0+3f_p$.

\section{Conclusion}

This work theoretically and experomentally studies metasurfaces with spatially-discrete, traveling-wave modulation (SD-TWM). A modified Floquet analysis of the SD-TWM metasurface was presented based on a new boundary condition that has been derived for SD-TWM structures. The analysis allows the separation of the macroscopic and microscopic spatial variations of the field on the metasurface. The macroscopic spatial variation is enforced by the modulation wavelength, while the microscopic spatial variation accounts for the field variation across each constituent stixel. The analysis presented in this paper provides an accurate model of the metasurface and serves as a guide for designing practical space-time metasurfaces. Most importantly, the analysis provides physical insight into SD-TWM metasurfaces and reveals various new phenomena including subharmonic frequency translation in specular or deflected directions. Such phenomenon is not possible with the continuous traveling-wave modulation commonly assumed in previous studies. 

In this paper, the operation of an SD-TWM metasurface is categorized into three regimes. When the spatial modulation period is electrically large, the metasurface allows simultaneous frequency translation and angular deflection. Tuning the spatial modulation period allows the metasurface to steer the reflected beam, and even exhibit retroreflection. When the spatial modulation period is electrically small, the metasurface exhibits subharmonic frequency translation. In this case, all the radiated harmonics are reflected in the specular direction for low frequencies of modulation. When the spatial modulation period is on the order of a wavelength, subharmonic frequency translation and deflection can be achieved. It is also shown that the frequency of the deflected wave can be changed by modifying the temporal modulation waveform.

To verify the analysis, a proof-of-principle metasurface was designed and fabricated at X-band frequencies. Measurements show good agreement with theoretical predictions in the three operating regimes of the metasurface. The designed metasurface provides a new level of reconfigurability. Multiple functions including beamsteering, retroreflection, serrodyne frequency translation, and subharmonic frequency translation are achieved with the ultra-thin ($0.06\lambda$) metasurface by appropriately tailoring the SD-TWM waveform. The designed metasurface could find various applications in next-generation communication, imaging and radar systems.

\begin{acknowledgments}
This work was supported by DSO National Laboratories under contract DSOCO15027, and the %Air Force Office of Scientific Research (AFOSR) under grant FA9550-15-1-0101 and the 
AFOSR MURI program FA9550-18-1-0379.
\end{acknowledgments}

%\bigbreak

%\appendix

%\clearpage

%\subsection{\label{app:subsec}A subsection in an appendix}

\nocite{*}
\bibliography{NpathMetasurface}
%\clearpage
%\bibliography{XQ10356}% Produces the bibliography via BibTeX.

\end{document}